\documentclass{emulateapj}
\usepackage{rotating}
\newcommand{\mbh}{$M_{\rm BH}$}
\newcommand{\mss}{$M_{\rm sph, \star}$}
\newcommand{\ms}{$M_{\rm sph}$}
\newcommand{\ls}{$L_{\rm sph}$}

\newcommand{\msv}{$M_{\rm sph, dyn}$}
\newcommand{\s}{$\sigma$}

\begin{document}

\title{A LOCAL BASELINE OF THE BLACK HOLE MASS SCALING RELATIONS FOR ACTIVE GALAXIES. I.\\
METHODOLOGY AND RESULTS OF PILOT STUDY}

\shorttitle{A LOCAL BASELINE OF THE \mbh~SCALING RELATIONS FOR ACTIVE GALAXIES. I.}
\shortauthors{Bennert et al.}

\author{Vardha Nicola Bennert\altaffilmark{1},
Matthew W. Auger\altaffilmark{1}, 
Tommaso Treu\altaffilmark{1,2}, 
Jong-Hak Woo\altaffilmark{3}, 
Matthew A. Malkan\altaffilmark{4}}

\altaffiltext{1}{Department of Physics, University of California, Santa
Barbara, CA 93106-9530; bennert@physics.ucsb.edu, mauger@physics.ucsb.edu, tt@physics.ucsb.edu}

\altaffiltext{2}{Sloan Fellow, Packard Fellow}

\altaffiltext{3}{Department of Astronomy, Seoul National University, Korea; woo@astro.snu.ac.kr}

\altaffiltext{4}{Department of Physics and Astronomy, University of California,
Los Angeles, CA 90095-1547; malkan@astro.ucla.edu}

\shortauthors{Bennert et al.}

\begin{abstract}
We present high-quality Keck/LRIS longslit spectroscopy of a pilot
sample of 25 local active galaxies selected from the SDSS (0.02 $\le$
$z$ $\le$ 0.1; \mbh$>10^{7}$M$_{\odot}$) to study the
relations between black hole mass (\mbh) and host-galaxy properties.
We determine stellar kinematics of the host galaxy, deriving
stellar-velocity dispersion profiles and rotation curves from three
spectral regions (including CaH\&K, MgIb
triplet, and CaII triplet).  In addition, we perform surface photometry
on SDSS images, using a newly developed code for joint
multi-band analysis. BH masses are estimated from the width of the H$\beta$
emission line and the host-galaxy free 5100\AA~AGN luminosity.
Combining results from spectroscopy and imaging allows us to study
four \mbh~scaling relations: \mbh-\s, \mbh-\ls,
\mbh-\mss, and \mbh-\msv. 
We find the following results. First,
stellar-velocity dispersions determined from aperture
spectra (e.g.~SDSS fiber spectra or unresolved data from distant
galaxies) can be biased, depending on aperture size, AGN contamination, and host-galaxy morphology. 
However, such a bias cannot
explain the offset seen in the \mbh-\s~relation at higher redshifts. 
Second, while the CaT region is the cleanest
to determine stellar-velocity dispersions, both the MgIb region,
corrected for FeII emission, and the 
CaHK region, although often swamped by the
AGN powerlaw continuum and emission lines, 
can give results accurate to within a few percent. 
Third, the \mbh~scaling relations of our pilot sample
agree in slope and scatter with those of other local active and
inactive galaxies.  In the next papers of the
series we will quantify the scaling relations, exploiting
the full sample of $\sim100$ objects.
\end{abstract}


\section{INTRODUCTION}
\label{sec:intro}
The empirical relations
between the mass of the central supermassive black hole (BH)
and the properties of the spheroid
(ellipticals and classical bulges of spirals)
such as stellar-velocity dispersion $\sigma$
\citep{fer00,geb00}, stellar mass \citep[e.g.,][]{mar03}, 
and luminosity \citep[e.g.,][]{har04}
discovered in the local Universe
have been interpreted
as an indication of a close connection between
the growth of the BH and the formation and evolution of galaxies
\citep[e.g.,][]{kau00,vol03,cio07,hop07,dim08,hop09}.  
In this framework, active galactic nuclei (AGNs)
are thought to represent a stage
in the evolution of galaxies in which
the supermassive BH
is actively growing through accretion.

To understand the origin of the BH mass (\mbh) scaling relations,
our group has been studying their evolution with cosmic time
\citep{tre04,woo06,tre07,woo08,ben10}. 
To distinguish mechanisms causing evolution in $\sigma$ (e.g.,
dissipational merger events) and \ls~(e.g., through
passive evolution due to aging of the stellar population,
or dissipationless mergers), we simultaneously study both the
\mbh-$\sigma$ and \mbh-\ls~relations for a sample of low-luminous AGNs, Seyfert-1 galaxies,
at $z \simeq 0.4$ and 0.6 (lookback time 4-6 Gyrs).
Our results reveal an offset with
respect to the local relationships which cannot be accounted for by
known systematic uncertainties. 
The evolutionary trend we find \citep[e.g., \mbh/\ls $\propto$ $(1+z)^{1.4 \pm 0.2}$, including
selection effects;][]{ben10}
suggests that BH growth
precedes spheroid assembly. Several other studies 
have found results in
qualitative agreement with ours, over different ranges in black hole
mass and redshifts, and with different observing techniques 
\citep[e.g.,][]{wal04,shi06,mcl06,pen06b,sal07,wei07,rie08,rie09,gu09,jah10,dec10,mer10}.

However, to study the evolution of the \mbh~scaling
relations, it is crucial to
understand slope and scatter of the local relations.
In particular, an open question is whether quiescent
galaxies and active galaxies follow the same relations,
as expected if the nuclear activity was 
just a transient phase in the life-cycle of galaxies.
Recently, \citet{woo10} presented the 
\mbh-\s~relation for a sample of 24 active galaxies
in the local Universe,
for which the BH mass was derived via reverberation mapping (RM)
\citep[e.g.,][]{wan99, kas00, kas05, ben06, ben09b}.
They find a slope ($\beta=3.55\pm0.60$) and intrinsic scatter 
($\sigma_{\rm int}=0.43\pm0.08$)
which are indistinguishable from that of quiescent galaxies 
\citep[e.g.,][]{fer05,gue09} within the uncertainties,
supporting the scenario in which active galaxies are an evolutionary
stage in the life cycle of galaxies.

While the great advantage of such a study is the 
multi-epoch data which provide more reliable measurements of the BH mass,
such a quality comes at the expense of quantity.
Studies based on larger samples drawn from the Sloan Digital Sky Survey (SDSS)
infer the BH mass indirectly from single-epoch spectra \citep[e.g.,][]{gre06a,shen08}.
They hinted at a shallower \mbh-\s~relation 
than that observed for quiescent samples, but the available
dynamic range is too small to be conclusive.
In particular, the relation 
above 10$^{7.5}$ M$_{\odot}$ is very poorly known,
which has profound implications for evolutionary studies 
that by necessity focus on this mass range.

However, there is another uncertainty in the \mbh-\s~relation
that arises when measuring $\sigma$ 
from fiber-based SDSS data \citep[e.g.,][]{gre06a,shen08} and also
 from the unresolved ``aperture spectra'' for more distant galaxies
\citep[e.g., our studies on the evolution of 
the \mbh-\s~relation;][J.-H. Woo et al. 2011, in preparation]{woo06,woo08}.
Local active galaxies seem to span a range of morphologies
\citep[e.g.,][]{mal98,hun04,kim08,ben09a}
and a significant fraction  ($>$15/40) of 
our distant sample of Seyfert-1 galaxies
have morphologies of Sa or later \citep{ben10}.
Given the diversity of morphologies of AGN hosts, it is most likely that
there is a degree of rotational support:
If the disk is seen edge-on, the disk rotation can bias $\sigma$
towards higher values. However, since the
disk is kinematically cold, it can also result in the opposite effect,
i.e.~biassing $\sigma$ towards smaller values, if the disk is seen face on
\citep[e.g.,][]{woo06}.
Either way, it questions the connection between the ``global''
dispersion measured by those experiments and the spheroid-only dispersion
which may in fact scale more tightly with BH mass. 

More generally, measuring $\sigma$ in type-1 AGNs
is complicated by the presence of strong emission lines
and a continuum that dilutes the starlight.
$\sigma$ can be measured from 
different spectral regions with different merits and
complications \citep[][for inactive galaxies see also \citealt{bar02}]{gre06b}.
Finally, there is are different $\sigma$ measurements in use:
e.g., the luminosity-weighted line-of-sight velocity dispersion within the spheroid effective radius 
\citep[$\sigma_{\rm reff}$, e.g.,][]{geb00,geb03},
and the central velocity dispersions normalized to an aperture
of radius equal to 1/8 of the galaxy effective radius
\citep[$\sigma_{\rm 1/8 reff}$, e.g.,][]{fer00,fer05}.

Shedding light on the issues outlined above is essential to
understand what aspects of galaxy formation and AGN activity are
connected, but it requires spatially-resolved kinematic information for
a large sample of local AGNs. For this purpose,
we selected a sample of $\sim$100 local 
(0.02 $\le$ $z$ $\le$ 0.1) Seyfert-1 galaxies from the SDSS (DR6)
with \mbh$>10^{7}$M$_{\odot}$ and obtained
high-quality longslit spectra with Keck/LRIS.
From the Keck spectra, we derive
the BH mass and measure the spatially-resolved stellar-velocity dispersion
from three different spectral regions
(around CaH+K\,$\lambda\lambda$3969,3934\AA;
around MgIb $\lambda\lambda\lambda$5167,5173,5184\AA~triplet;
and around Ca II$\lambda\lambda\lambda$8498,8542,8662\AA~triplet).
The spectra are complemented by archival SDSS images (g', r', i', z')
on which we perform surface photometry using a newly developed code
to determine the spheroid effective radius, spheroid luminosity,
and the host-galaxy free 5100\AA~luminosity of the AGN
(for an accurate BH mass measurement).
Our code allows a joint multi-band analysis
to disentangle the AGN which dominates
in the blue from the host galaxy that dominates
in the red.
The resulting multi-filter spheroid luminosities 
allow us to estimate spheroid stellar masses.

Combining the results from spectroscopic and imaging
analysis, we can study four different BH mass scaling relations
(namely \mbh-\s, \mbh-\ls, \mbh-\mss, and \mbh-\msv).
In this paper, we focus on the methodology and present
results for a pilot sample of 25 objects.
The full sample will be discussed in the upcoming papers of this series.
The paper is organized as follows.
We summarize sample selection and sample
properties in \S~\ref{sec:sample}; 
observations and data reduction in 
\S~\ref{sec:obs}. \S~\ref{sec:quan} describes the derived quantities,
such as spatially-resolved stellar-velocity dispersion and velocity,
aperture stellar-velocity dispersion, BH masses, surface photometry,
and  spheroid masses. 
In \S~\ref{sec:comp}, we describe comparison samples
drawn from literature, consisting of local inactive and
active galaxies.
We present and discuss our results in
\S~\ref{sec:res}, including the BH mass scaling relations.
We conclude with a summary in \S~\ref{sec:sum}. 
In Appendix~\ref{sec:mattscode}, we describe the details of
a python-based code developed by us to determine
surface-photometry from multi-filter SDSS images. 
Throughout the paper, we assume
a Hubble constant of $H_0$ = 70\,km\,s$^{-1}$\,Mpc$^{-1}$,
$\Omega_{\Lambda}$ = 0.7 and $\Omega_{\rm M}$ = 0.3. 

\section{SAMPLE SELECTION}
\label{sec:sample}
\begin{figure*}[ht!]
\begin{center}
\includegraphics[scale=0.7]{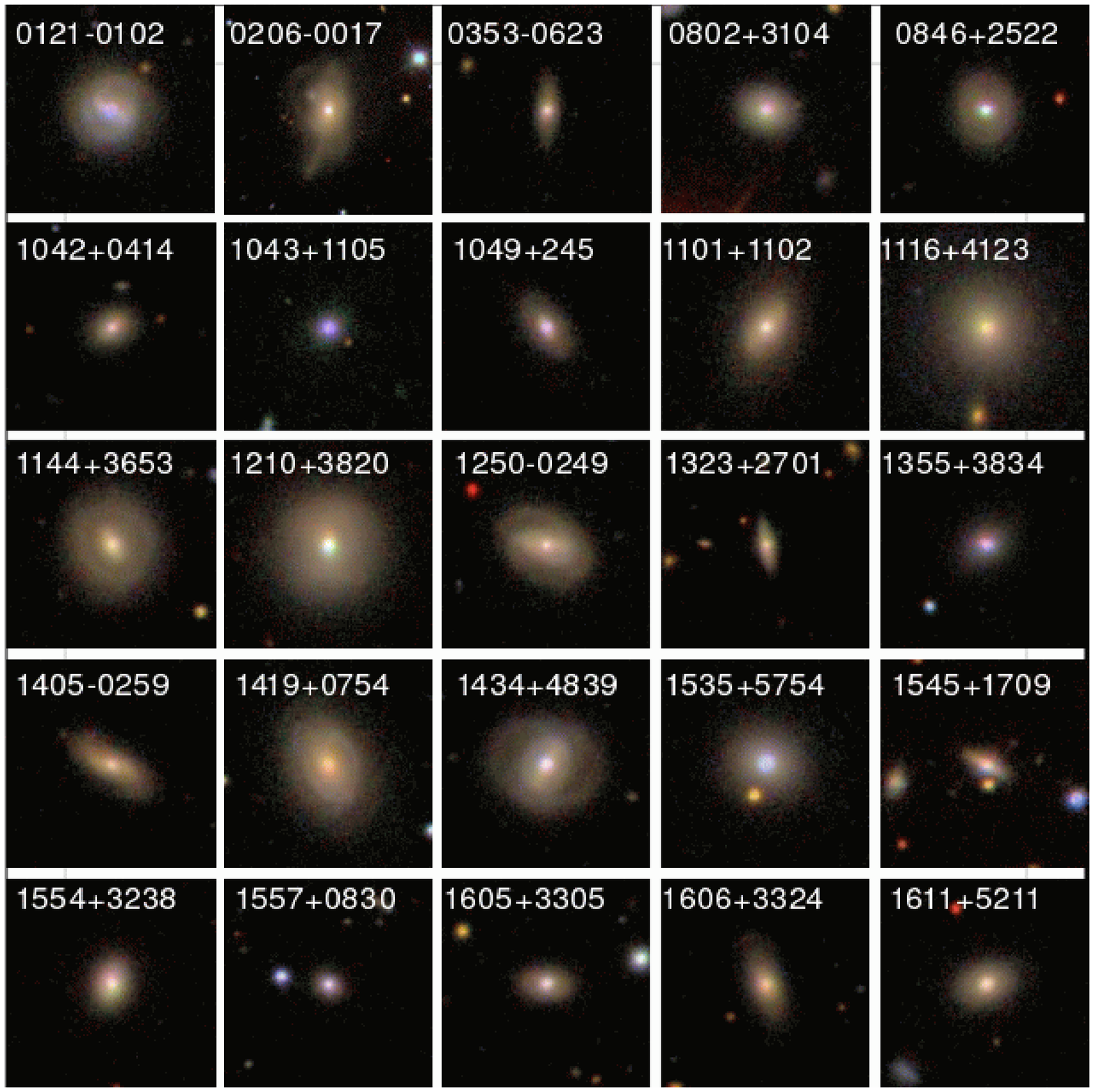}
\end{center}
\caption{Postage stamp SDSS-DR7 multi-filter images. 
North is up, east is to the left. The size of the field-of-view is 50\arcsec$\times$50\arcsec~(corresponding to
$\sim$20$\times$20kpc up to $\sim$76$\times$76kpc for the redshift range covered by the objects
presented here). For
0206-0017, we show the central 100\arcsec$\times$100\arcsec.}
\label{sdss}
\end{figure*}

Making use of the SDSS DR6 data release,
we selected type-1 AGNs with \mbh$>10^{7}$M$_{\odot}$, 
as estimated from the spectra
based on their optical luminosity and H$\beta$ full-width-at-half-maximum 
(FWHM) \citep{mcg08}.
We restricted the redshift range to $0.02-0.1$ to ensure that both the
Ca triplet and a bluer wavelength region are accessible to measure
stellar kinematics and that the objects are well resolved.
This results in a list of 332 objects from which targets were selected
based on visibility during the assigned Keck observing time.
Moreover, we visually inspected all spectra to make sure
that the BH mass measurement is reliable and that
there are no spurious outliers lacking broad emission lines
($\sim$5\% of the objects).
A total of 111 objects were observed with Keck between January 2009 and
March 2010. 
Here, we present the methodology of our approach and the results for our pilot sample of
25 objects. Their properties are summarized
in Table~\ref{sample}. Fig.~\ref{sdss} shows postage stamp SDSS-DR7 images.
The results for the full sample will be presented in the forthcoming
papers of this series.

All 25 objects were covered by the
VLA FIRST (Faint Images of the Radio Sky at Twenty-cm) survey\footnote{See 
VizieR Online Data Catalog, 8071 (R. H. Becker et al., 2003)}, but only
10 have detected counterparts within a radius of 5\arcsec.
Out of these 10, seven are listed in \citet{li08} with only one being
radio-loud. Thus, the majority ($>$$\sim$85\%) of our objects are radio-quiet.
Note that none of the objects has HST images available.
As our sample was selected from the SDSS, most objects are included
in studies that discuss the local BH mass function
\citep{gre07} or BH fundamental plane \citep{li08}.
In addition, 1535+5754 (Mrk\,290) has a reverberation-mapped BH mass from
\citet{den10}. We will compare the BH masses derived
in these studies with ours when we present the full sample.
For a total of eight objects, stellar-velocity dispersion measurements
from aperture spectra exist in the literature
(mainly derived from SDSS fiber data: six in \citealt{gre06a} and five in
\citealt{shen08}, with three overlapping, and one object in \citealt{nel96} 
determined from independent spectra, but included in both SDSS studies).
We briefly compare the stellar-velocity dispersions derived in these studies
with ours in \S~\ref{subsec:ap}, 
but will get back to it in more detail 
when we present the full sample.

\begin{deluxetable*}{lcccccccc}
\tabletypesize{\scriptsize}
\tablecolumns{9}
\tablewidth{0pc}
\tablecaption{Sample Properties}
\tablehead{
\colhead{Object} & \colhead{SDSS Name} & \colhead{$z$} & \colhead{$D_{\rm L}$} & \colhead{Scale} & \colhead{RA (J2000)} & \colhead{DEC (J2000)} 
& \colhead{$i^\prime$} & \colhead{Alternative Name(s)}\\
& & & (Mpc) & (kpc/arcsec) & & & (mag) \\
\colhead{(1)} & \colhead{(2)} & \colhead{(3)}  & \colhead{(4)} & \colhead{(5)} & \colhead{(6)}  & \colhead{(7)} & \colhead{(8)} & \colhead{(9)}}
\startdata
0121-0102 & SDSSJ012159.81-010224.4 & 0.0540  & 240.8 & 1.05  &  01 21 59.81  & -01 02 24.4  & 14.32 & Mrk\,1503	       \\ %L11    
0206-0017 & SDSSJ020615.98-001729.1 & 0.0430  & 190.2 & 0.85  &  02 06 15.98  & -00 17 29.1  & 13.24 & Mrk\,1018, UGC\,01597	       \\ %L2	       
0353-0623 & SDSSJ035301.02-062326.3 & 0.0760  & 344.1 & 1.44  &  03 53 01.02  & -06 23 26.3  & 16.10 &   \\ %L6        
0802+3104 & SDSSJ080243.40+310403.3 & 0.0410  & 181.1 & 0.81  &  08 02 43.40  & +31 04 03.3  & 15.06 &   \\ %L1        
0846+2522 & SDSSJ084654.09+252212.3 & 0.0510  & 226.9 & 1.00  &  08 46 54.09  & +25 22 12.3  & 15.16 &   \\ %L4        
1042+0414 & SDSSJ104252.94+041441.1 & 0.0524  & 233.4 & 1.02  &  10 42 52.94  & +04 14 41.1  & 15.82 &   \\ %L32    
1043+1105 & SDSSJ104326.47+110524.2 & 0.0475  & 210.8 & 0.93  &  10 43 26.47  & +11 05 24.3  & 16.06 &  		       \\ %L33       
1049+2451 & SDSSJ104925.39+245123.7 & 0.0550  & 245.4 & 1.07  &  10 49 25.39  & +24 51 23.7  & 15.52 &   \\ %L34    
1101+1102 & SDSSJ110101.78+110248.8 & 0.0355  & 156.2 & 0.71  &  11 01 01.78  & +11 02 48.8  & 14.67 & MRK\,728 	       \\ %L35      
1116+4123 & SDSSJ111607.65+412353.2 & 0.0210  & 91.4  & 0.43  &  11 16 07.65  & +41 23 53.2  & 14.08 & UGC\,06285	       \\ %L13         
1144+3653 & SDSSJ114429.88+365308.5 & 0.0380  & 167.5 & 0.75  &  11 44 29.88  & +36 53 08.5  & 14.50 &         \\ %L15         
1210+3820 & SDSSJ121044.27+382010.3 & 0.0229  & 99.8  & 0.46  &  12 10 44.27  & +38 20 10.3  & 13.89 &         \\ %L43       
1250-0249 & SDSSJ125042.44-024931.5 & 0.0470  & 208.5 & 0.92  &  12 50 42.44  & -02 49 31.5  & 14.47 &         \\ %L46      
1323+2701 & SDSSJ132310.39+270140.4 & 0.0559  & 249.6 & 1.09  &  13 23 10.39  & +27 01 40.4  & 16.25 &   \\ %L49  
1355+3834 & SDSSJ135553.52+383428.5 & 0.0501  & 222.7 & 0.98  &  13 55 53.52  & +38 34 28.5  & 15.72 & Mrk\,0464	       \\ %L50    
1405-0259 & SDSSJ140514.86-025901.2 & 0.0541  & 241.2 & 1.05  &  14 05 14.86  & -02 59 01.2  & 15.15 &     \\ %L51	    
1419+0754 & SDSSJ141908.30+075449.6 & 0.0558  & 249.1 & 1.08  &  14 19 08.30  & +07 54 49.6  & 14.01 &         \\ %L53       
1434+4839 & SDSSJ143452.45+483942.8 & 0.0365  & 160.7 & 0.73  &  14 34 52.45  & +48 39 42.8  & 14.29 & NGC\.5683	       \\ %L54    
1535+5754 & SDSSJ153552.40+575409.3 & 0.0304  & 133.2 & 0.61  &  15 35 52.40  & +57 54 09.3  & 14.52 & Mrk\,290 	       \\ %L57    
1545+1709 & SDSSJ154507.53+170951.1 & 0.0481  & 213.5 & 0.94  &  15 45 07.53  & +17 09 51.1  & 15.66 &  		       \\ %L58       
1554+3238 & SDSSJ155417.42+323837.6 & 0.0483  & 214.5 & 0.95  &  15 54 17.42  & +32 38 37.6  & 14.88 &   \\ %L59     
1557+0830 & SDSSJ155733.13+083042.9 & 0.0465  & 206.2 & 0.91  &  15 57 33.13  & +08 30 42.9  & 16.31 &   \\ %L60    
1605+3305 & SDSSJ160502.46+330544.8 & 0.0532  & 237.1 & 1.04  &  16 05 02.46  & +33 05 44.8  & 15.66 &   \\ %L61  
1606+3324 & SDSSJ160655.94+332400.3 & 0.0585  & 261.7 & 1.13  &  16 06 55.94  & +33 24 00.3  & 15.45 &   \\ %L62  
1611+5211 & SDSSJ161156.30+521116.8 & 0.0409  & 180.6 & 0.81  &  16 11 56.30  & +52 11 16.8  & 15.17 &  \\ %L63     
\enddata						  
\tablecomments{ 					  
Col. (1): Target ID used throughout the text (based on RA and DEC). 
Col. (2): Full SDSS name.			  
Col. (3): Redshift from SDSS-DR7.			  
Col. (4): Luminosity distance in Mpc, based on redshift and the adopted cosmology.
Col. (5): Scale in kpc/arcsec, based on redshift and the adopted cosmology.
Col. (6): Right Ascension. 				  
Col. (7): Declination. 
Col. (8): $i^\prime$ AB magnitude from SDSS-DR7 photometry (``modelMag\_i'').
Col. (9): Alternative name(s) from the NASA/IPAC Extragalactic Database (NED).}
\label{sample}
\end{deluxetable*}

\section{SPECTROSCOPY: OBSERVATIONS AND DATA REDUCTION}
\label{sec:obs}
We here summarize only the spectroscopic observations and data reduction.
The photometric data consist of SDSS archival images and the data reduction
is summarized in Appendix~\ref{sec:mattscode}.

All objects were observed with the Low Resolution Imaging Spectrometer
(LRIS) at Keck I, using a 1\arcsec~wide longslit, the D560 dichroic, 
the 600/4000 grism in the blue, and the 831/8200 grating in the red
(central wavelength = 8950\AA).
In addition to inferring the BH mass from the width of the broad H$\beta$ line,
this setup allows us to simultaneously cover
three spectral regions commonly used to determine
stellar-velocity dispersions.
In the blue, we cover
the region around the CaH+K\,$\lambda\lambda$3969,3934\AA~(hereafter CaHK)
and around the MgIb $\lambda\lambda\lambda$5167,5173,5184\AA~triplet (hereafter MgIb);
in the red, we cover the Ca II$\lambda\lambda\lambda$8498,8542,8662\AA~triplet 
(hereafter CaT).
The instrumental spectral resolution is $\sim$90 km\,s$^{-1}$ in the blue
and $\sim$45 km\,s$^{-1}$ in the red.

The long slit was aligned with the host galaxy 
major axis as determined from SDSS (``expPhi\_r''),
allowing us to measure 
the stellar-velocity dispersion profile and rotation curves.
Observations were carried out on January 21 2009
(clear, seeing 1-1.5\arcsec), 
January 22 2009 (clear, seeing $\sim$1.1\arcsec), 
April 15 2009 (scattered clouds, seeing $\sim$1\arcsec),
and April 16 2009 (clear, seeing $\sim$0.8\arcsec;
see also Table~\ref{observation}).

\begin{deluxetable*}{lcccccc}
\tabletypesize{\scriptsize}
\tablecolumns{7}
\tablewidth{0pc}
\tablecaption{Details of Observation and Reduction}
\tablehead{
\colhead{Object} & \colhead{PA} & \colhead{Exp. time} & \colhead{Date} & \colhead{S/N$_{\rm blue}$} & \colhead{S/N$_{\rm red}$} & \colhead{FeII sub.}\\
& & (sec) & \\
\colhead{(1)} & \colhead{(2)} & \colhead{(3)}  & \colhead{(4)} & \colhead{(5)} & \colhead{(6)} & \colhead{(7)}}
\startdata
0121-0102      &    65.6    &	1200  &    01-21-09  & 111.7 &   85.8 &  yes\\ %L11 
0206-0017      &    176.0   &	1200  &    01-22-09  & 152.7 &  142.8 &  no\\ %L2
0353-0623      &    171.2   &	1200  &    01-22-09  &  50.2 &   40.2 &  yes\\ %L6
0802+3104      &     82.9   &	1200  &    01-21-09  &  79.1 &   72.8 &  yes\\ %L1
0846+2522      &     50.9   &	1200  &    01-22-09  & 103.6 &   89.2 &  no\\ %L4
1042+0414      &    126.2   &	1200  &    04-16-09  &  53.8 &   52.1 &  yes\\ %L32
1043+1105$^a$  &    128.2   &	600   &    04-16-09  &  22.9 &   17.4 &  no\\ %L33
1049+2451      &     29.9   &	600   &    04-16-09  &  52.2 &   43.1 &  yes\\ %L34
1101+1102      &    147.5   &	600   &    04-16-09  &  32.2 &   37.5 &  yes\\ %L35
1116+4123      &     11.7   &	850   &    04-15-09  &  46.3 &   62.4 &  no\\ %L13
1144+3653      &     20.7   &	600   &    04-16-09  &  58.7 &   62.4 &  no\\ %L15
1210+3820      &      0.8   &	600   &    04-16-09  & 113.1 &  132.7 &  yes\\ %L43
1250-0249      &     73.9   &	1200  &    04-16-09  &  37.3 &   45.8 &  yes\\ %L46
1323+2701      &      8.1   &	700   &    04-16-09  &  25.0 &   28.7 &  no\\ %L49
1355+3834$^a$  &     78.0   &   300   &    04-16-09  &  34.7 &   34.4 &  no\\ %L50
1405-0259      &     64.8   &	1600  &    04-16-09  &  54.4 &   72.2 &  yes\\ %L51
1419+0754      &     19.3   &	900   &    04-16-09  &  58.8 &   80.1 &  yes\\ %L53
1434+4839      &    152.1   &	600   &    04-16-09  &  49.6 &   61.3 &  yes\\ %L54
1535+5754      &    103.8   &	1200  &    04-15-09  & 180.8 &  169.6 &  yes\\ %L57
1545+1709      &     60.0   &	1200  &    04-15-09  &  83.7 &   91.3 &  no\\ %L58
1554+3238      &    169.1   &	1200  &    04-15-09  &  83.6 &   95.0 &  yes\\ %L59
1557+0830$^a$  &     58.6   &   1200  &    04-15-09  &  54.9 &   55.6 &  yes\\ %L60
1605+3305      &     90.2   &	1200  &    04-15-09  &  80.0 &   80.9 &  yes\\ %L61
1606+3324      &     20.8   &	1200  &    04-15-09  &  44.9 &   59.1 &  yes\\ %L62
1611+5211      &    114.3   &	1200  &    04-15-09  &  72.3 &   84.4 &  no\\ %L63
\enddata
\tablecomments{
Col. (1): Target ID.
Col. (2): Position angle of major axis, along which the long slit was placed
(taken from SDSS-DR7 ``expPhi\_r'').
Col. (3): Total exposure time in seconds.
Col. (4): Date of Observations (month-day-year).
Col. (5): S/N in total blue spectrum (per pix, at rest wavelength 5050-5450\AA), aperture size $\sim$7\arcsec.
Col. (6): S/N in total red spectrum (per pix, at rest wavelength 8480-8690\AA), aperture size  $\sim$7\arcsec.
Col. (7): Subtraction of broad FeII emission.
\\
$^a$ Note that for these three objects, the spectra
did not allow a robust measurement of the stellar kinematics due to AGN contamination and we only
present BH mass and results from surface photometry.}
\label{observation}
\end{deluxetable*}

Note that all data included in this paper were obtained
before the LRIS red upgrade. The rest of our  sample
($\sim$75 objects) benefited from the upgrade with
higher throughput and lower fringing (data obtained from June 2009 onwards)
and will be presented in an upcoming paper (C. E. Harris et al. 2011, in preparation).
A total of 25 objects were observed with the old red LRIS chip.
For three objects, the spectra did not allow a robust measurement of the stellar
kinematics, due to dominating AGN continuum and emission lines.
However, we were able to determine BH mass and surface photometry.
Thus our ``imaging'' sample consists of 25 objects,
our ``spectroscopic'' sample of 22 objects.

The data were reduced using a python-based script which
includes the standard reduction steps such as bias subtraction,
flat fielding, and cosmic ray rejection.
Arc-lamps were used for wavelength calibration
in the blue spectral range,
sky emission lines in the red.
A0V {\it Hipparcos} stars, observed immediately after
a group of objects close in coordinates (to minimize
overhead), were used to correct for telluric absorption
and perform relative flux calibration.

From these final reduced 2D spectra, we extracted 1D spectra
in the following manner.
For the blue, a central spectrum with a width of 1.08\arcsec~
(8 pixels) was
extracted to measure the H$\beta$ width
for BH mass determination, i.e.~encompassing the broad-line region (BLR) emission
given a typical seeing of $\simeq$1\arcsec~and a slit width of 1\arcsec.
To measure the stellar-velocity
dispersion and its variation as a function of radius,
we extracted a central spectrum with a width of 0.54\arcsec~(0.43\arcsec)
for the blue (red). Outer spectra were extracted by stepping out in both
directions, increasing the width of the extraction window by one pixel
at each step (above and below the trace)
choosing the stepsize such that there is no overlap with the previous spectrum.
If the S/N of the extracted spectrum fell below 10 pix$^{-1}$
(at rest wavelength 5050-5450\AA~in the blue, and
8480-8690\AA~in the red), 
the width of the extraction window was increased
until an S/N of at least 10 pix$^{-1}$ was reached. We only use spectra
with S/N $\geq$ 10 pix$^{-1}$.
The signal-to-noise ratio (S/N) of the final reduced total spectra
(extraction with aperture radius of $\sim$7\arcsec)
is on average $\sim$80 pix$^{-1}$ in the blue
(ranging from $\sim$30 pix$^{-1}$ to 190 pix$^{-1}$) and
$\sim$70 pix$^{-1}$ in the red (ranging from $\sim$20 pix$^{-1}$ to 170 pix$^{-1}$; Table~\ref{observation}).

\begin{figure}[ht!]
\begin{center}
\includegraphics[scale=0.3]{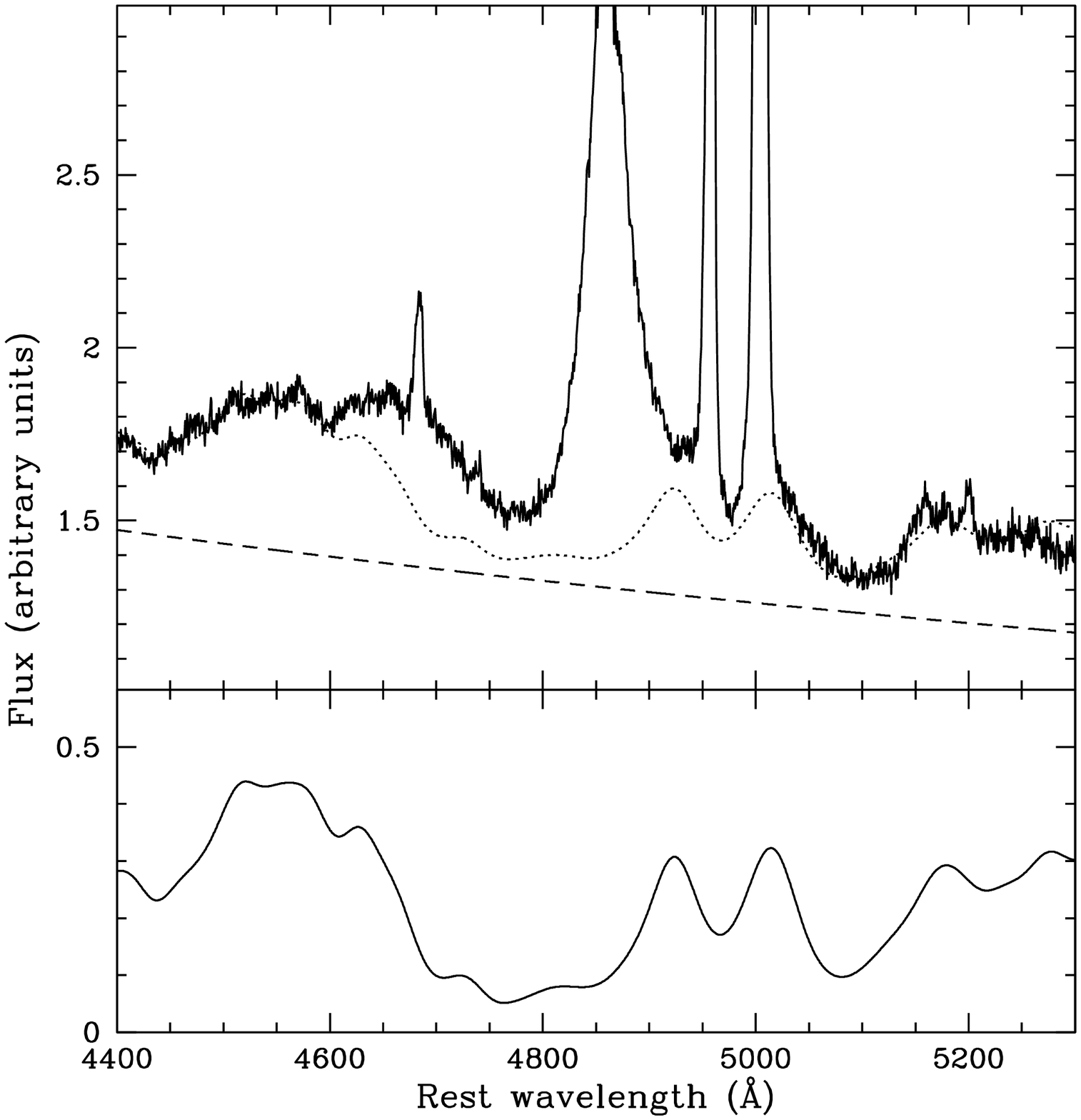}
\end{center}
\caption{Example of the broad FeII emission fit using a IZw1 template and a featureless continuum
for 0121-0102 (central spectrum; $\sim$0.54\arcsec$\times$1\arcsec). This object has the strongest FeII emission in our sample.
{\bf Upper panel:} Observed spectra with best-fit model (dotted line) composed of a broadened
FeII template and AGN+stellar continuum (dashed line). {\bf Lower panel:} Broadened FeII template.}
\label{feii_ex}
\end{figure}

\begin{figure*}[ht!]
\begin{center}
\includegraphics[scale=0.7]{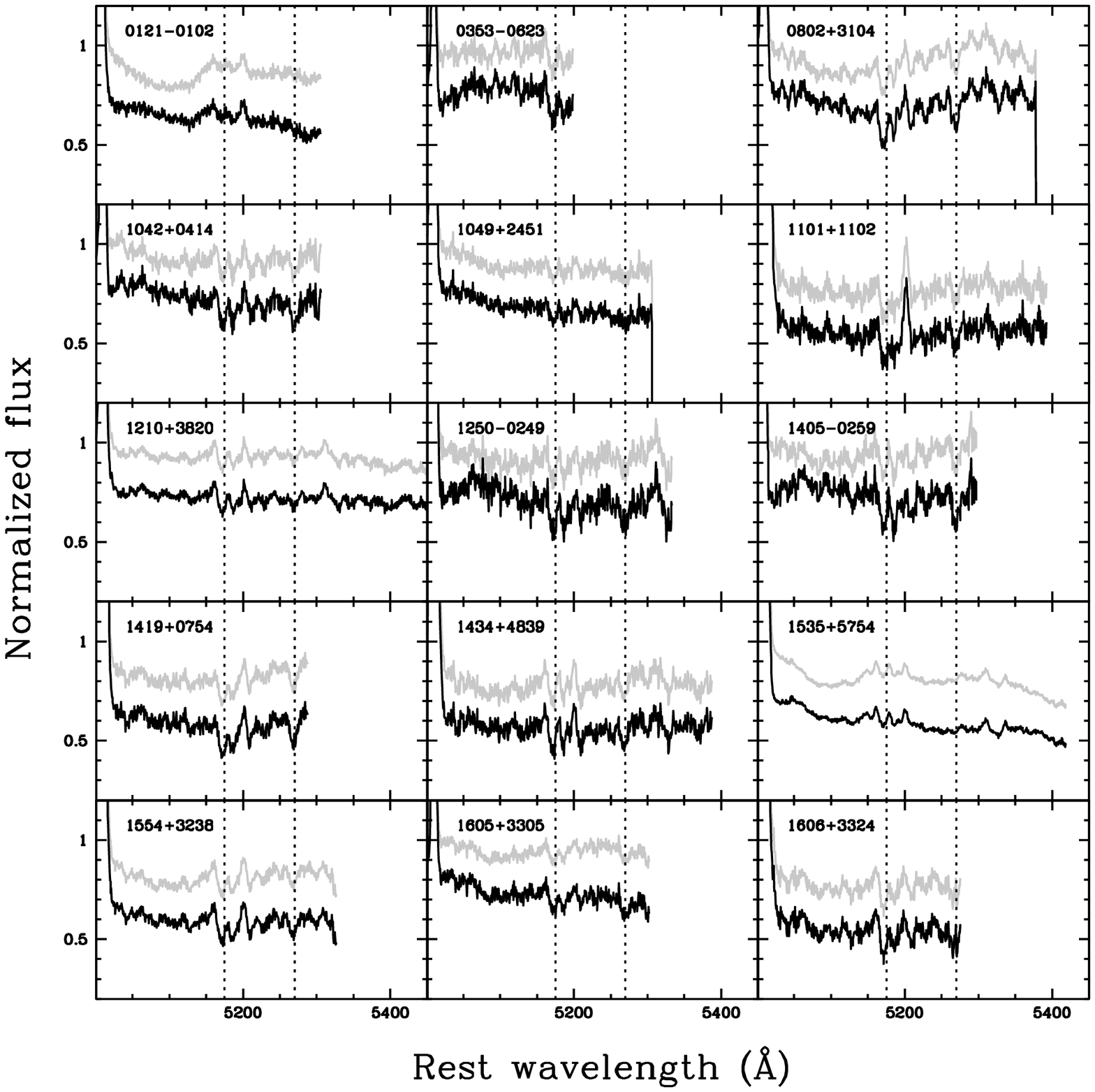}
\end{center}
\caption{Central spectra ($\sim$0.54\arcsec$\times$1\arcsec; normalized to average flux over plotted wavelength region), 
 before (gray) and after (black; shifted arbitrarily by -0.2 for comparison) 
broad FeII subtraction for the 15 objects for
which FeII was subtracted. The location of the stellar absorption lines 
MgIb\,$\lambda$5175\AA~and Fe\,$\lambda$5270\AA~is indicated by dotted lines. While some objects show prominent 
stellar absorption features, for others the region is swamped by AGN emission.
Note that the spectra end at the rest wavelength 5600\AA/(1+z).}
\label{feii_sub}
\end{figure*}

The majority of objects (15/22) display broad nuclear FeII emission in their spectra
($\sim$5150-5350\AA), complicating the $\sigma$ measurements
from the MgIb region.
For those objects, we fitted
a set of IZw1 templates, with various widths and strengths,
in addition to a featureless AGN continuum.
The best fit was determined by minimizing $\chi^2$ and then subtracted.
Details of this procedure are given in
\citet{woo06}.
We first derived the best fit for the central spectrum 
and then used the same FeII width also for the two outer spectra
that are still affected by FeII due to seeing effects and slit
width.
In Fig.~\ref{feii_ex}, we show an example of the FeII subtraction.
Fig.~\ref{feii_sub} compares the observed spectra to the FeII emission-subtracted
spectra.
Note that we did not correct the CaHK region of our spectrum for FeII emission,
since the broad FeII features near $\sim$3950 are weaker and broader \citep{gre06b}.

Details of the spectroscopic observations and data reduction are given in Table~\ref{observation}.

\section{DERIVED QUANTITIES}
\label{sec:quan}
In this section, we summarize the results we derive both
from the spectral analysis (stellar-velocity dispersion, velocity,
and H$\beta$ width) and image analysis (surface photometry
and stellar masses) as well as from combining results from
both (BH mass and dynamical masses). We will also distinguish
between different stellar-velocity dispersion measurements and define the nomenclature
we use.

\subsection{Spatially-Resolved Stellar-Velocity Dispersion And Velocity}
\label{subsec:spat}
The extracted spatially-resolved Keck spectra allow us to determine the stellar-velocity dispersion
as a function of distance from the center.
In the following, we will refer to these measurements as
$\sigma_{\rm spat}$, in contrast to velocity dispersion determined
from aperture spectra, as discussed in the next subsection.
The advantage of spatially-resolved spectra is twofold:
For one, the spatially-resolved
stellar velocity dispersions are not broadened 
by a rotating disk (if seen edge-on) and second,
the contamination by the AGN powerlaw continuum and broad
emission lines will only affect the nuclear spectra,
but not spectra extracted further out.

\begin{figure*}[ht!]
\begin{center}
\includegraphics[scale=0.7]{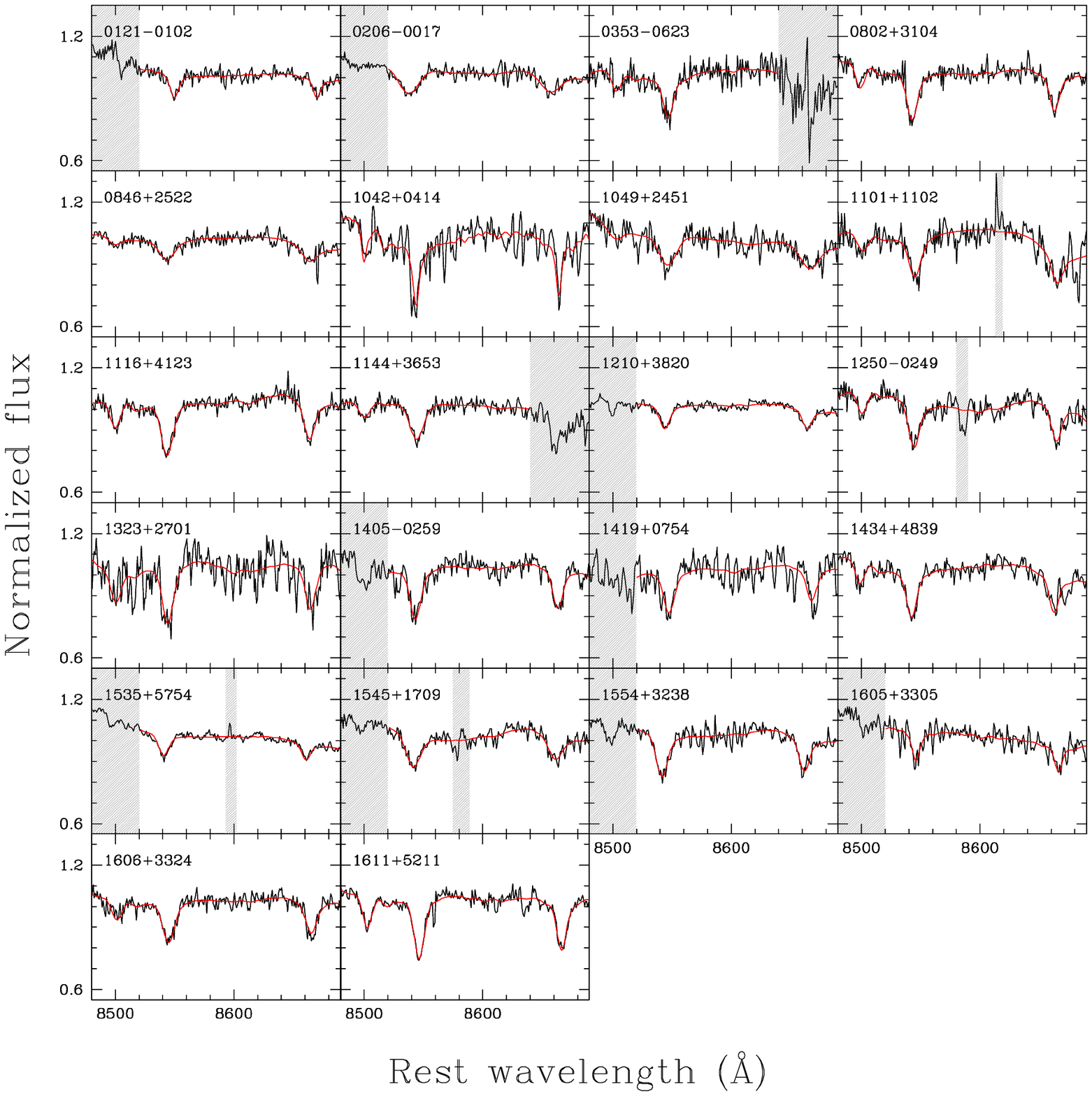}
\end{center}
\caption{Examples of stellar-velocity dispersion measurements in CaT region,
at a distance of $\sim$0.7\arcsec~from the center. The observed spectrum is shown in black, the model in red. The gray-shaded
area corresponds to regions that were not included in the fit, due to either AGN emission lines or
other spurious artifacts.}
\label{cat}
\end{figure*}

\begin{figure*}[ht!]
\begin{center}
\includegraphics[scale=0.7]{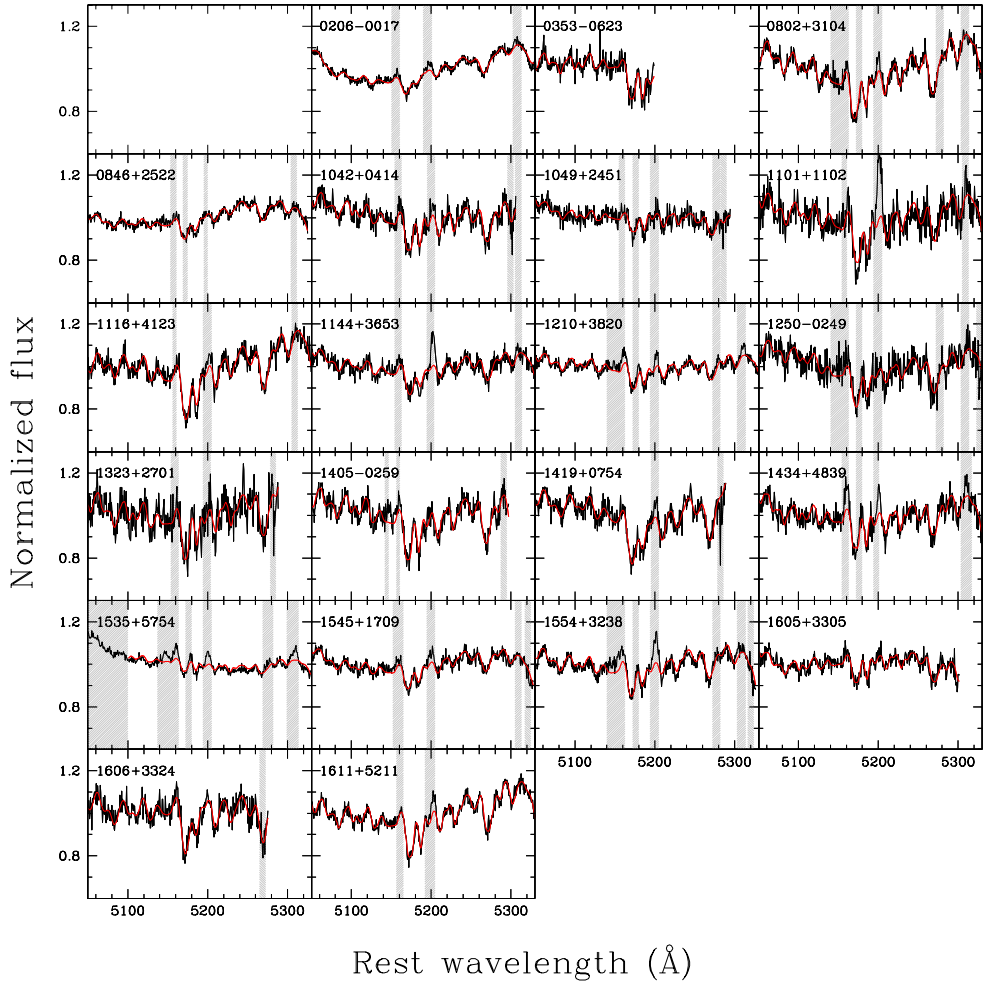}
\end{center}
\caption{Same as in Fig.~\ref{cat}, but for the MgIb region.
Note that the spectra end at the rest wavelength 5600\AA/(1+z).
(Object 0121-0102 is not shown
here since the S/N is too low to determine $\sigma_{\rm spat}$ from this region.)
}
\label{mg1b}
\end{figure*}

\begin{figure*}[ht!]
\begin{center}
\includegraphics[scale=0.7]{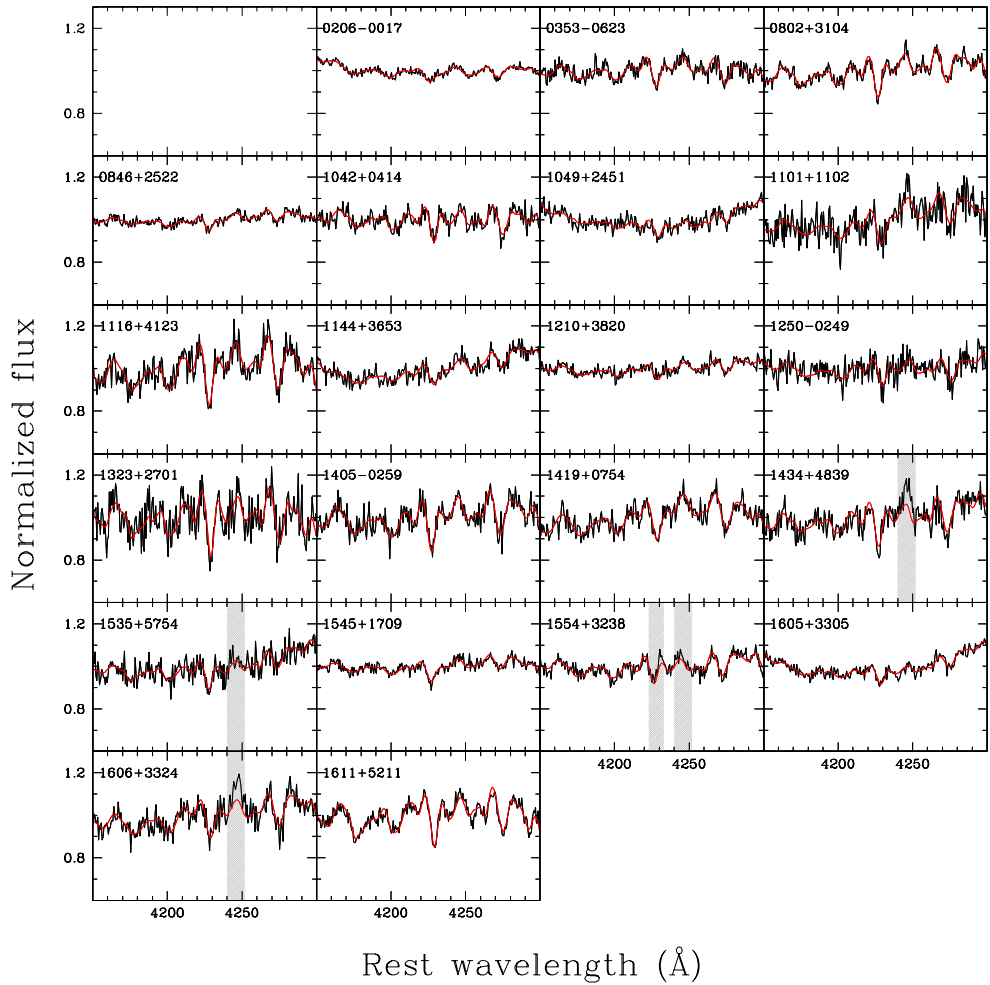}
\end{center}
\caption{Same as in Fig.~\ref{cat}, but in a region redwards of CaHK, excluding
the CaHK region due to AGN contamination. (Object 0121-0102 is not shown
here since the S/N is too low to determine $\sigma_{\rm spat}$ from this region.)}
\label{cahk}
\end{figure*}

\begin{figure*}[ht!]
\begin{center}
\includegraphics[scale=0.7]{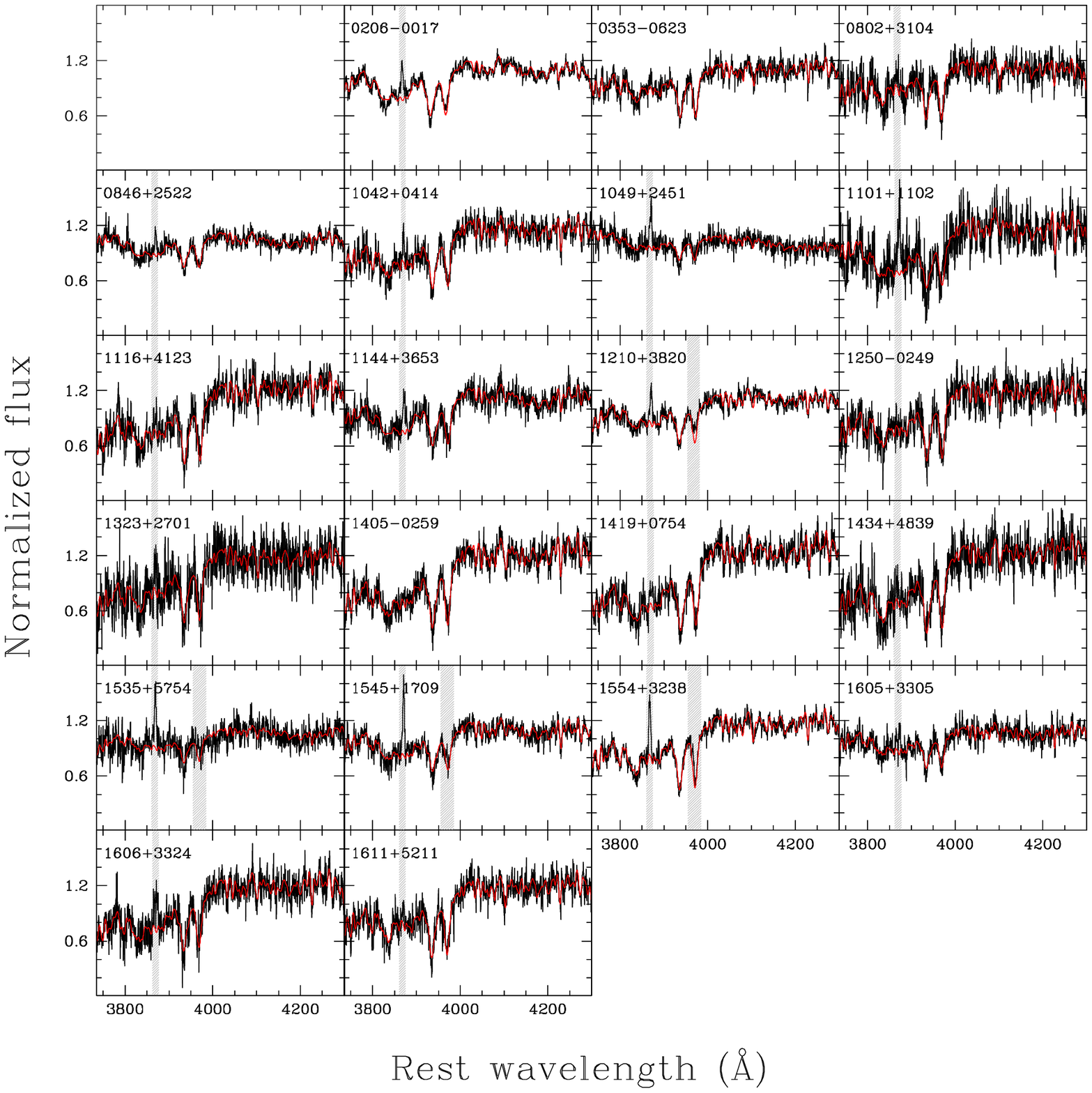}
\end{center}
\caption{Same as in Fig.~\ref{cahk} for a region including CaHK at $\sim$1.6\arcsec~from the center. 
(Object 0121-0102 is not shown
here since the S/N is too low to determine $\sigma_{\rm spat}$ from this region.)}
\label{cahknew}
\end{figure*}

A python-based code measures the stellar-velocity dispersion
from the extracted spectra,
using a linear combination of G\&K giant stars taken from the
Indo-US survey, broadened to a width ranging between 30-500 km\,s$^{-1}$.
In addition, a polynomial continuum of order 3-5 was fitted,
depending on the object and fitting region.
The code uses a Markov-chain
Monte Carlo (MCMC) simulation to find the best-fit velocity dispersion and error; 
see \citet{suy10} for more detailed description of the fitting technique.
Three different regions were fitted:
(i)  the region around CaT, 8480-8690\AA;
(ii) the region around MgIb
that also includes several Fe absorption lines,
5050-5450\AA~(i.e.~redwards of the [OIII]\,$\lambda$5006.85\AA~and
the HeI$\lambda\lambda$5015.8,5047.47\AA~AGN emission lines); and 
(iii) the region around the CaH+K lines, 3735-4300\AA~(i.e.~bluewards 
of the broad H$\gamma$ and [OIII]\,$\lambda$4363.21\AA~AGN emission lines).
In Fig.~\ref{cat}-\ref{cahknew}, examples of fits are shown for all three regions.

In region (i), the broad AGN emission line OI\,$\lambda$8446\AA~contaminates 
the spectra in the central region for some objects. 
In those cases, we excluded the first 
CaT line and only used the region 8520-8690.
In objects at higher redshifts, the third CaT line
can be affected by telluric absorption (although we attempted
to correct for this effect), and had to be excluded in some cases.

In region (ii), several AGN narrow-emission lines were masked,
if present, such as 
[Fe VI]\,$\lambda\lambda\lambda$5145.77, 5176.43, 5335.23\AA;
[Fe VII]\,$\lambda\lambda$5158.98, 5277.67\AA; 
[NI]\,$\lambda\lambda$5197.94,5200.41\AA;
and 
[CaV]\,$\lambda$5309.18\AA~\citep[wavelengths taken from][]{moo45,bow60}.
The blue spectra end at an observed wavelength of $\sim$5600\AA,
which corresponds to restframe $\sim$5200\AA~for our
highest redshifted object (z=0.076).

Region (iii) includes AGN emission from e.g., [FeV] $\lambda$4227.49\AA~and 
various broad HeI and Balmer lines,
that were masked during the fitting procedure.
Also, the CaH$\lambda$3969\AA~line is often filled by AGN emission
(H$\epsilon$). Thus, in most cases, we restricted the fitting region
to 4150-4300\AA~for the central spectra (see Fig.~\ref{cahk}), due to AGN contamination.
Only in the outer parts, the wavelength regime 3735-4300\AA~was fitted
(Fig.~\ref{cahknew}). In the following, we still generally refer to region (iii) as CaHK region,
even though it might not actually include the CaH+K line in the central spectra
where it is contaminated by AGN emission.

\begin{figure*}[ht!]
\includegraphics[scale=0.23]{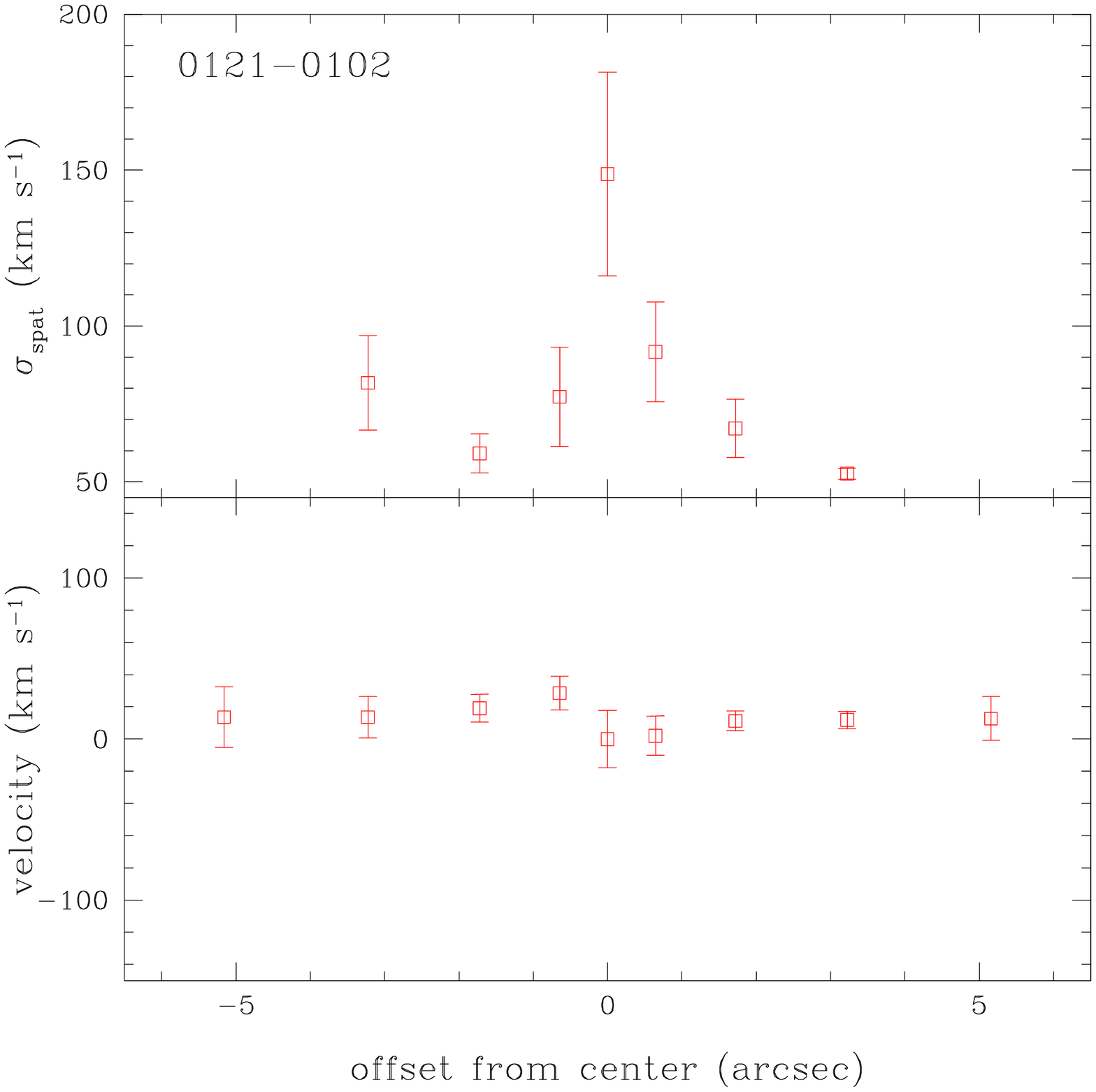}
\includegraphics[scale=0.16]{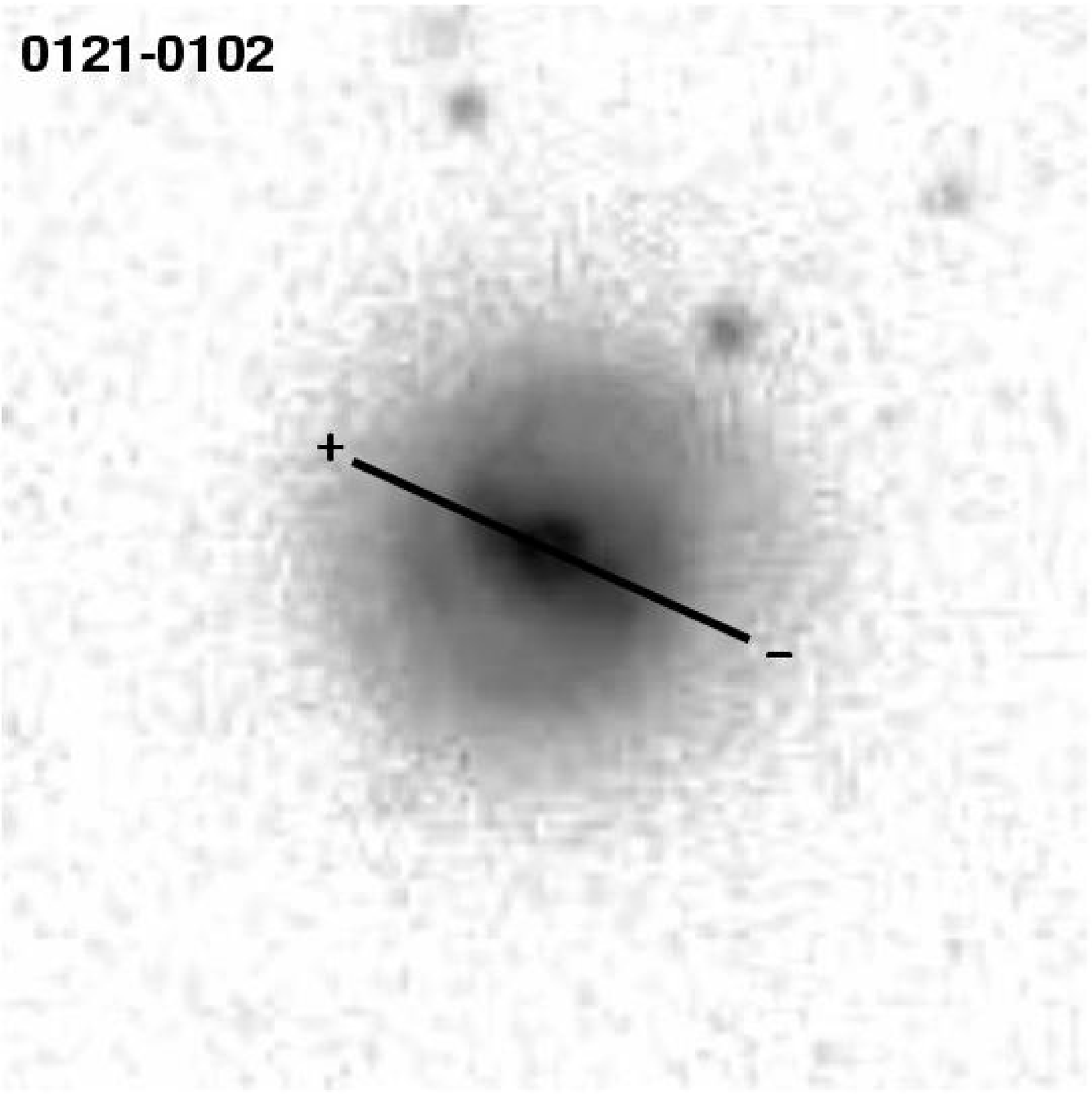}\hspace*{0.48cm}
\includegraphics[scale=0.23]{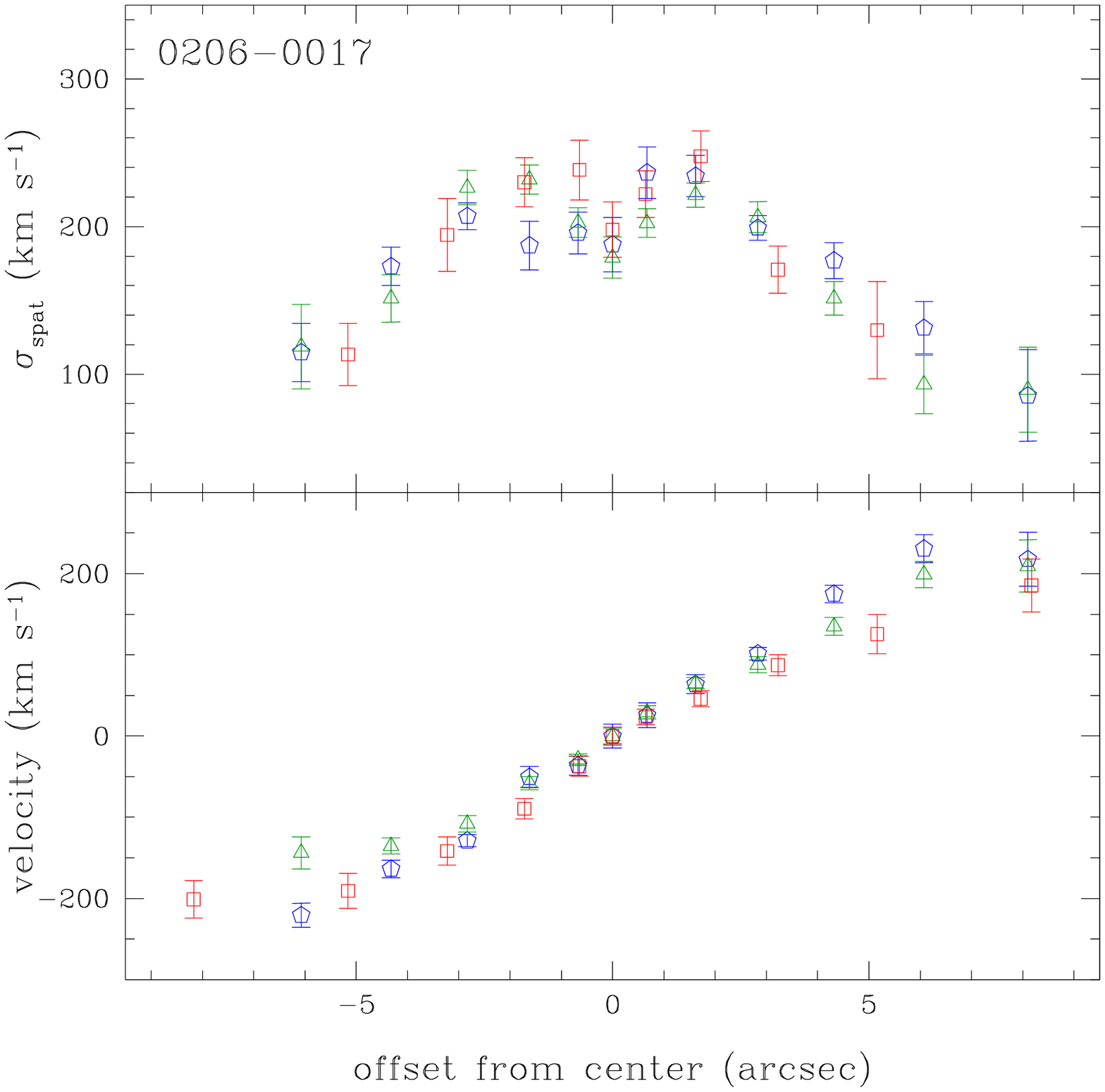}
\includegraphics[scale=0.16]{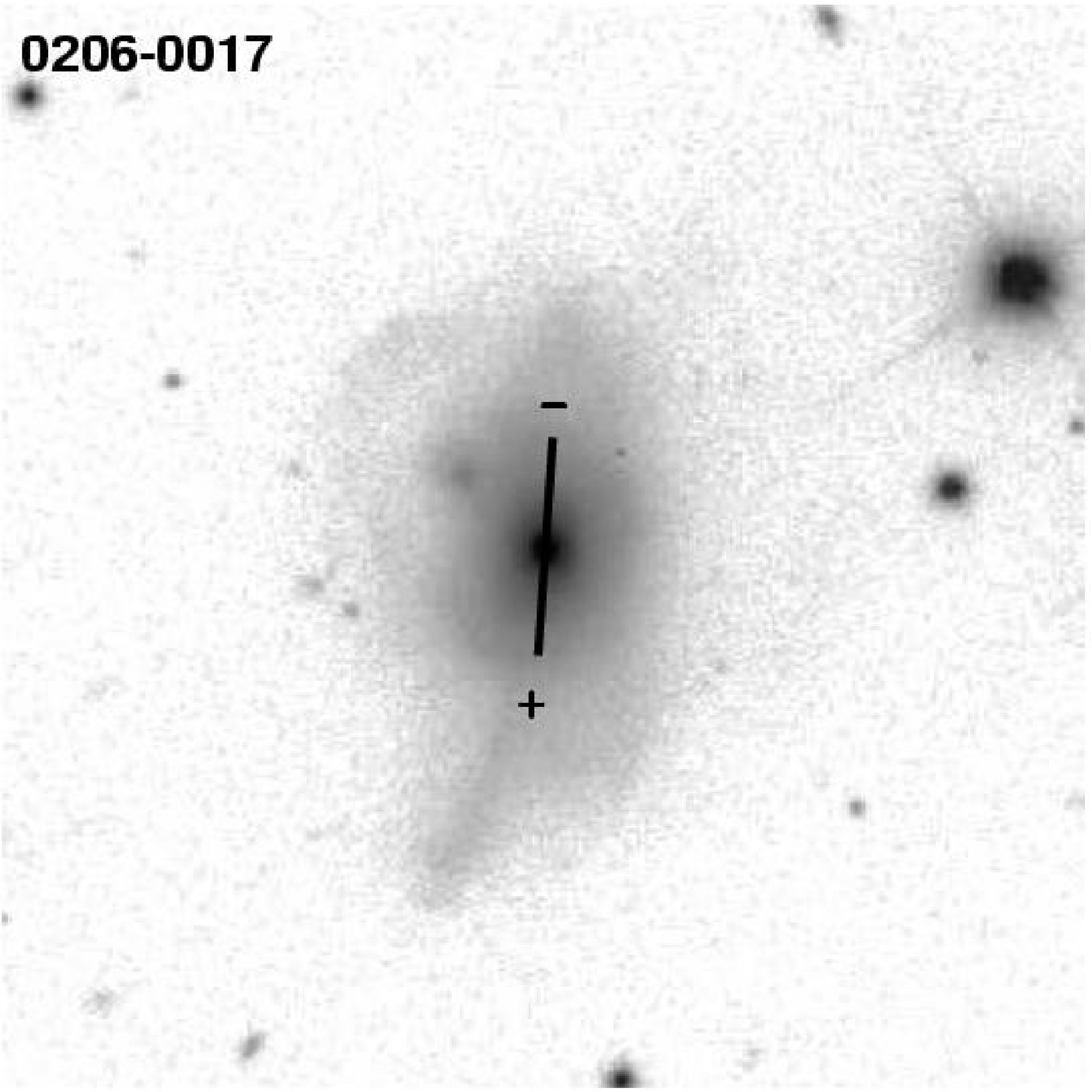}\\
\includegraphics[scale=0.23]{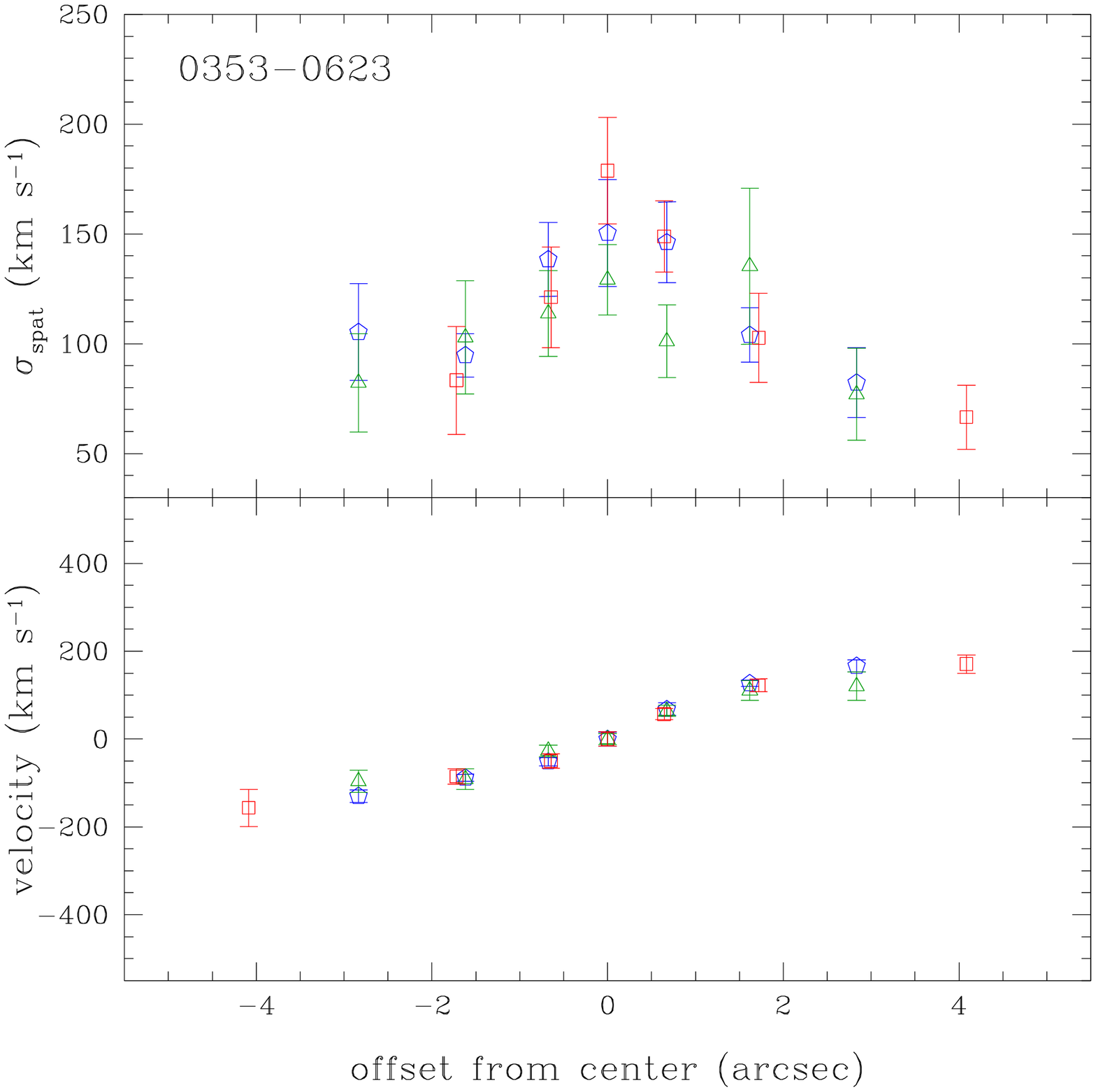}
\includegraphics[scale=0.16]{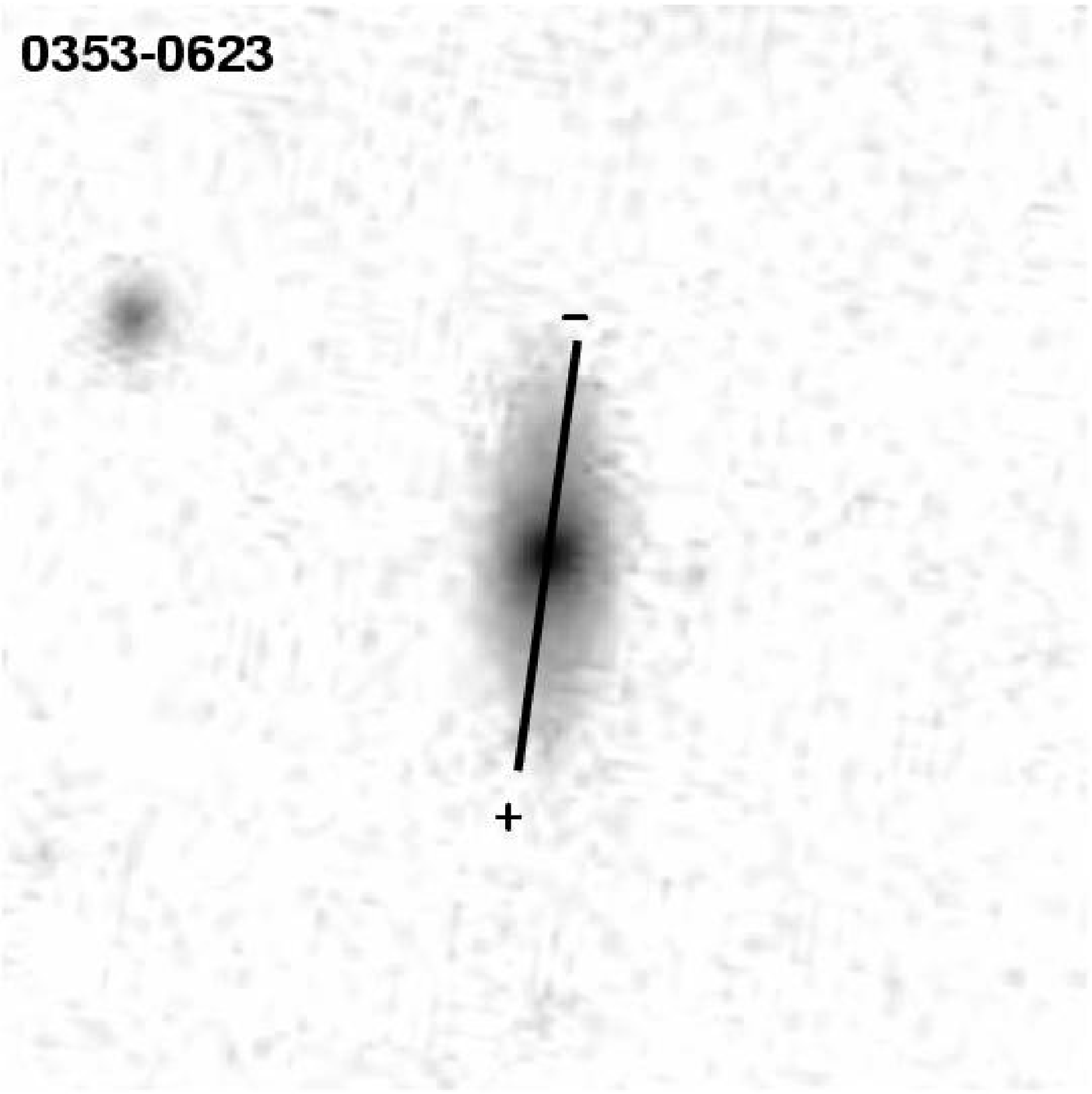}\hspace*{0.48cm}
\includegraphics[scale=0.23]{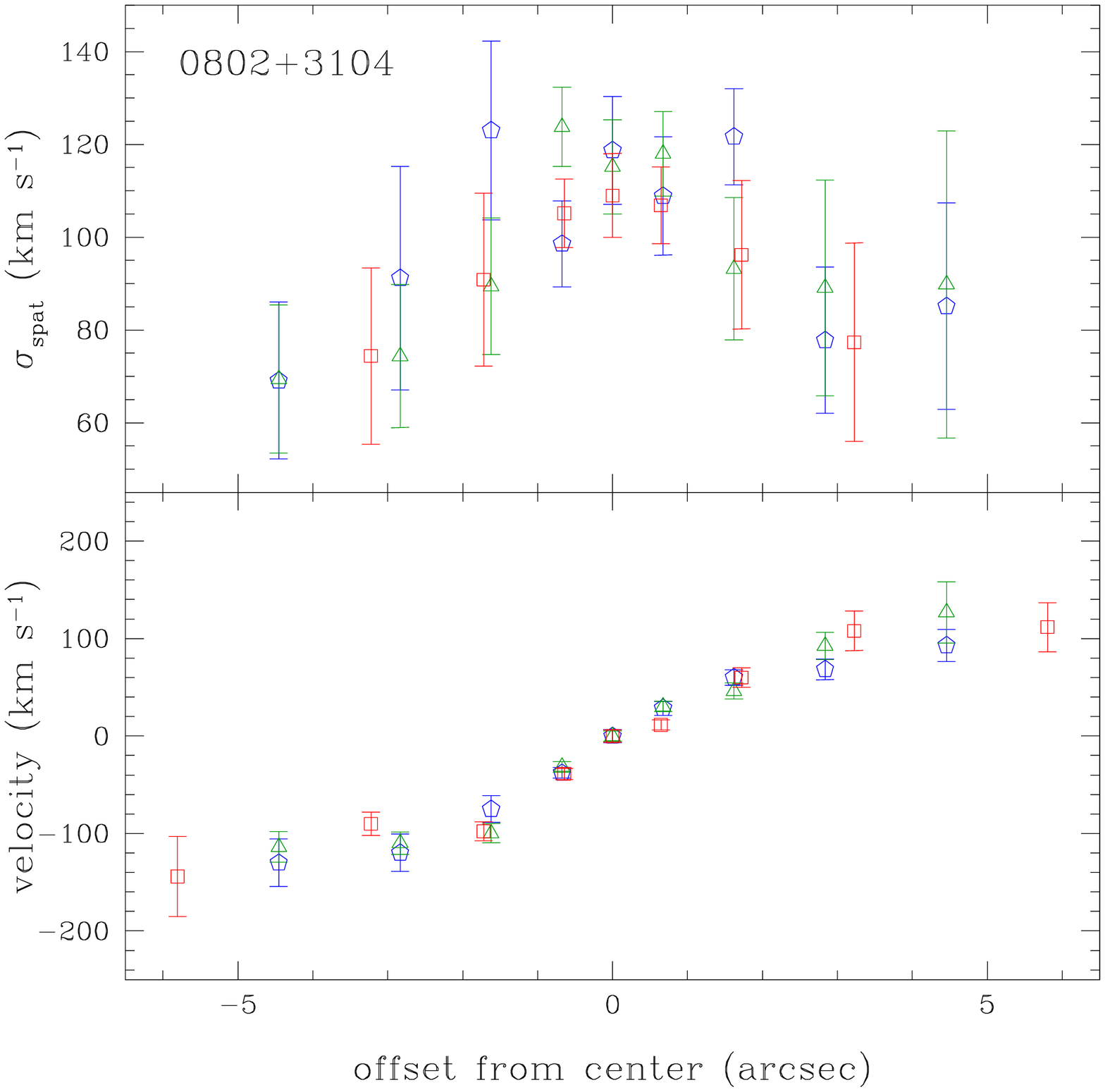}
\includegraphics[scale=0.16]{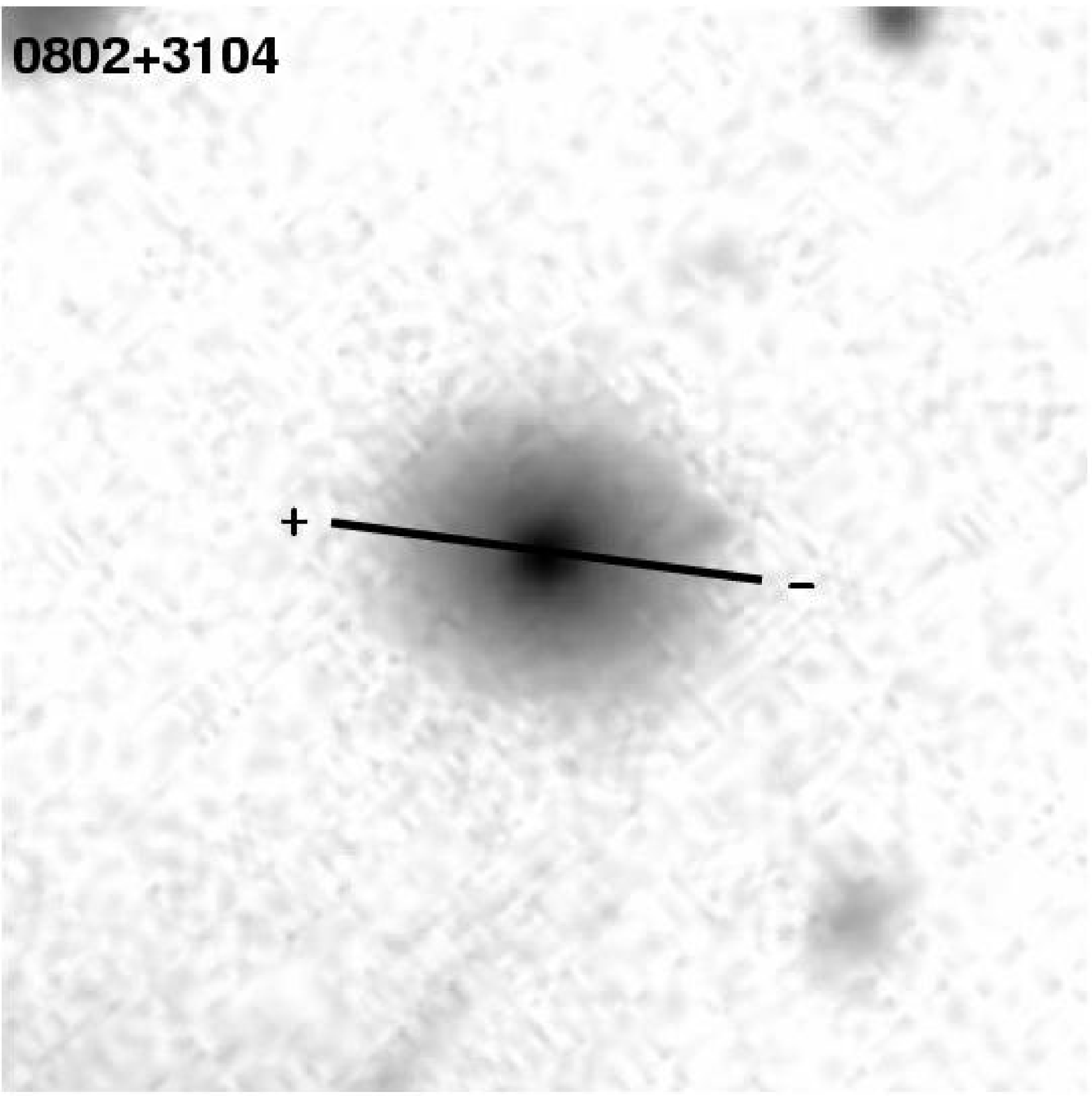}\\
\includegraphics[scale=0.23]{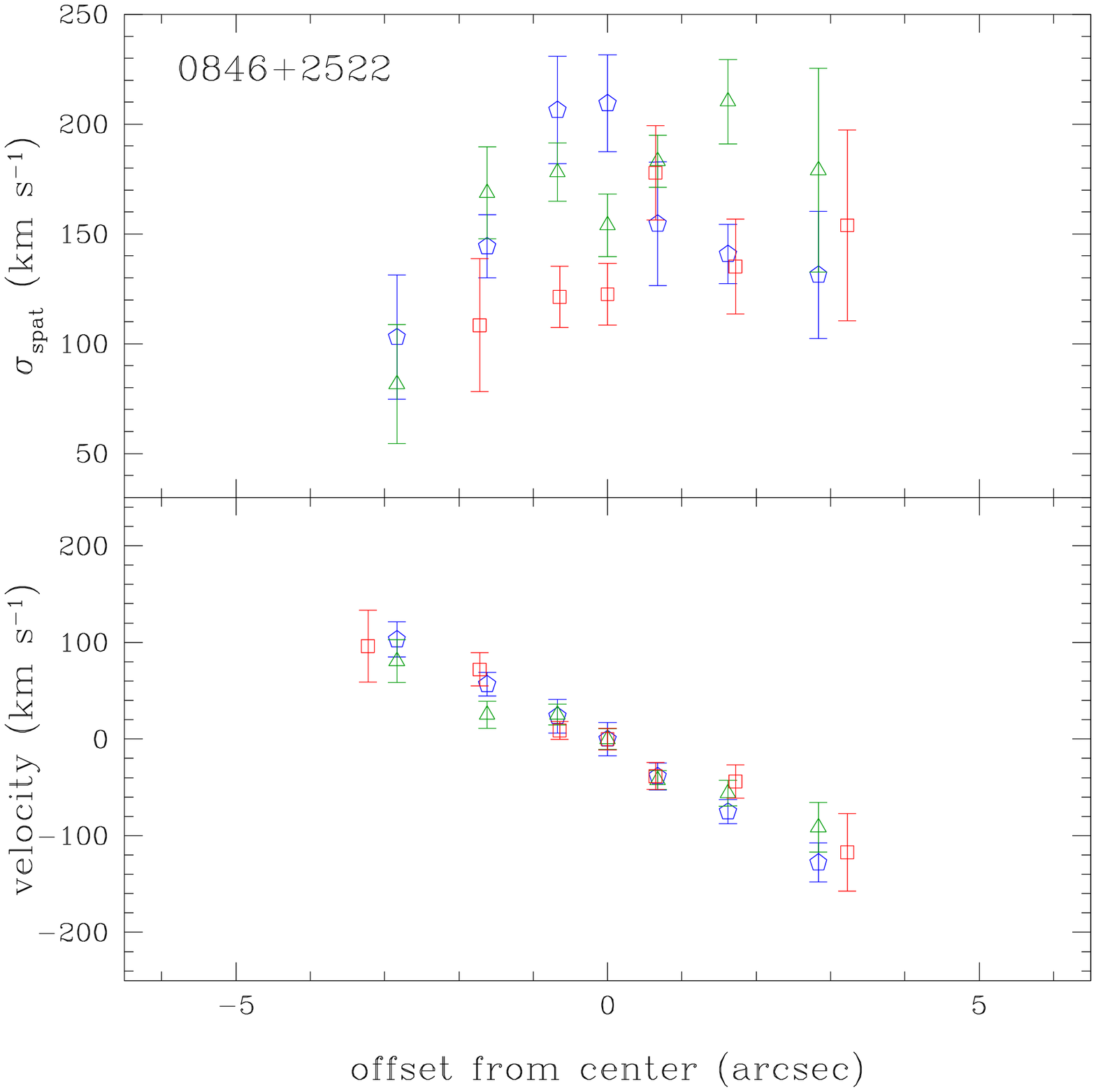}
\includegraphics[scale=0.16]{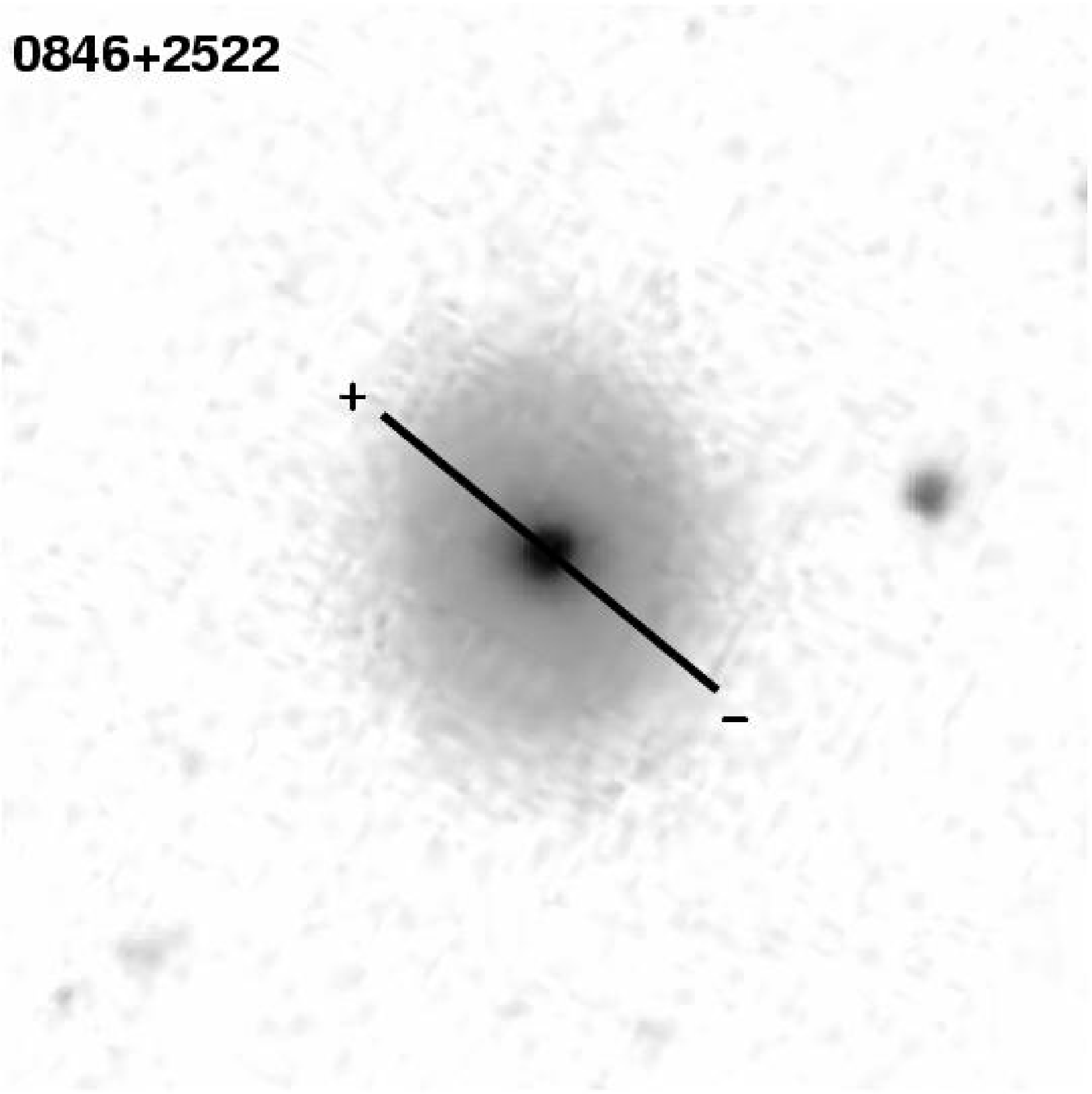}\hspace*{0.48cm}
\includegraphics[scale=0.23]{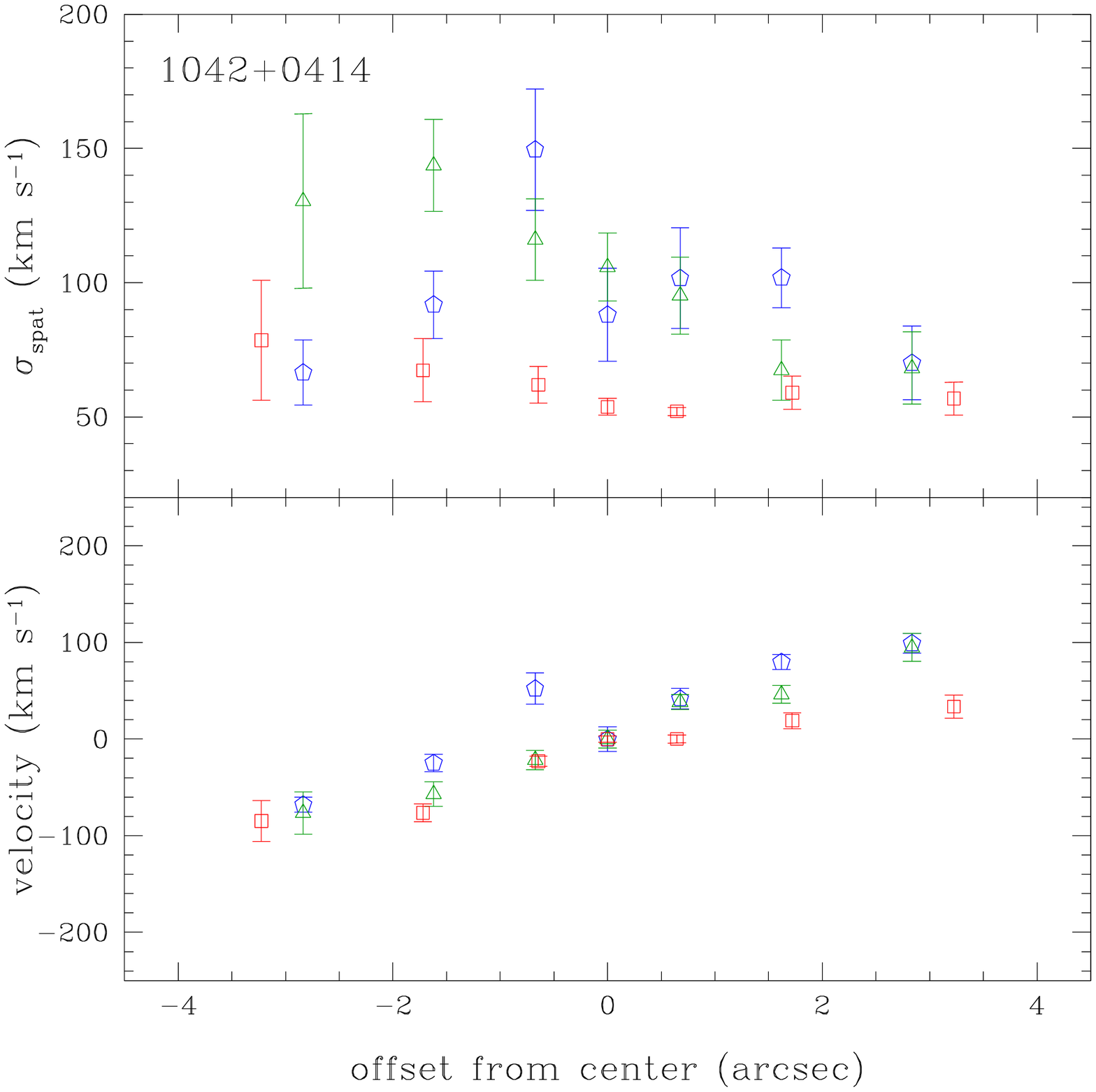}
\includegraphics[scale=0.16]{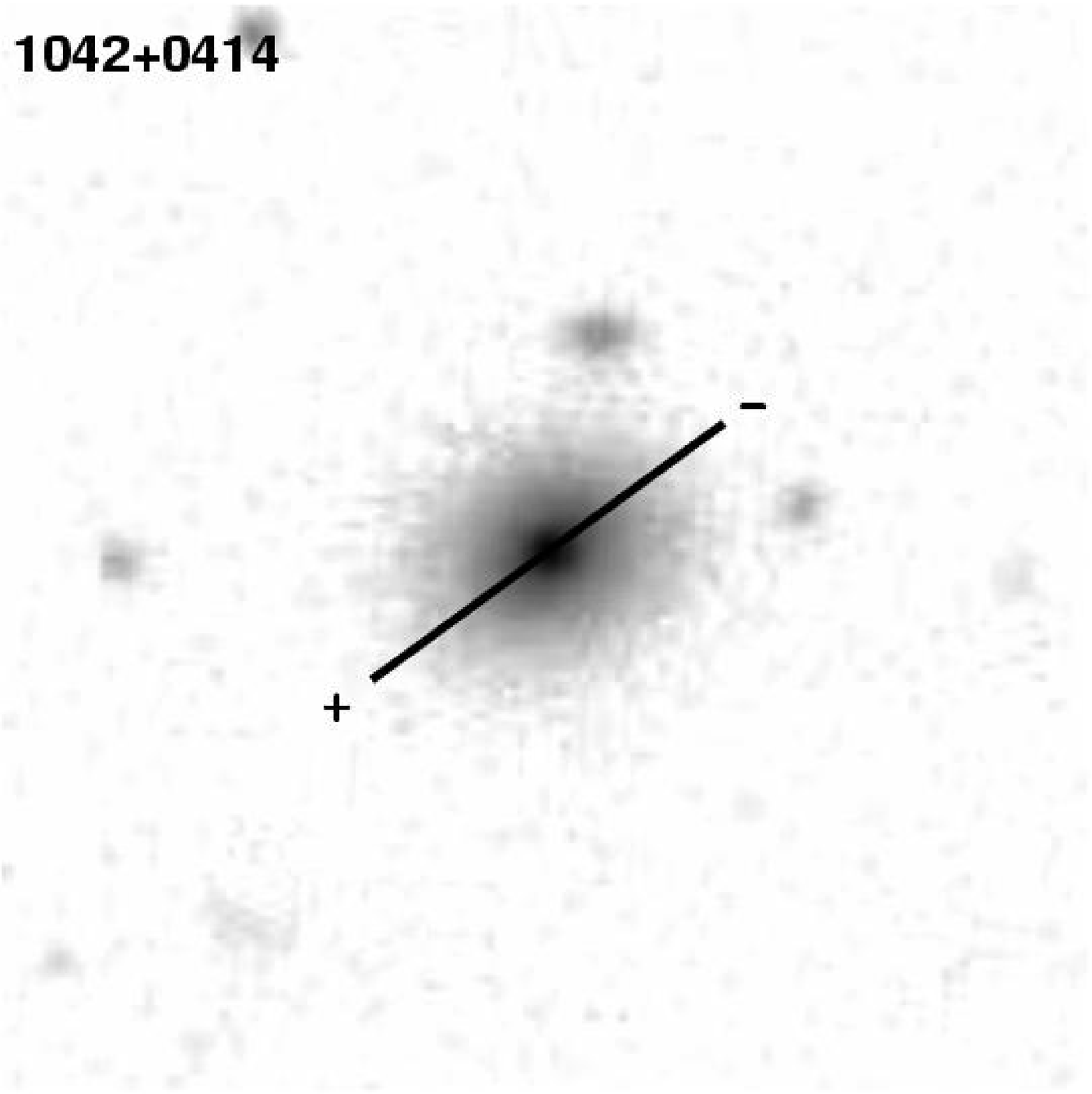}\\
\includegraphics[scale=0.23]{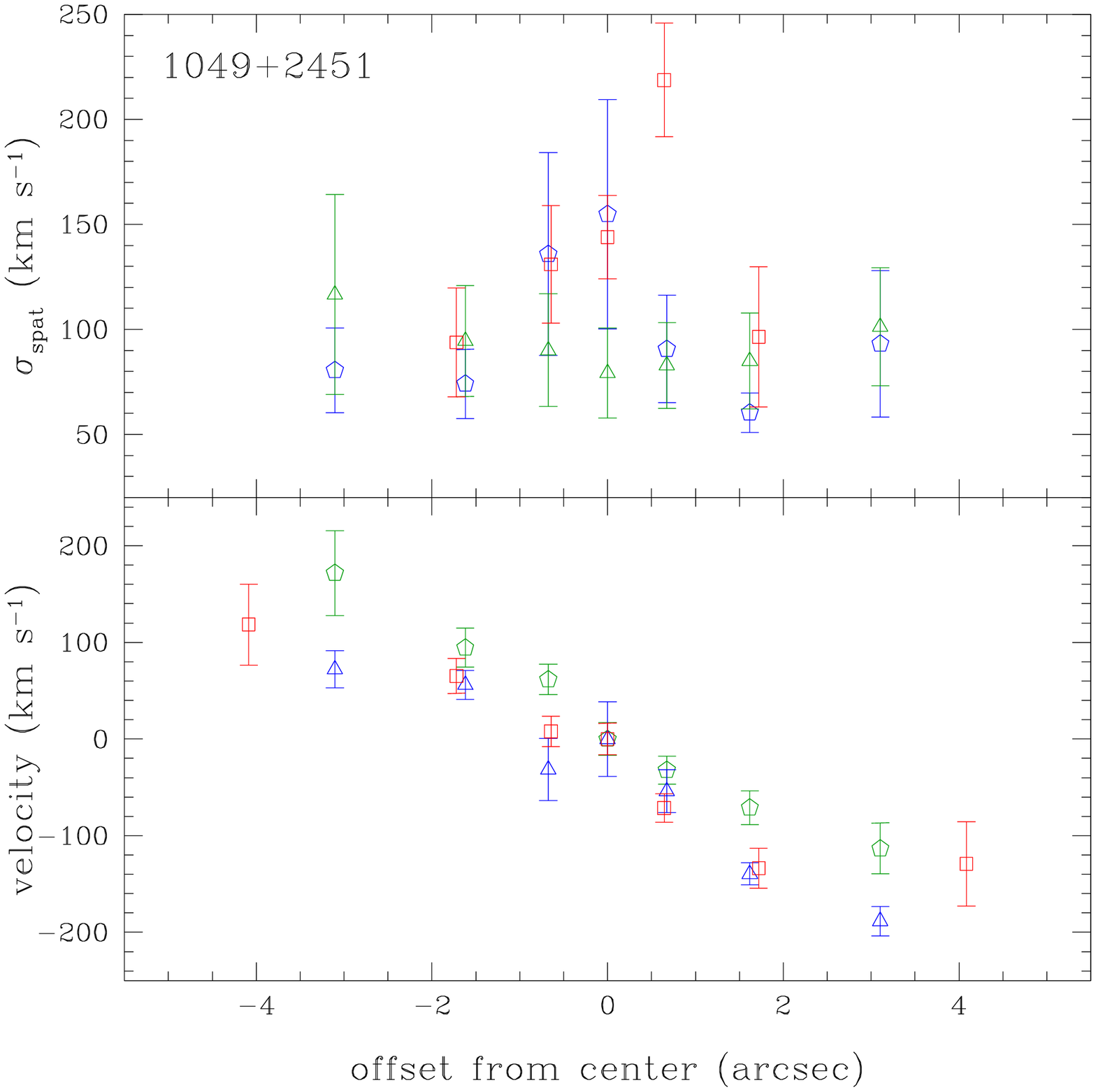}
\includegraphics[scale=0.16]{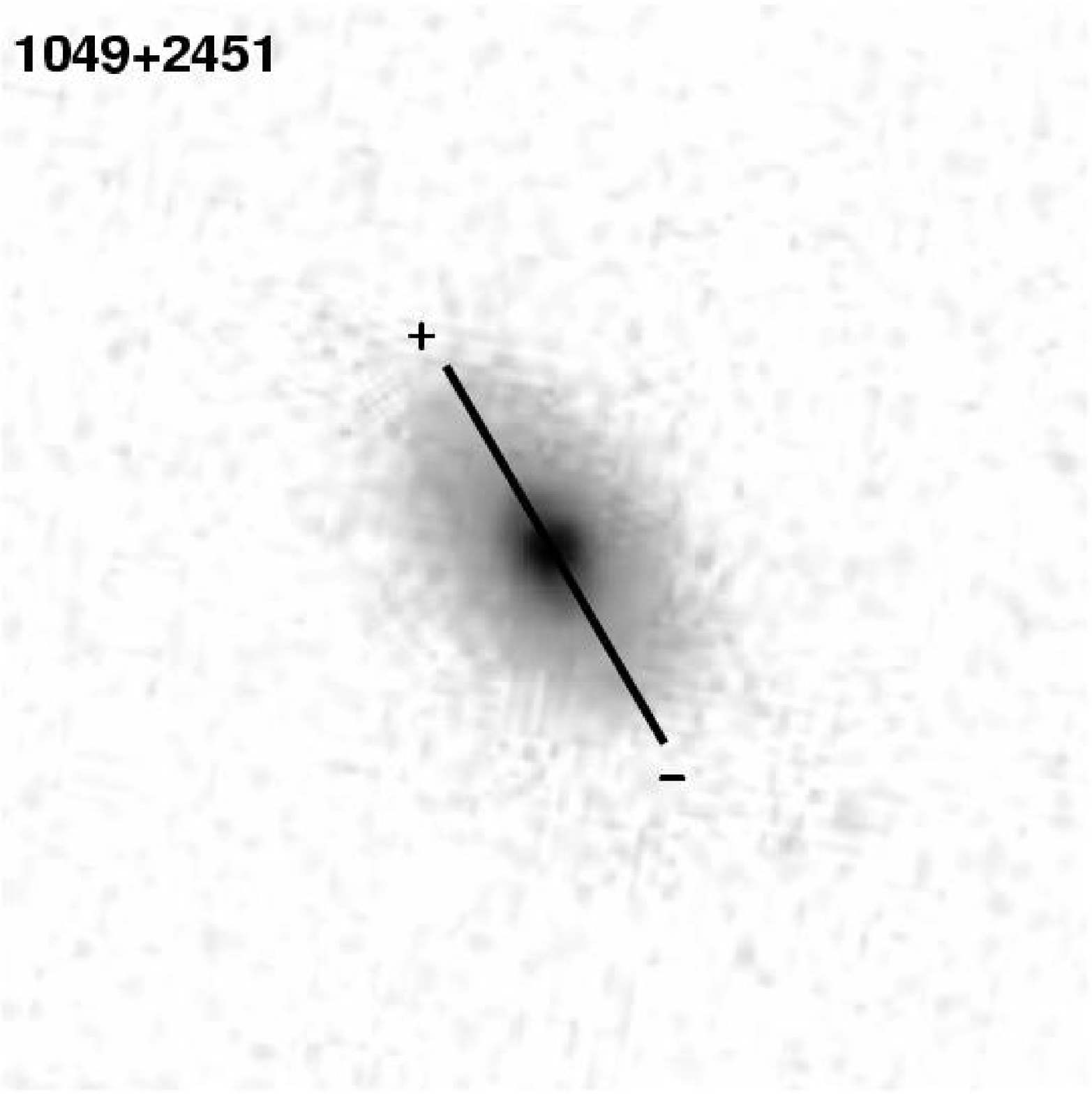}\hspace*{0.48cm}
\includegraphics[scale=0.23]{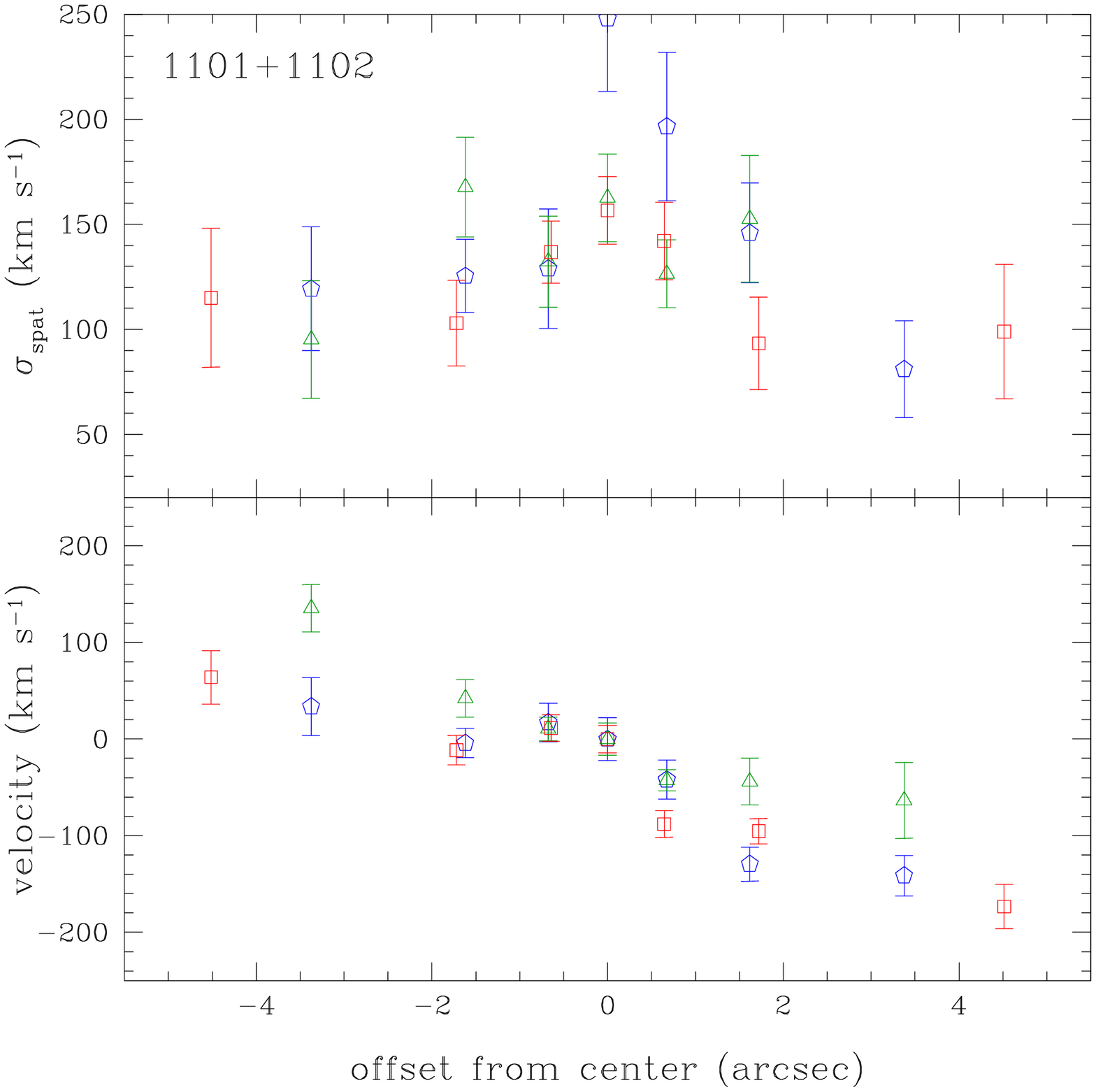}
\includegraphics[scale=0.16]{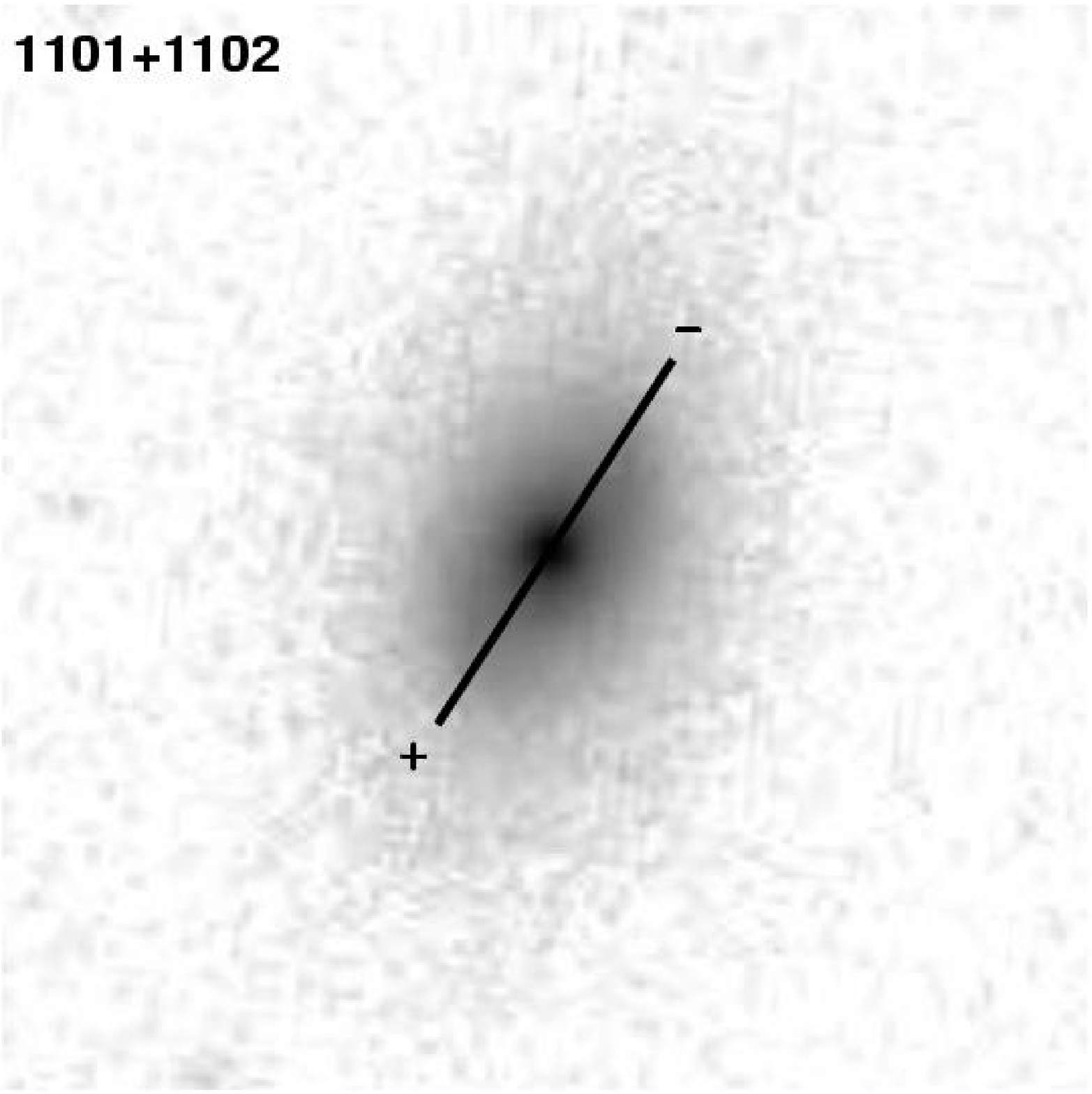}
\caption{Spatially-resolved stellar-velocity dispersions and velocities.
In the first and third column, we show spatially-resolved stellar-velocity dispersion (upper panel) 
and velocity curve (lower panel) as derived from CaHK (blue pentagons), MgIb (green triangles), 
and the CaT region (red squares).
In the second and fourth column, the corresponding SDSS-DR7 multi-filter
image is shown in gray scales. North is up, east is to the left; same dimensions as in Fig.~\ref{sdss}, with the position of the 
longslit indicated as black line (corresponding to 20\arcsec~ for clarity, but
the slit is 175\arcsec~long). The ``+'' and ``-'' signs indicate the extraction direction of the spectra, corresponding
to the x-axis in the plots.
Note that all figures are on the same gray scale to allow for comparison.}
\label{sigma_vel1}
\end{figure*}

\begin{figure*}[ht!]
\includegraphics[scale=0.23]{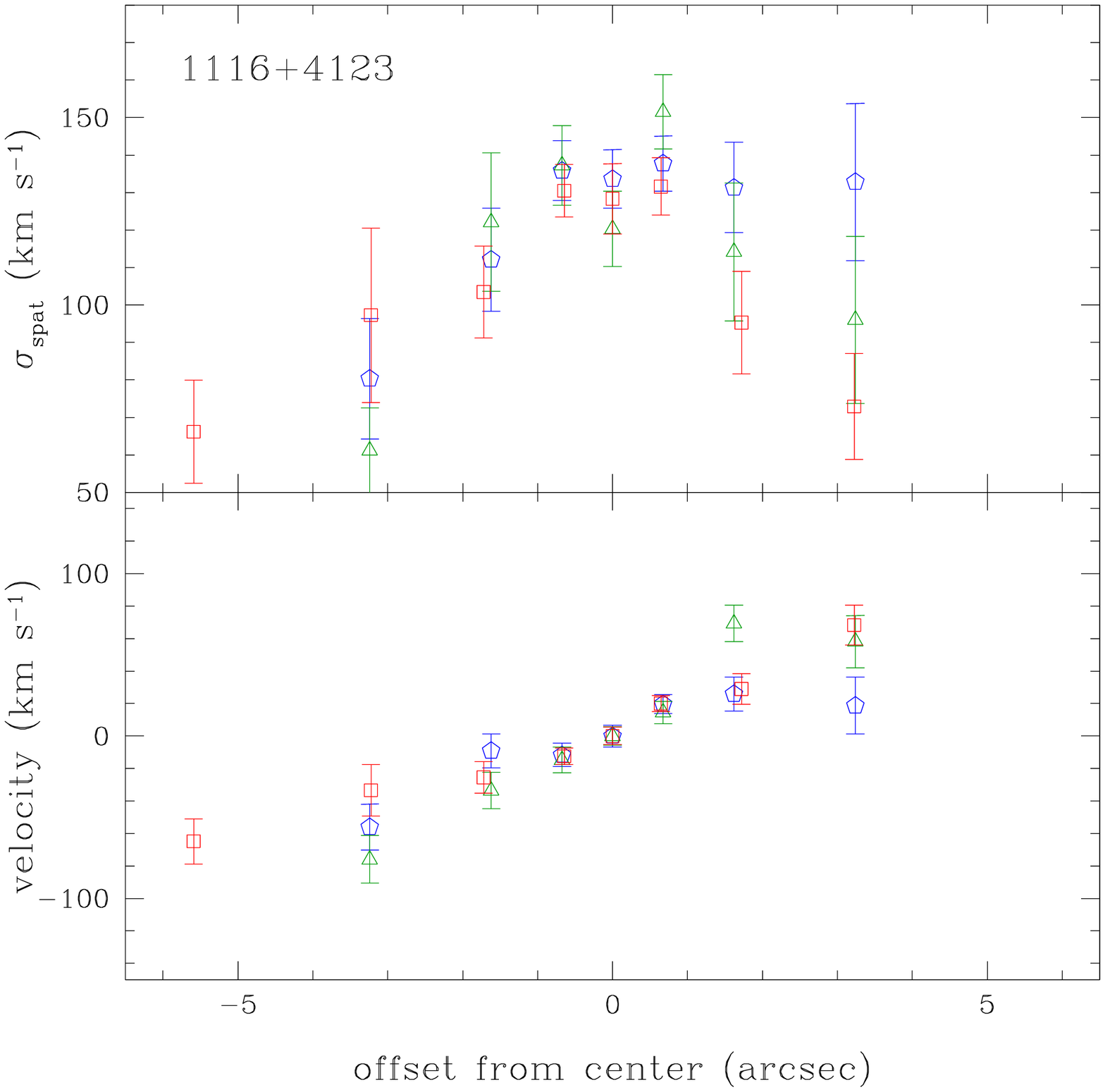}
\includegraphics[scale=0.16]{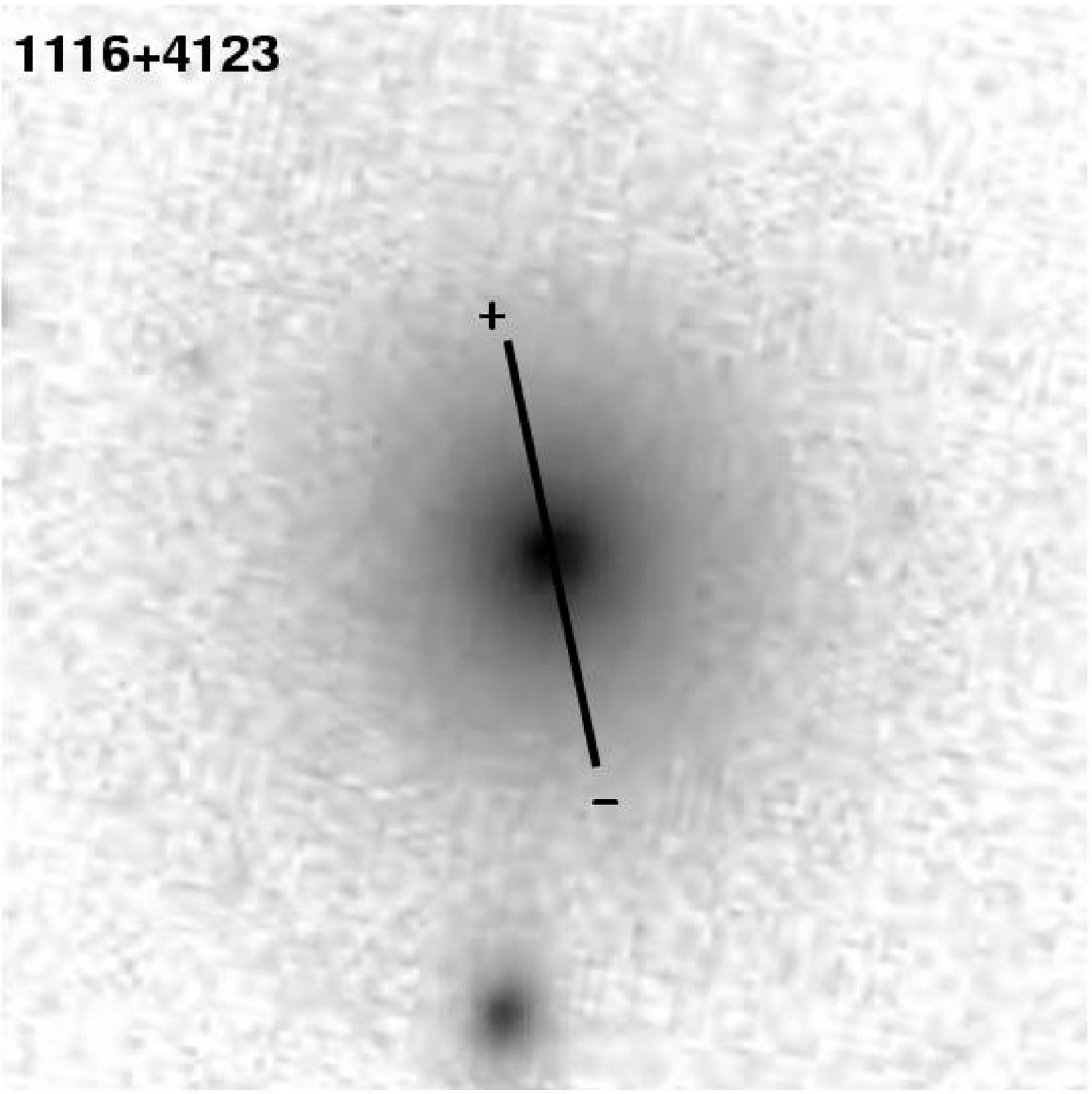}\hspace*{0.48cm}
\includegraphics[scale=0.23]{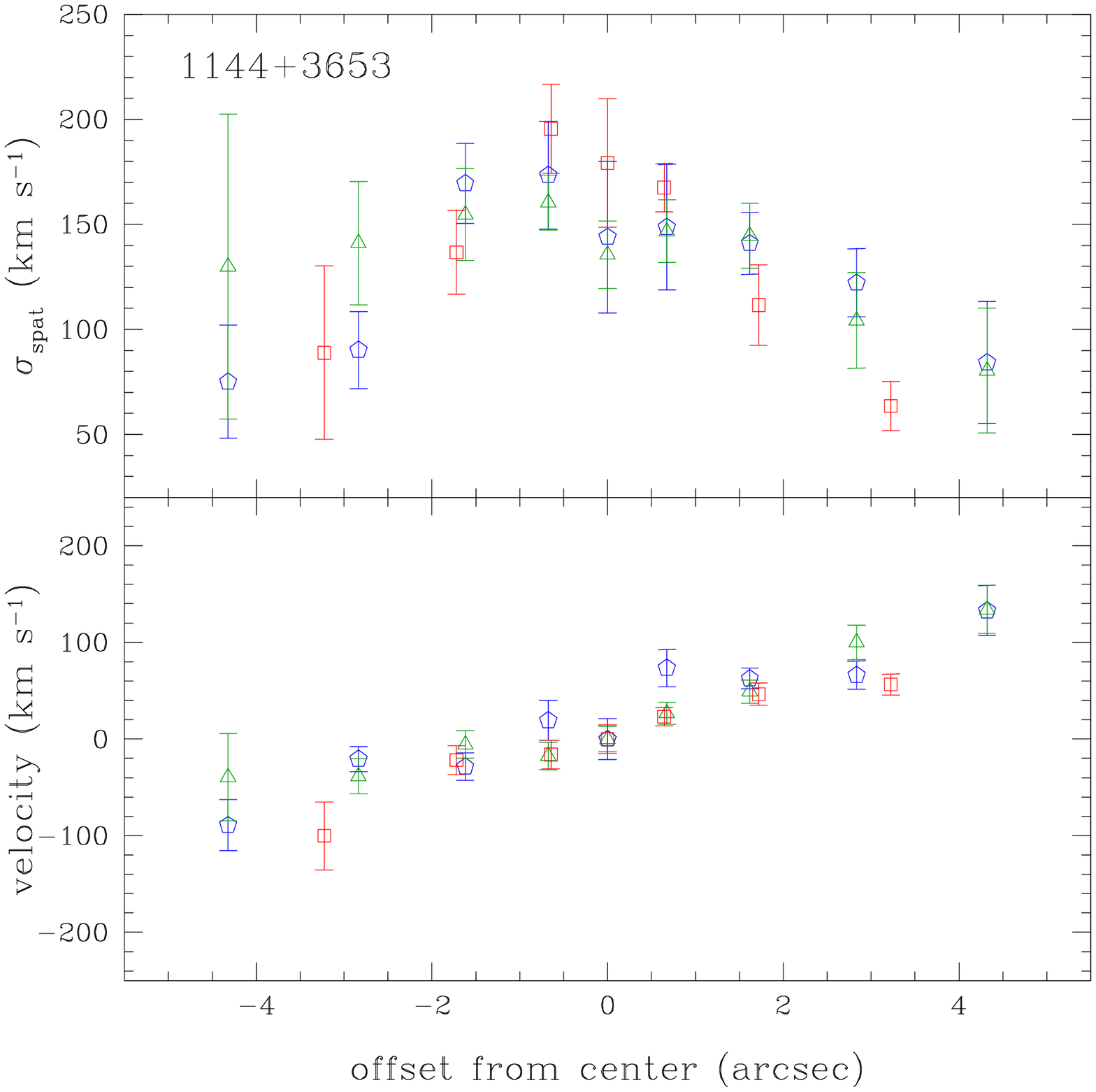}
\includegraphics[scale=0.16]{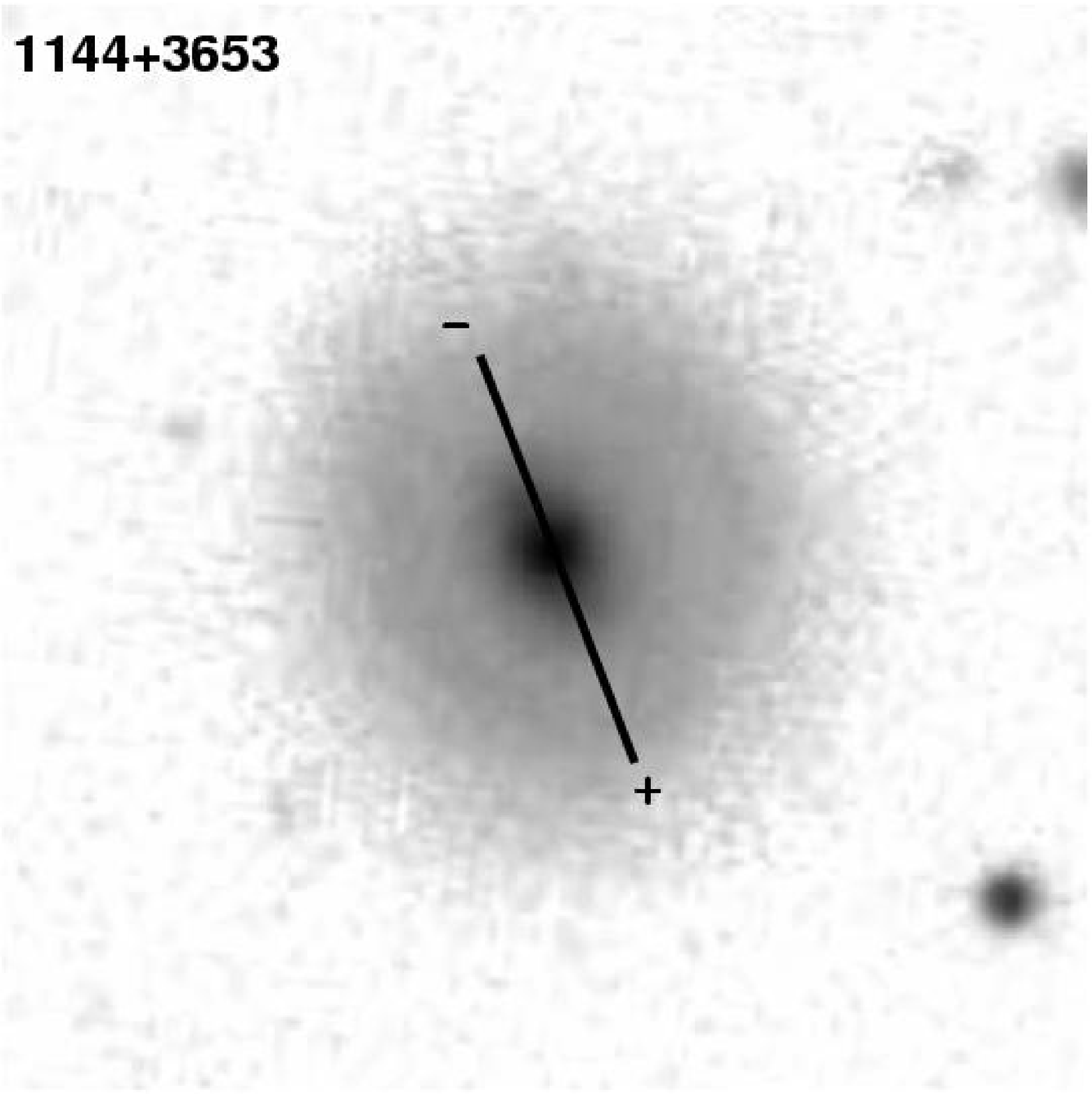}\\
\includegraphics[scale=0.23]{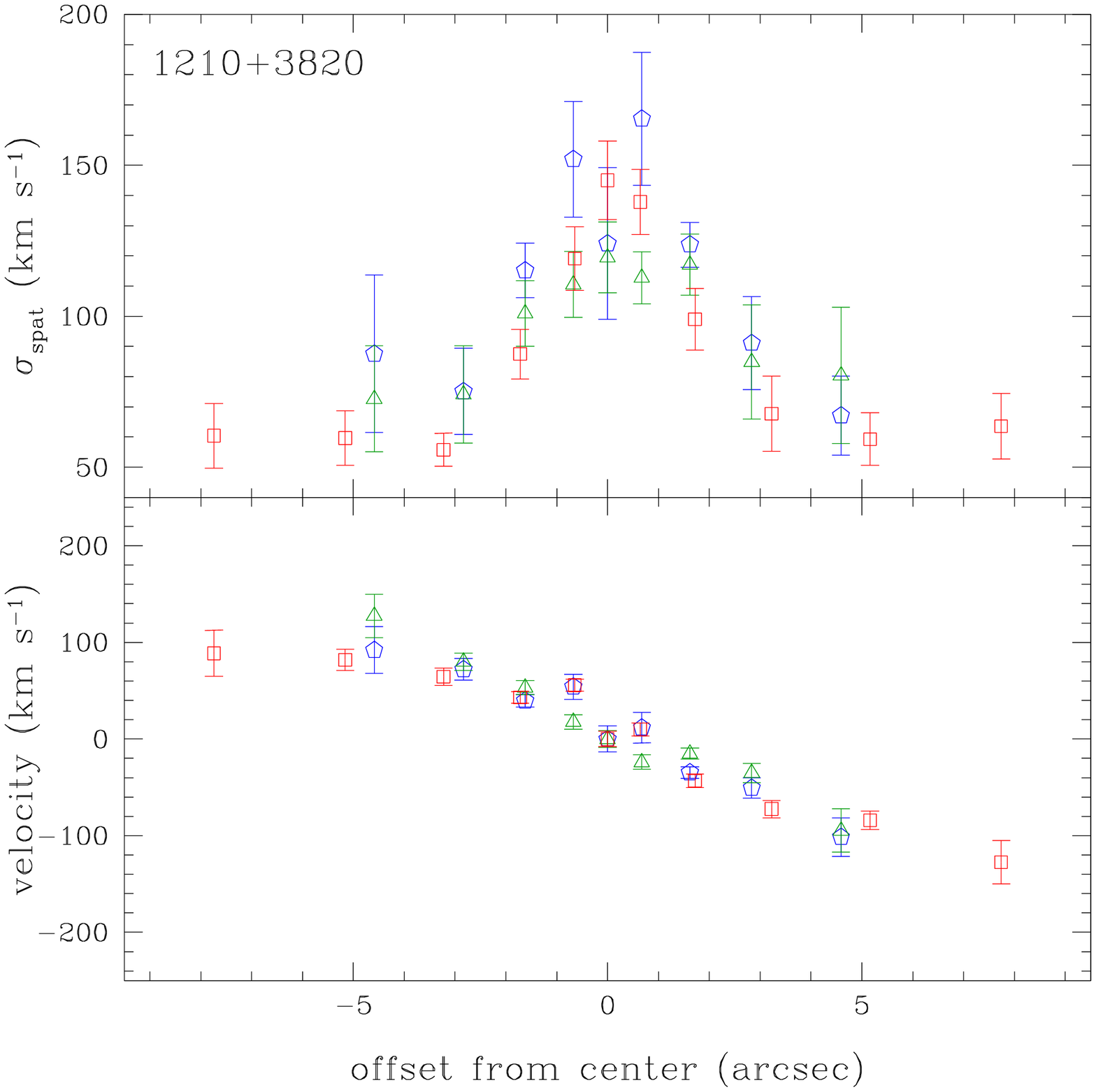}
\includegraphics[scale=0.16]{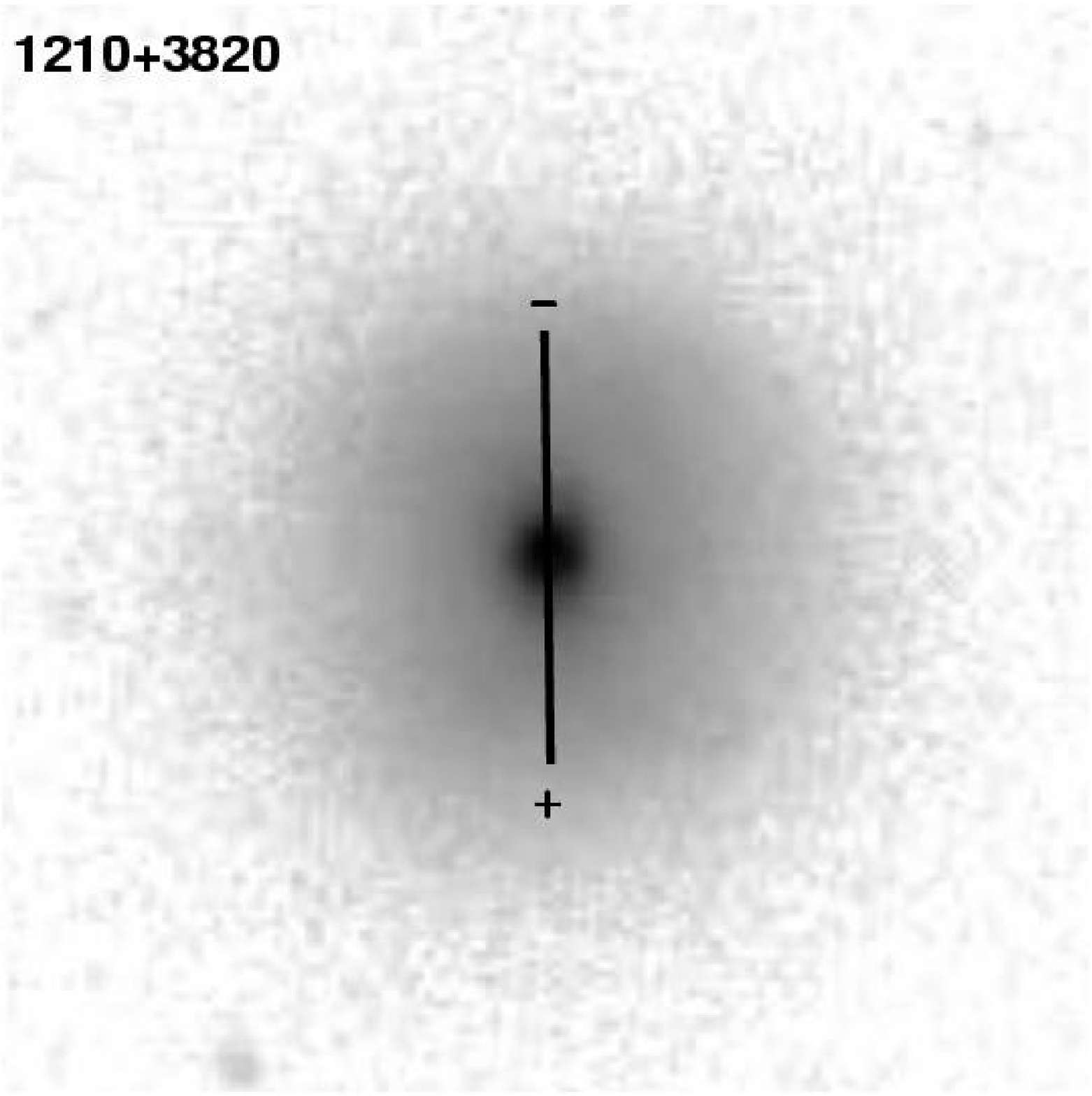}\hspace*{0.48cm}
\includegraphics[scale=0.23]{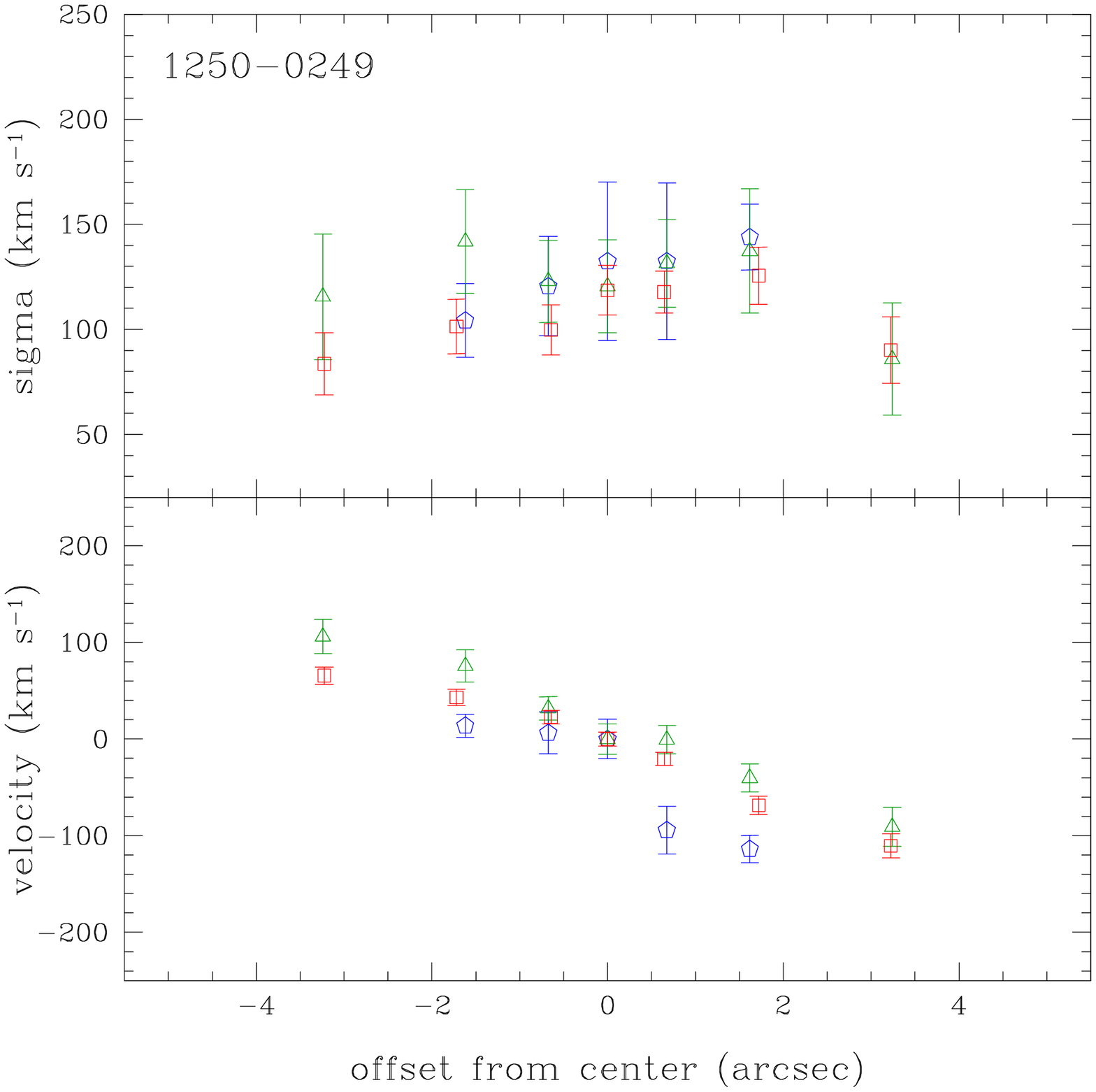}
\includegraphics[scale=0.16]{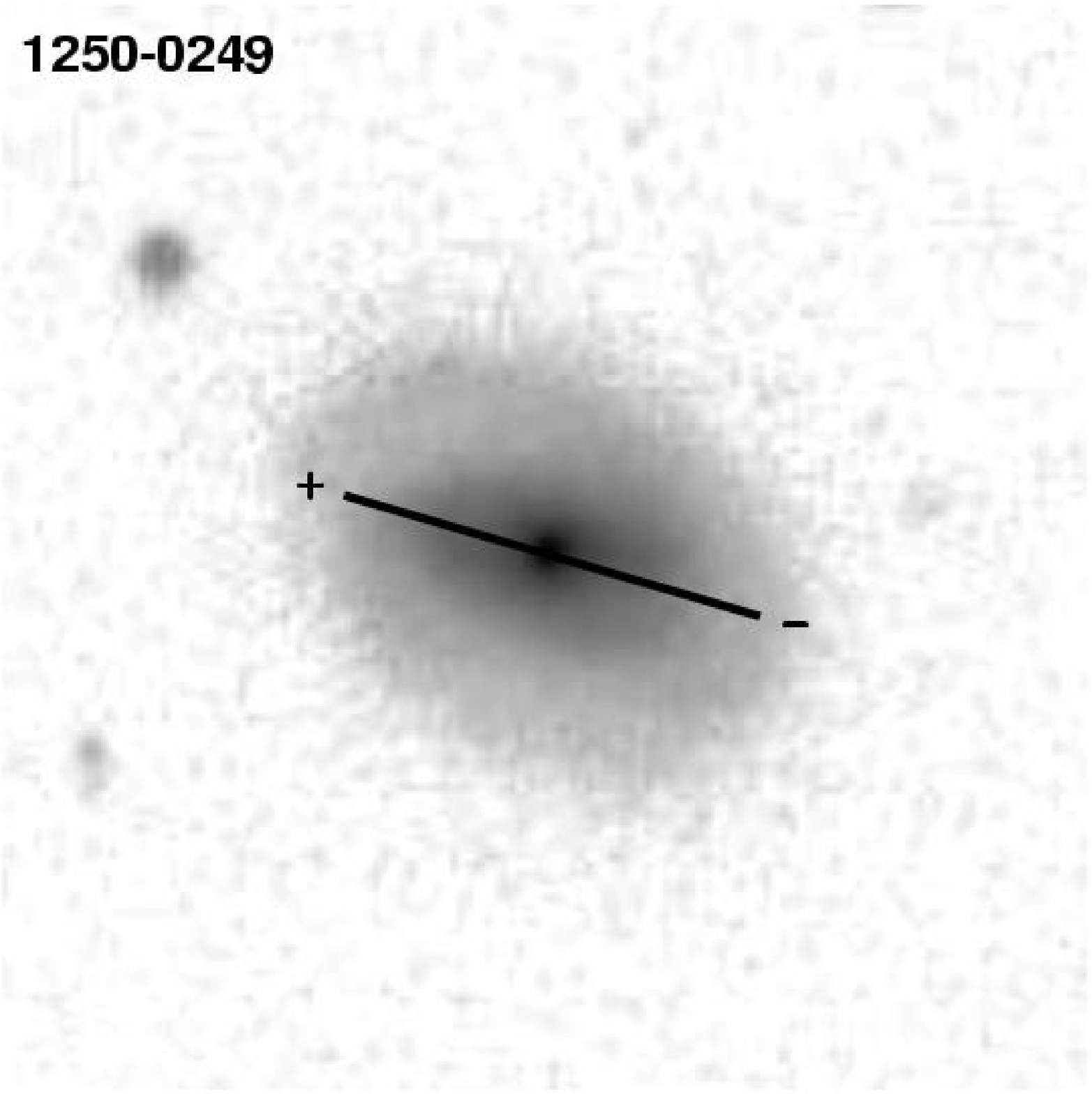}\\
\includegraphics[scale=0.23]{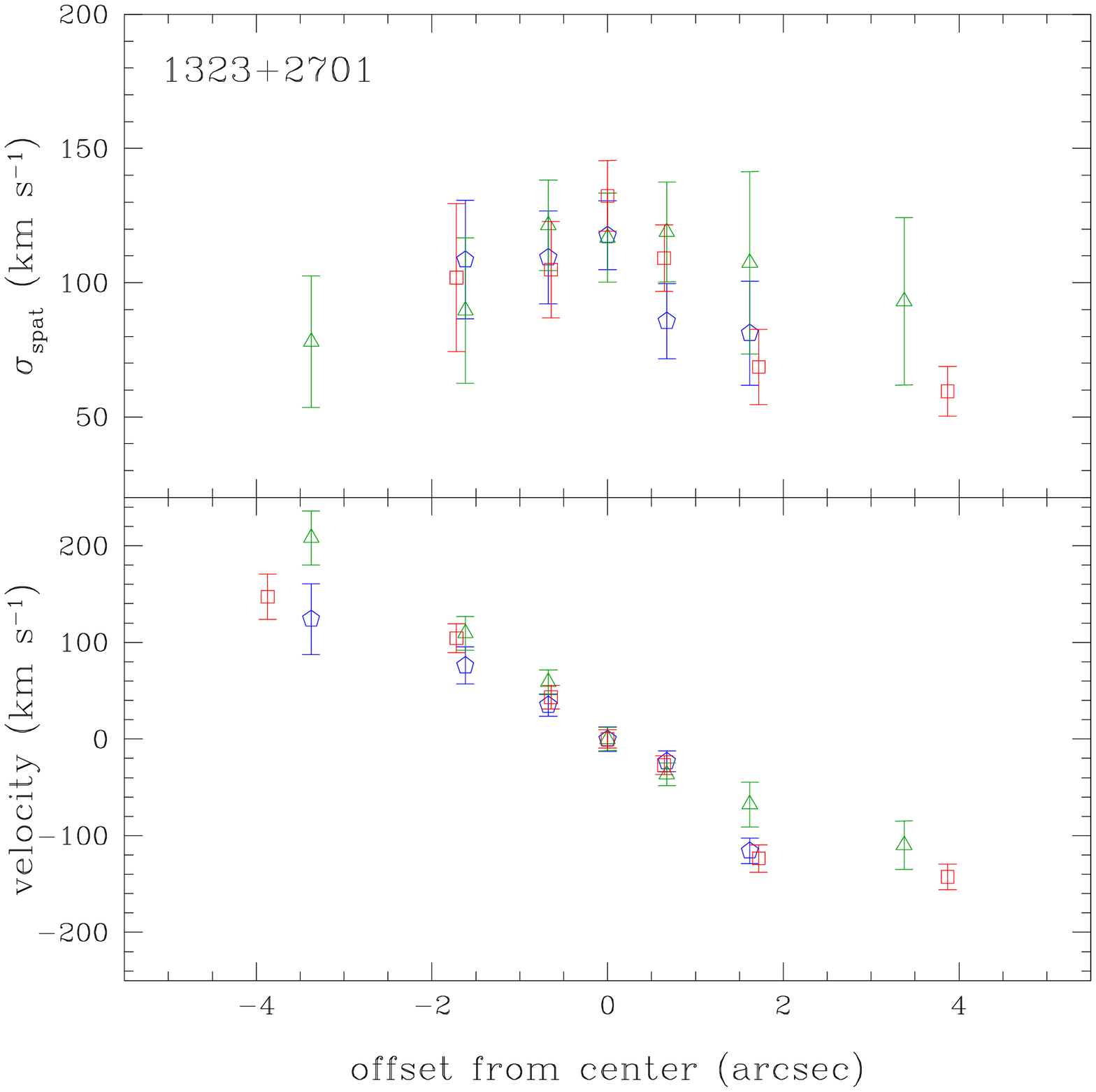}
\includegraphics[scale=0.16]{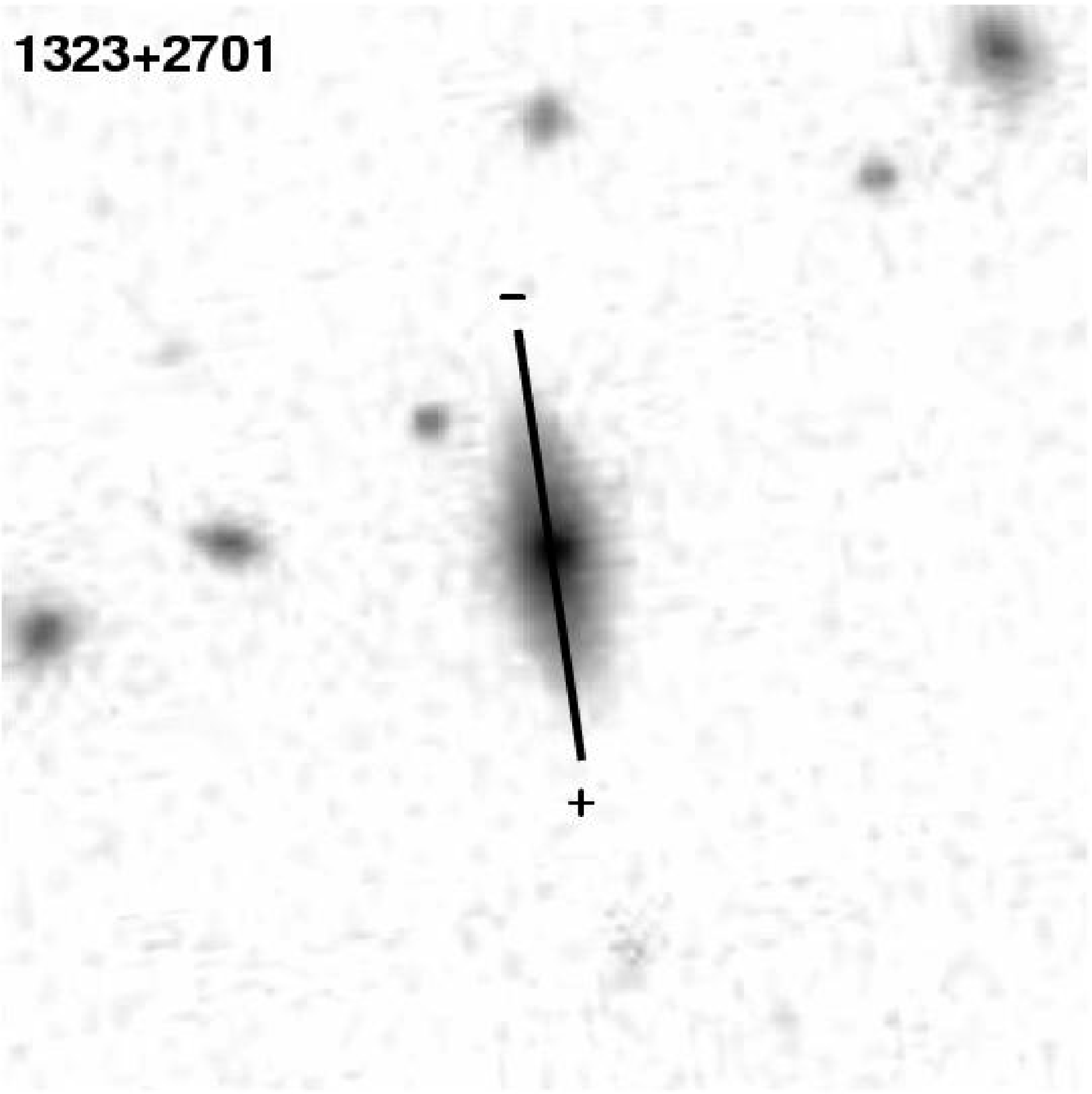}\hspace*{0.48cm}
\includegraphics[scale=0.23]{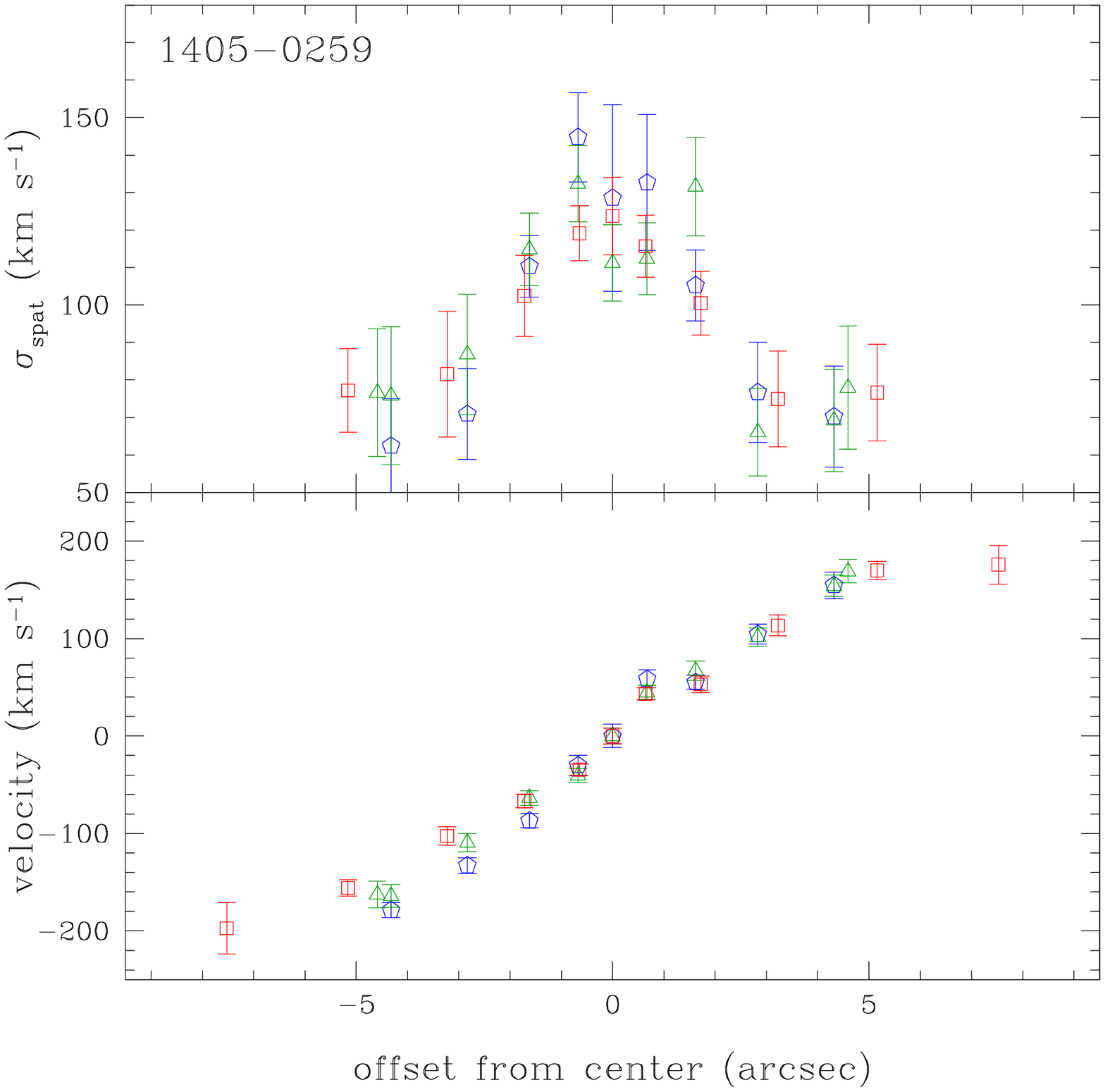}
\includegraphics[scale=0.16]{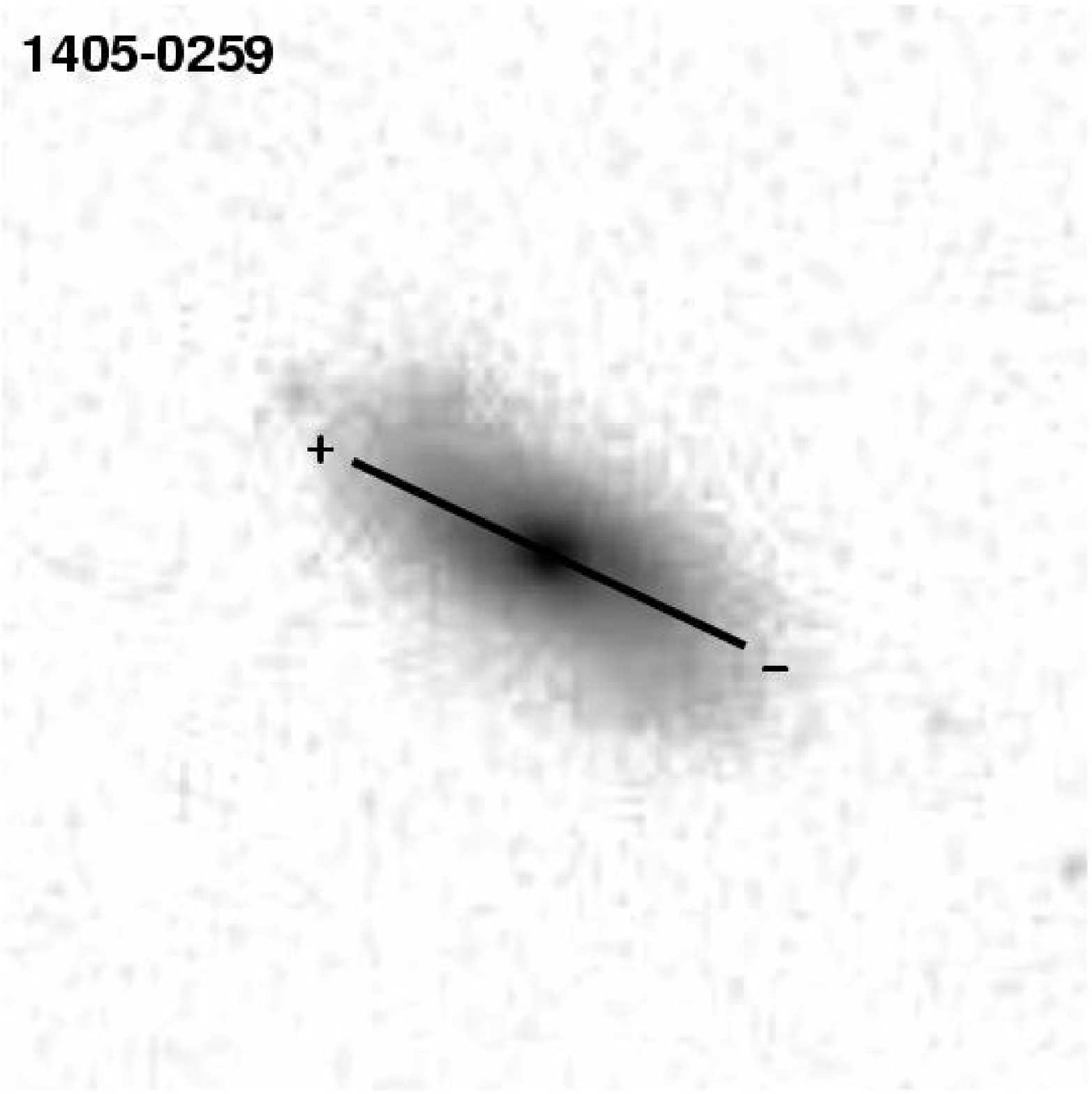}\\
\includegraphics[scale=0.23]{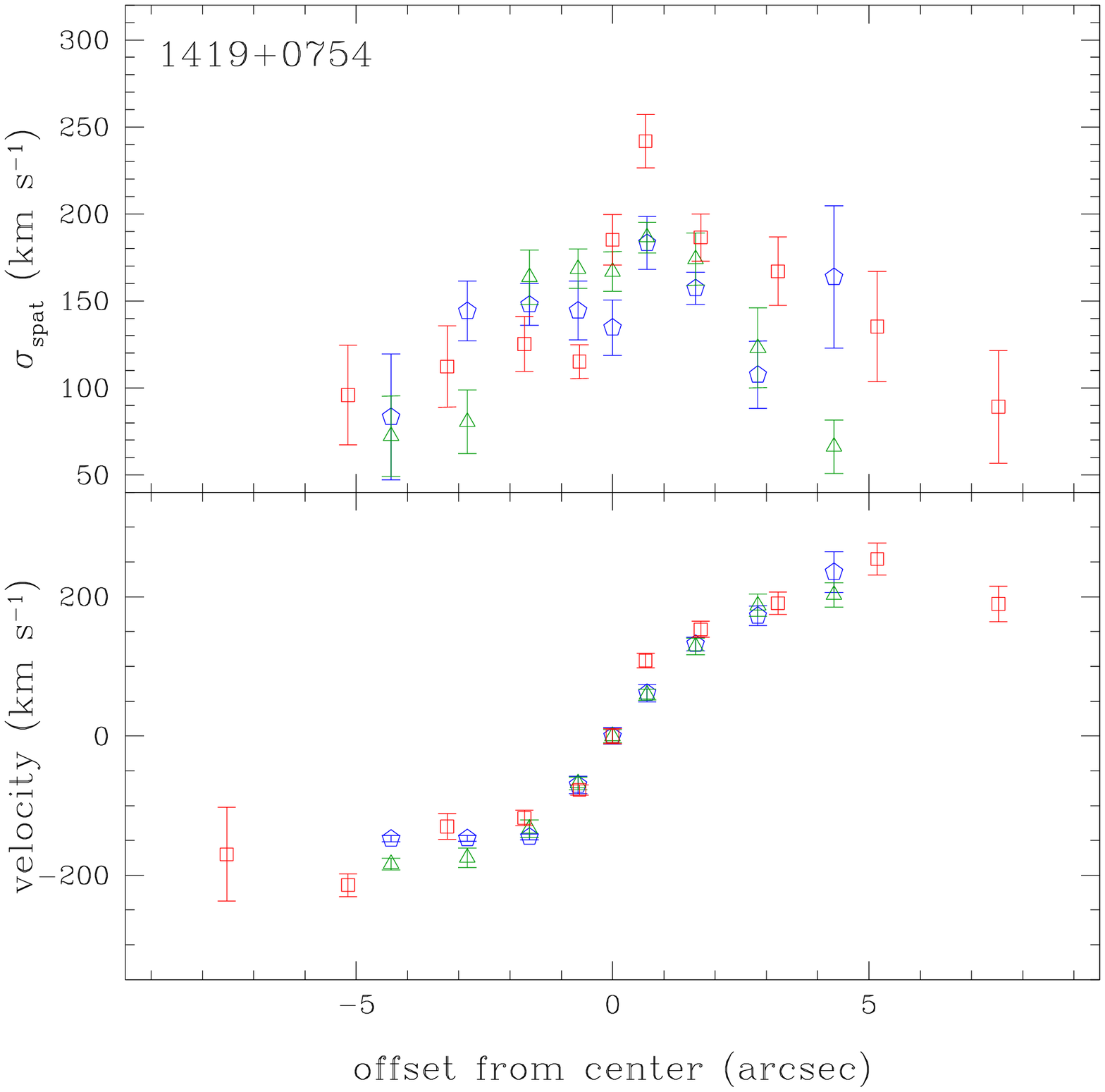}
\includegraphics[scale=0.16]{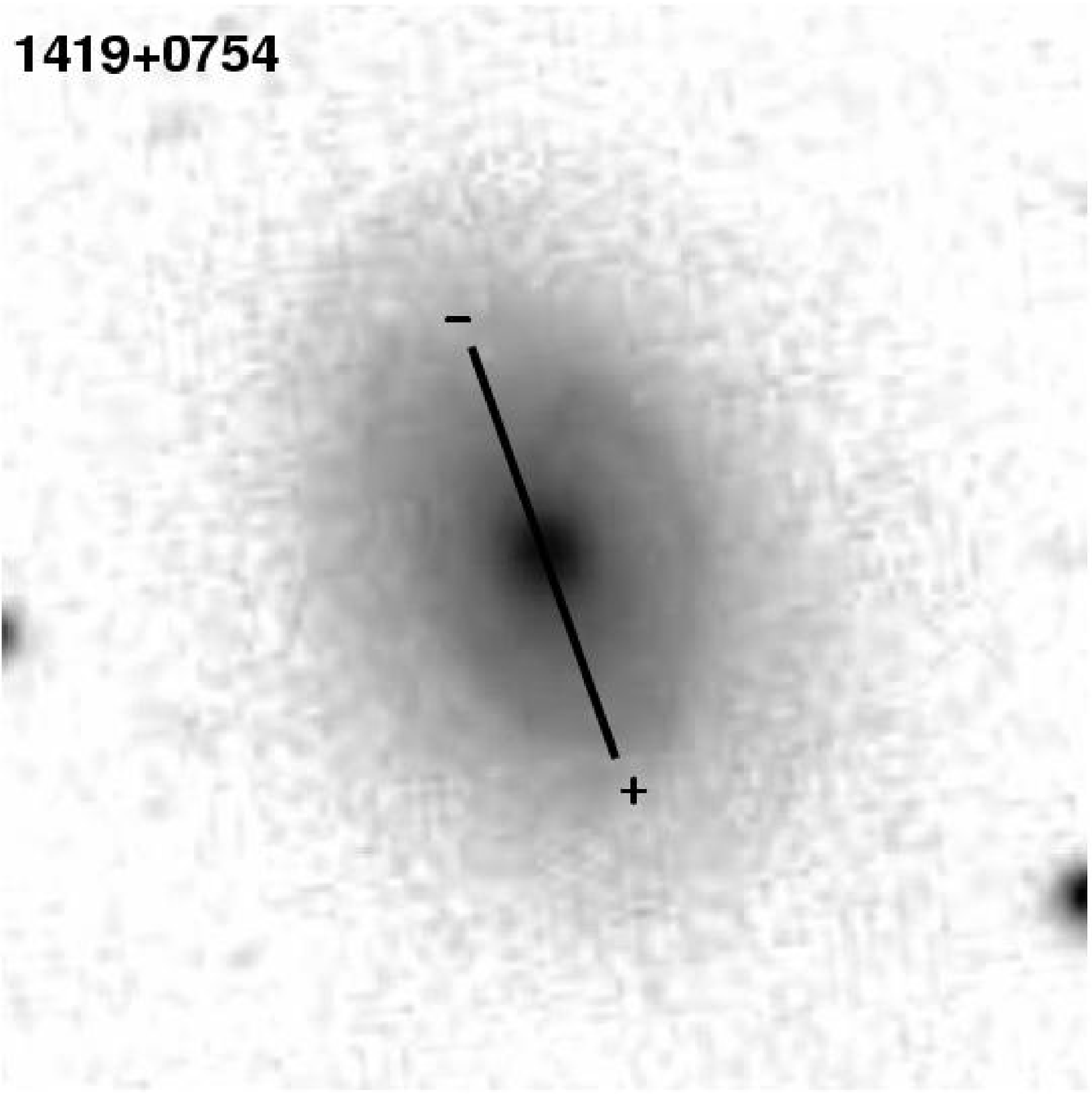}\hspace*{0.48cm}
\includegraphics[scale=0.23]{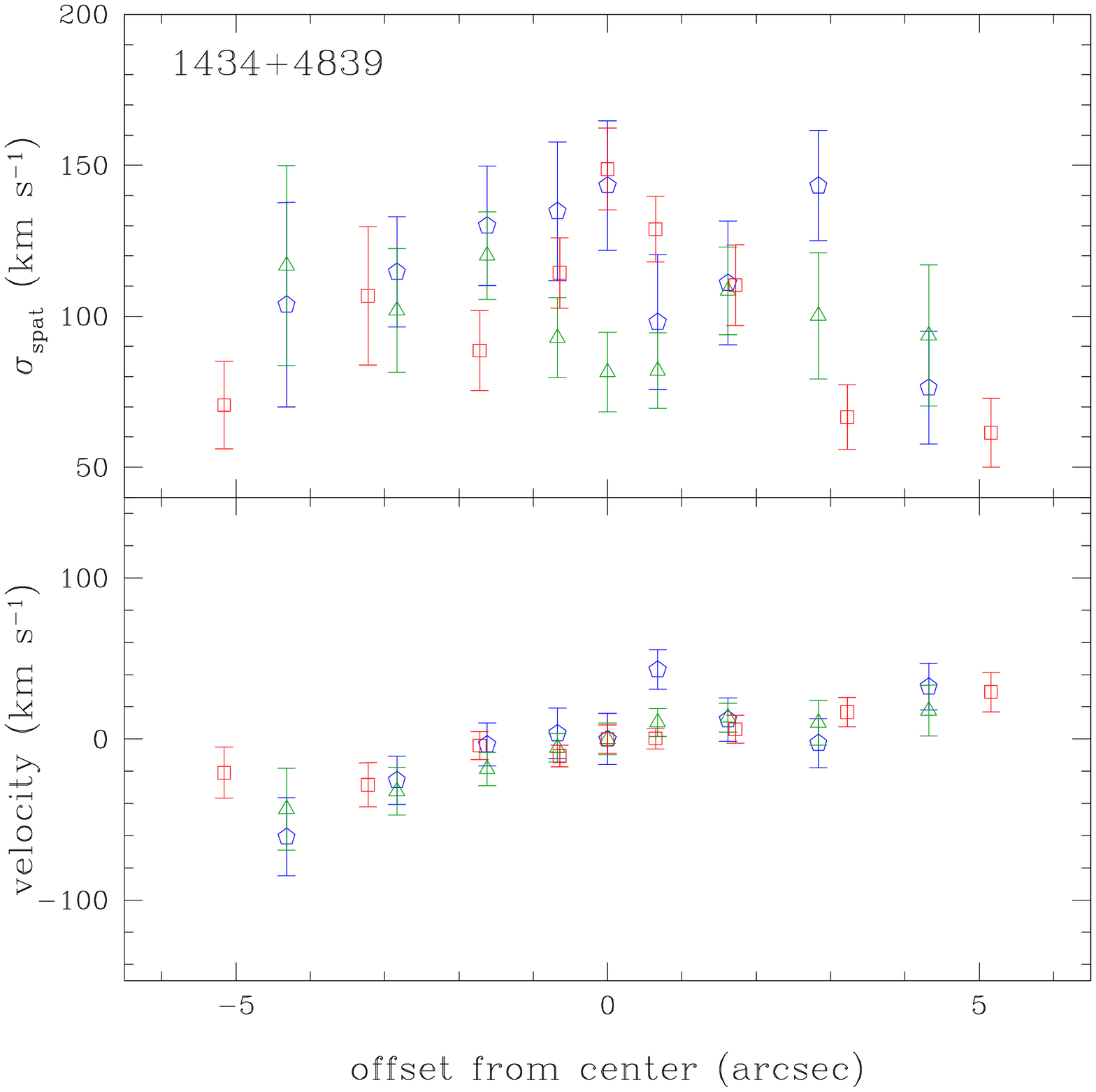}
\includegraphics[scale=0.16]{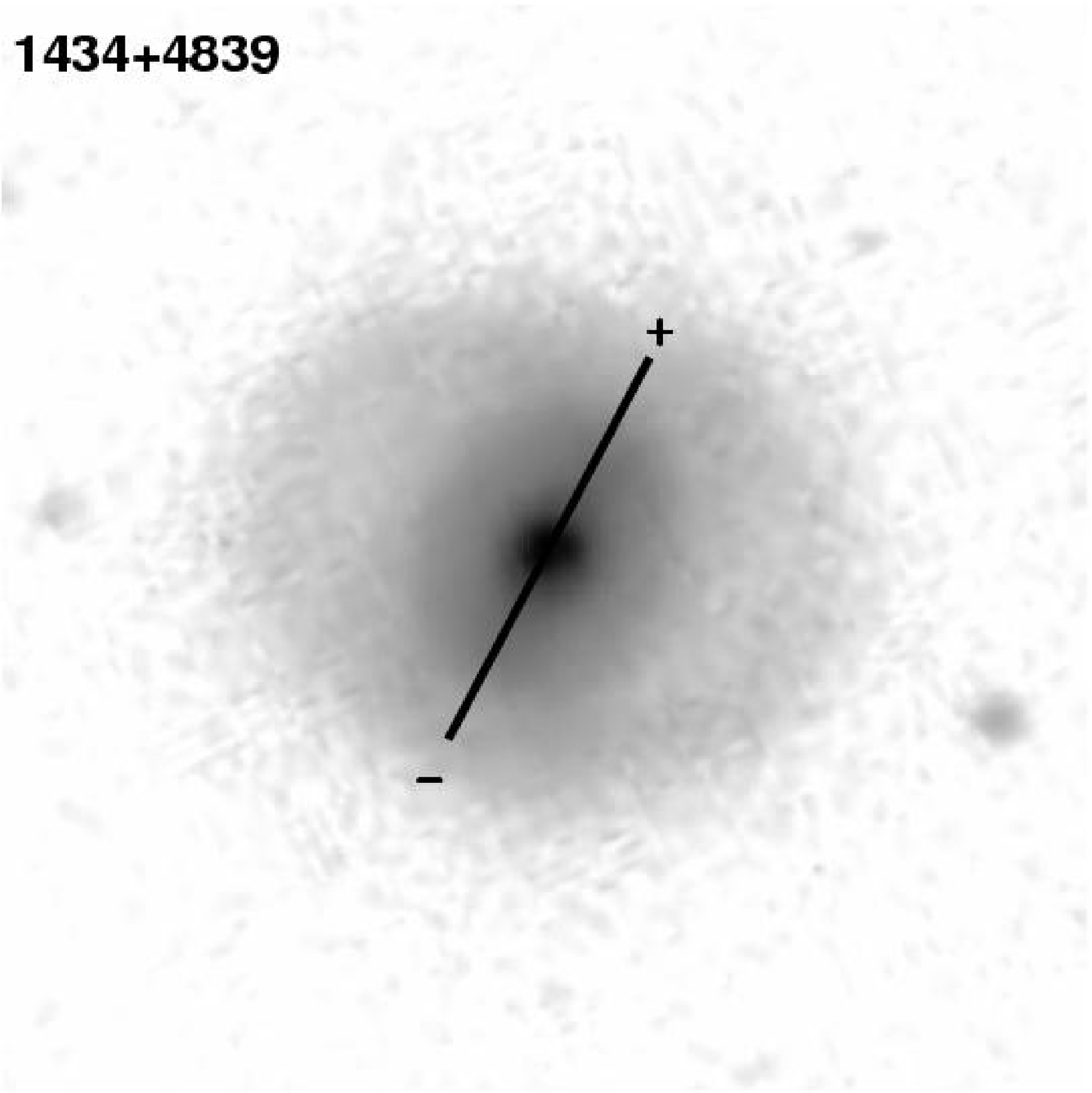}
\caption{The same as in Fig.~\ref{sigma_vel1}.}
\label{sigma_vel2}
\end{figure*}

\begin{figure*}[ht!]
\includegraphics[scale=0.23]{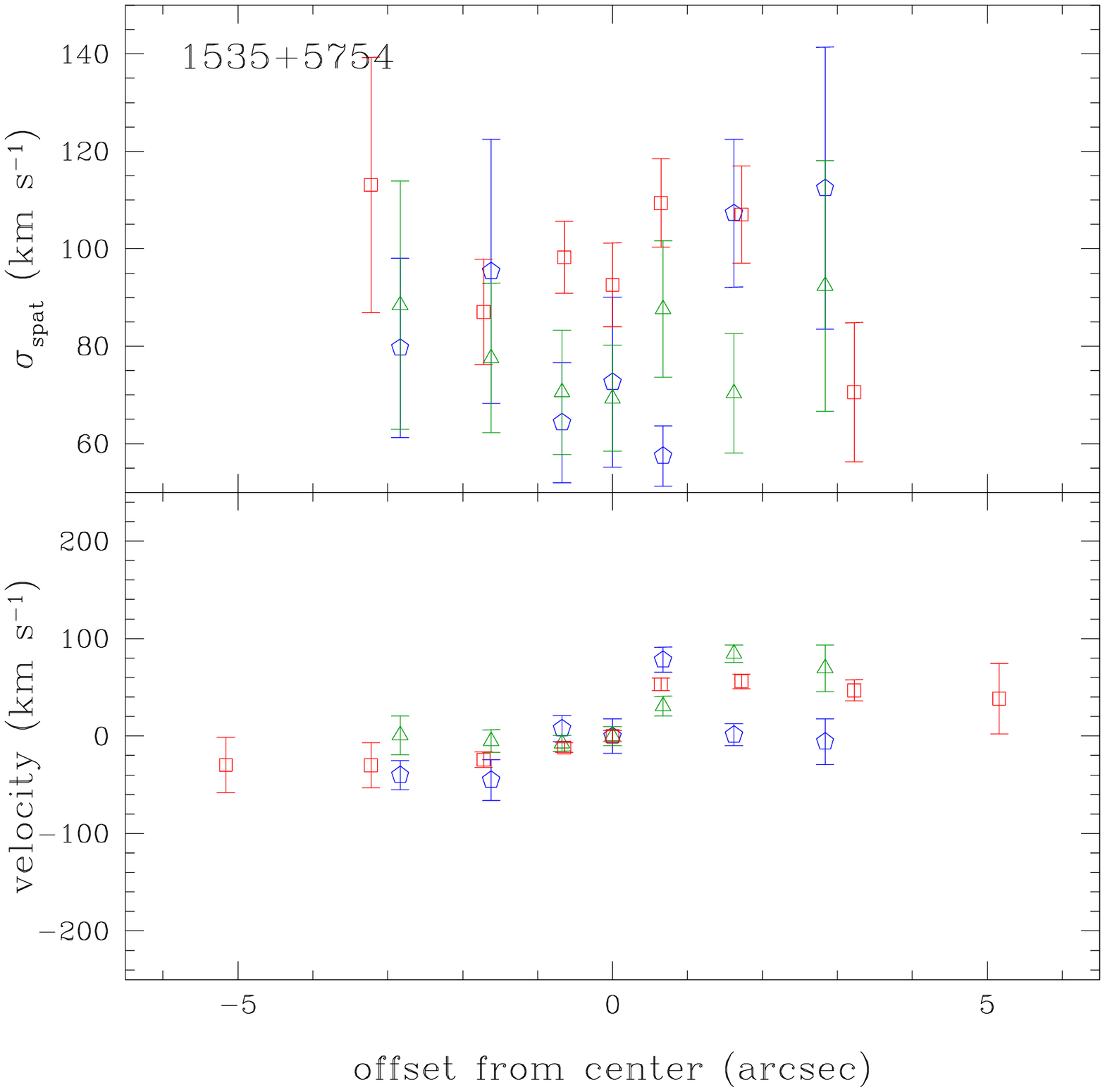}
\includegraphics[scale=0.16]{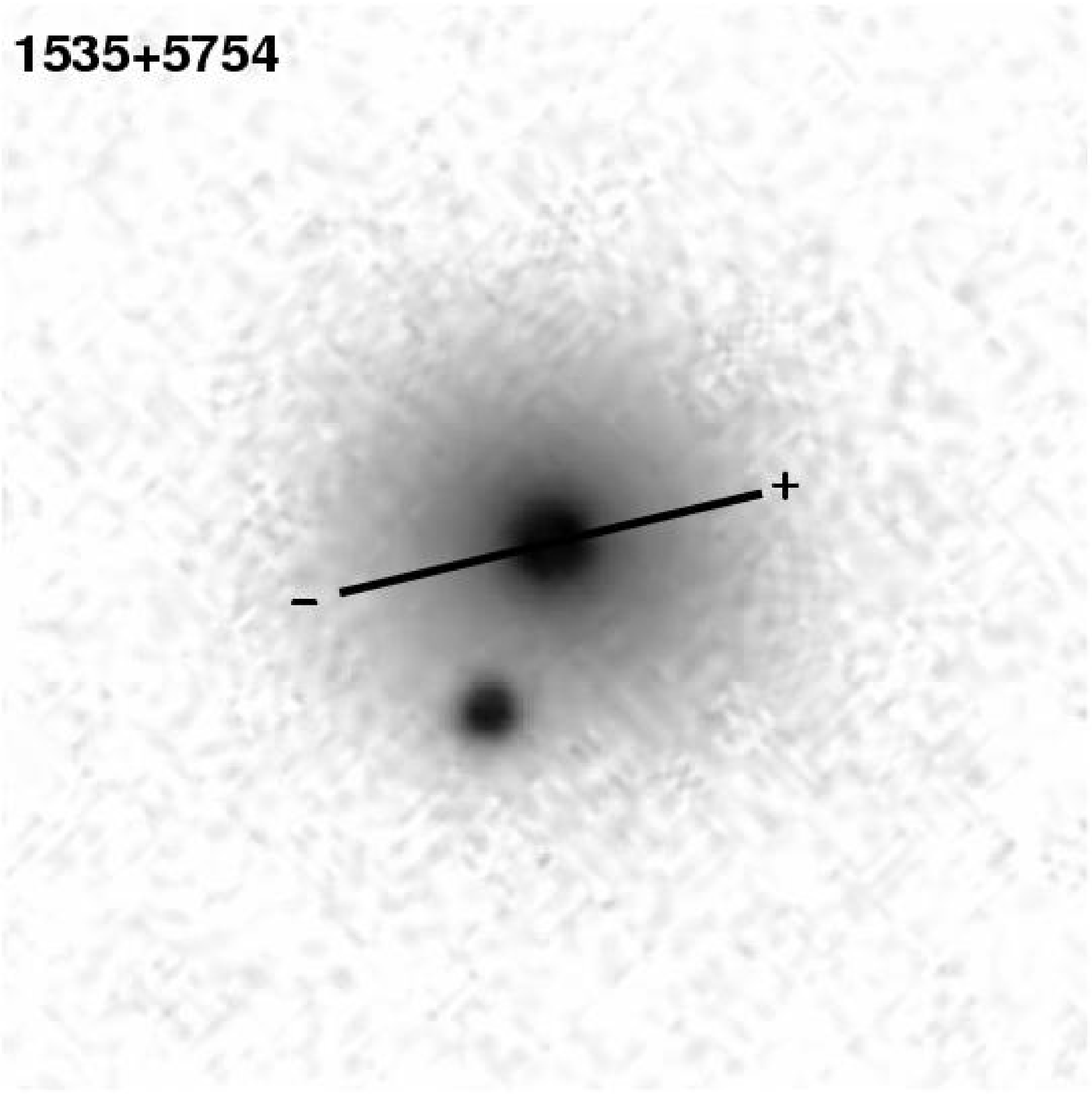}\hspace*{0.48cm}
\includegraphics[scale=0.23]{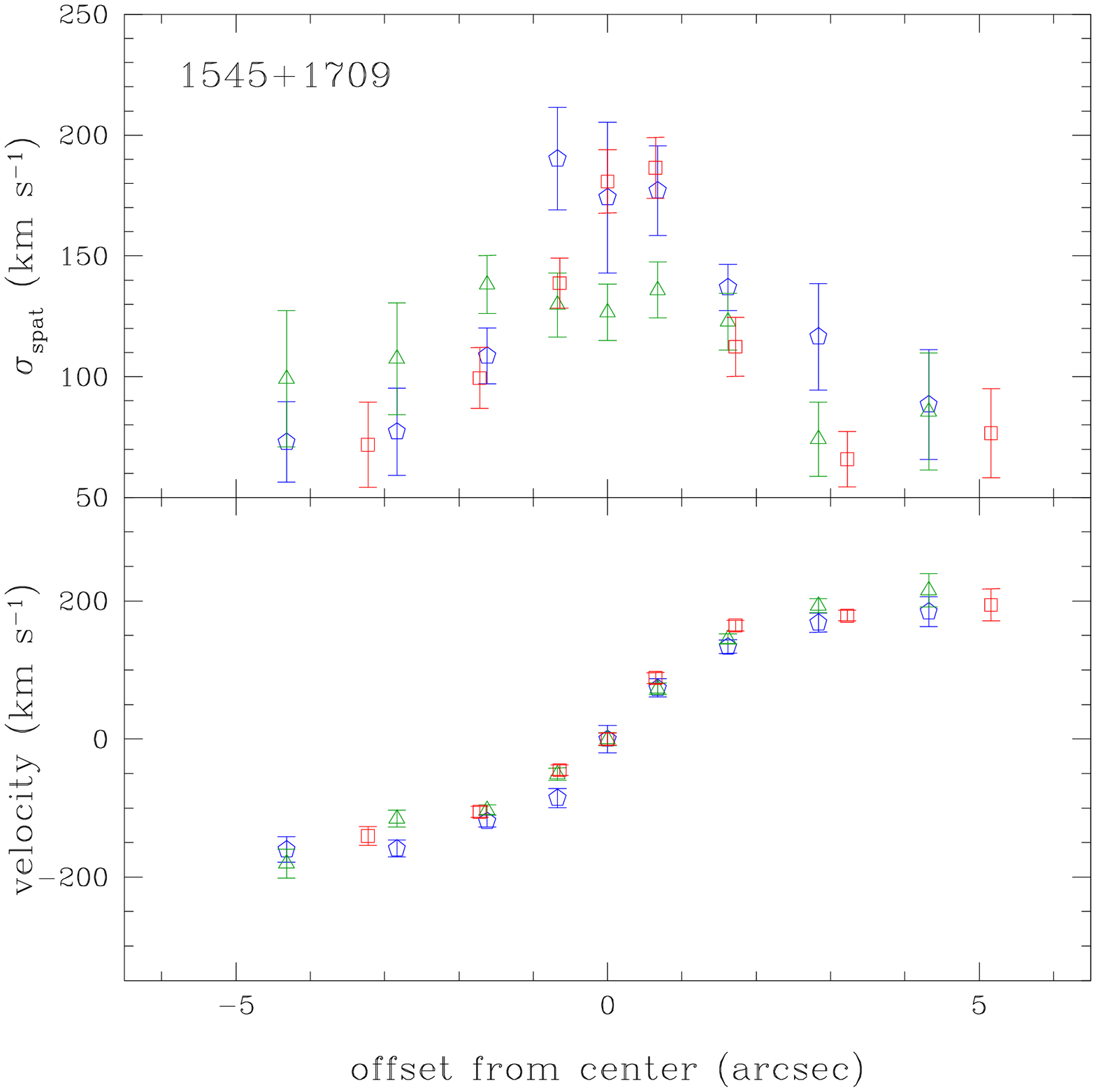}
\includegraphics[scale=0.16]{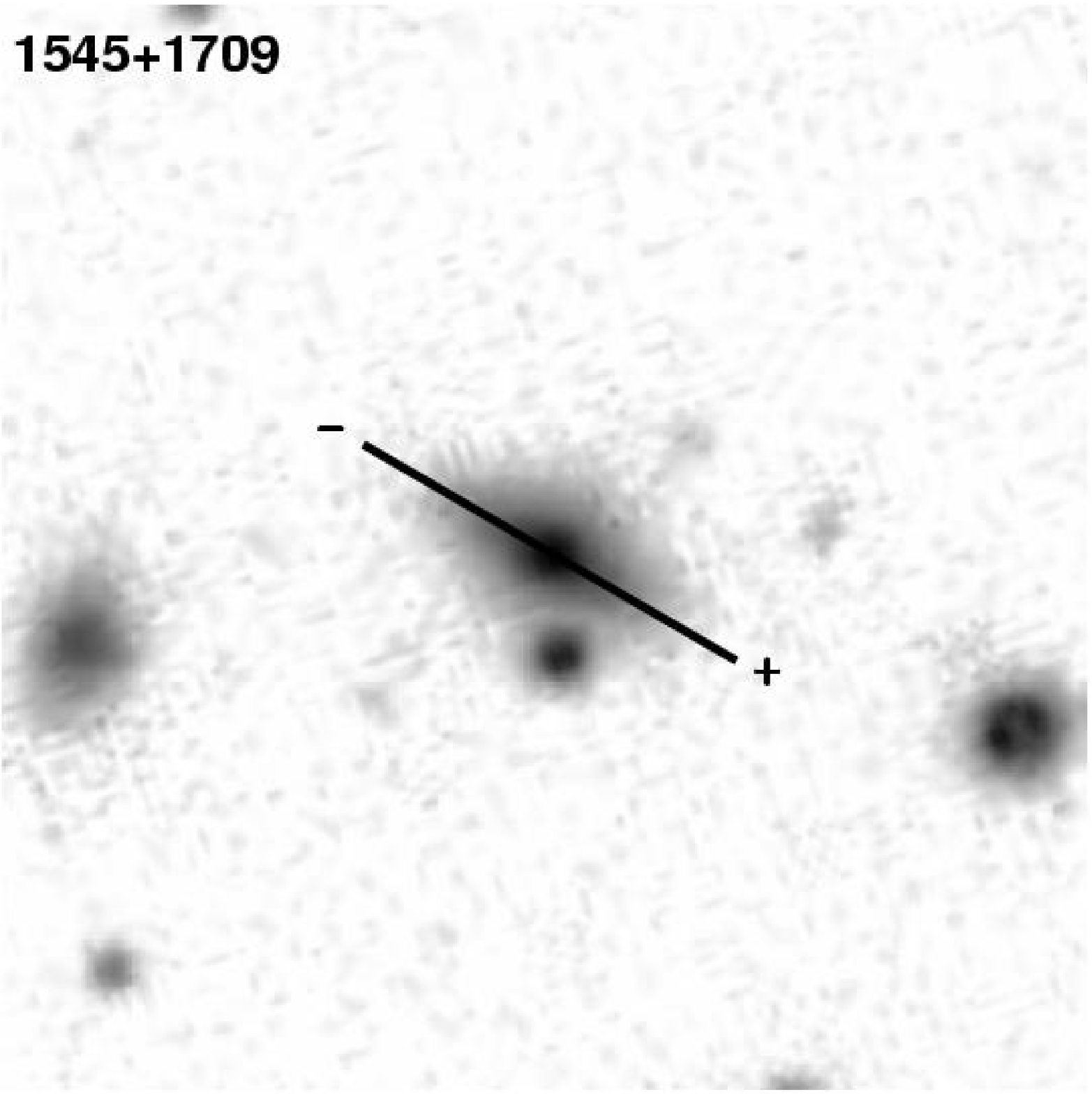}\\
\includegraphics[scale=0.23]{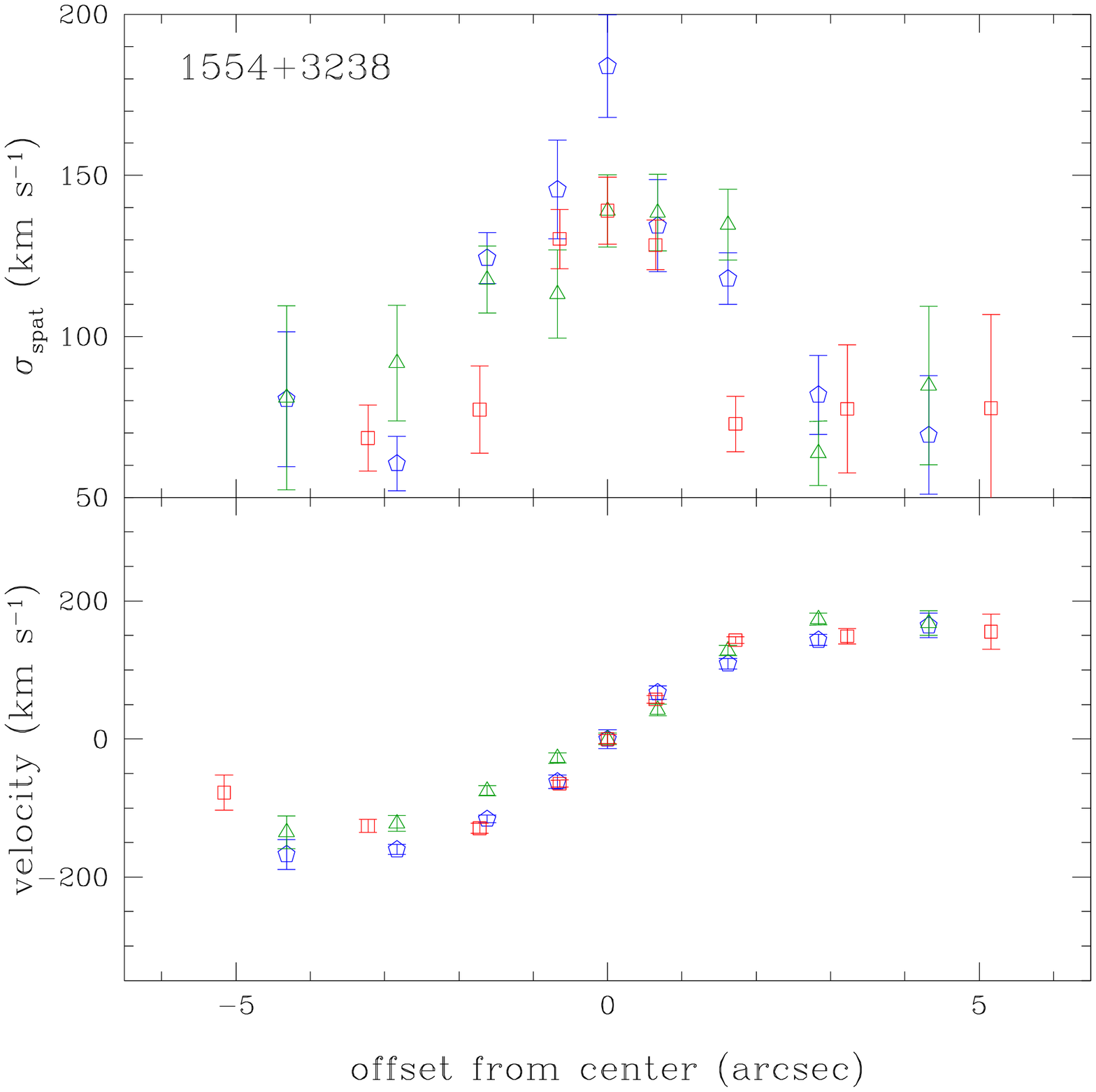}
\includegraphics[scale=0.16]{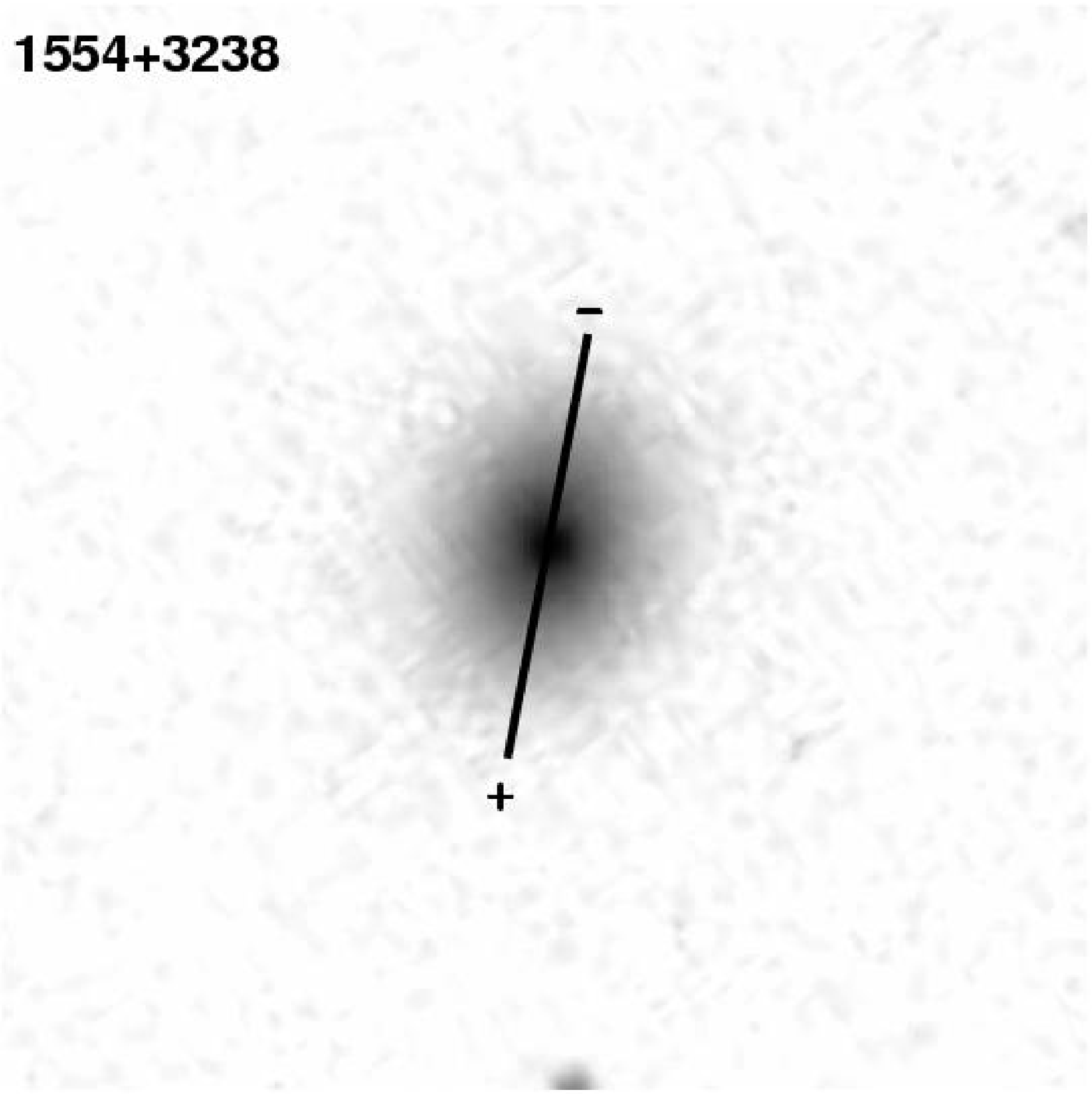}\hspace*{0.48cm}
\includegraphics[scale=0.23]{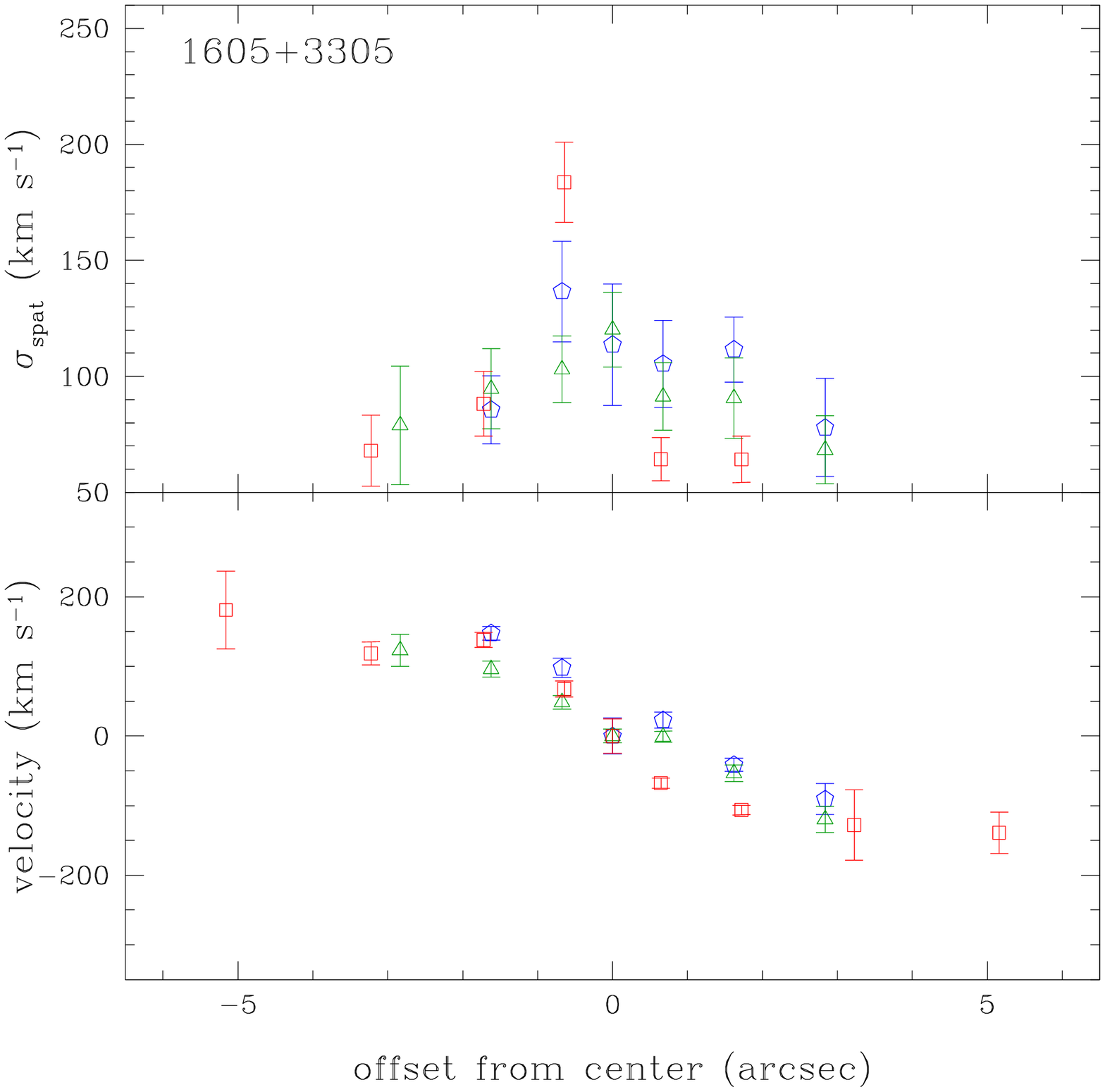}
\includegraphics[scale=0.16]{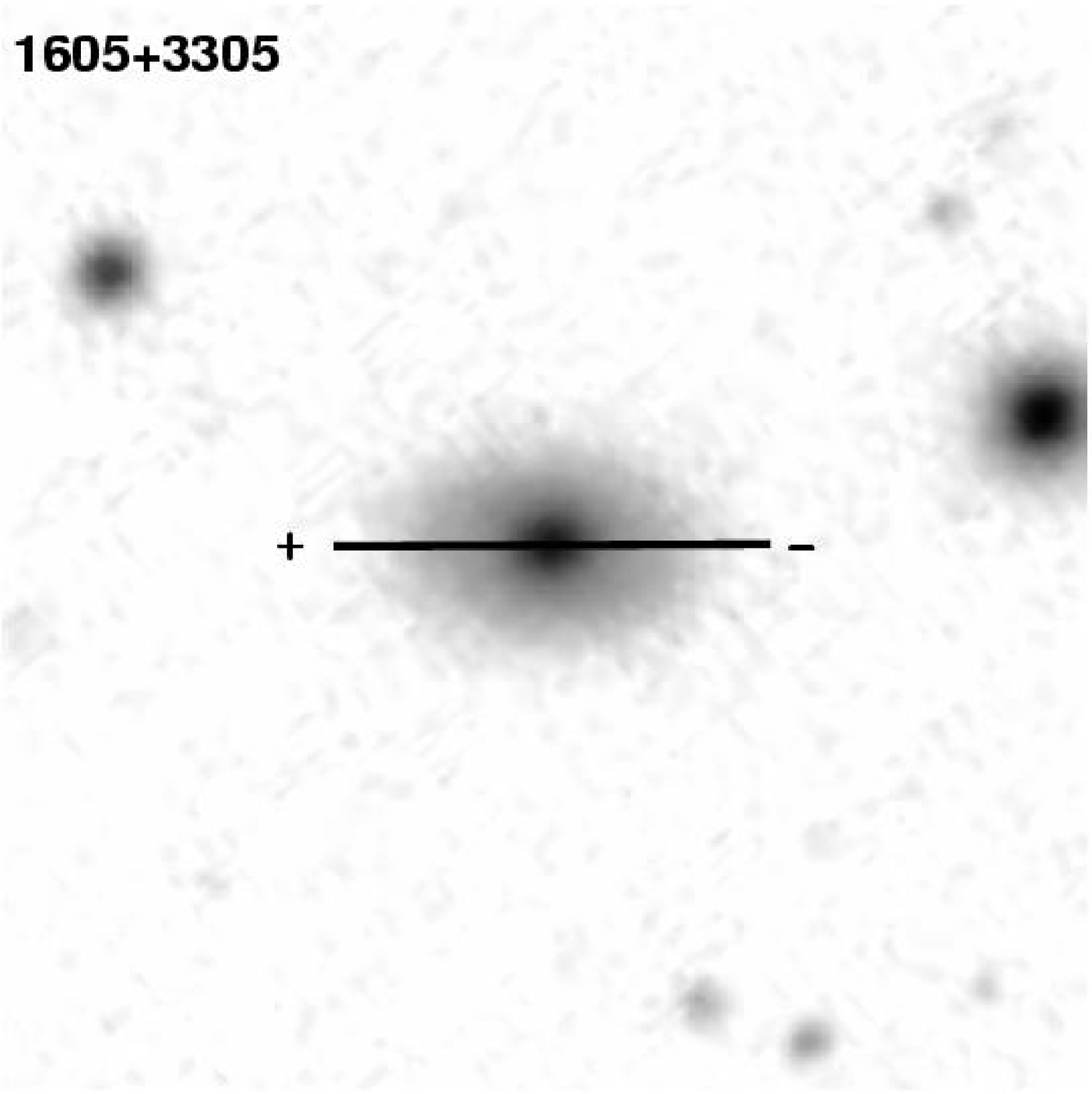}\\
\includegraphics[scale=0.23]{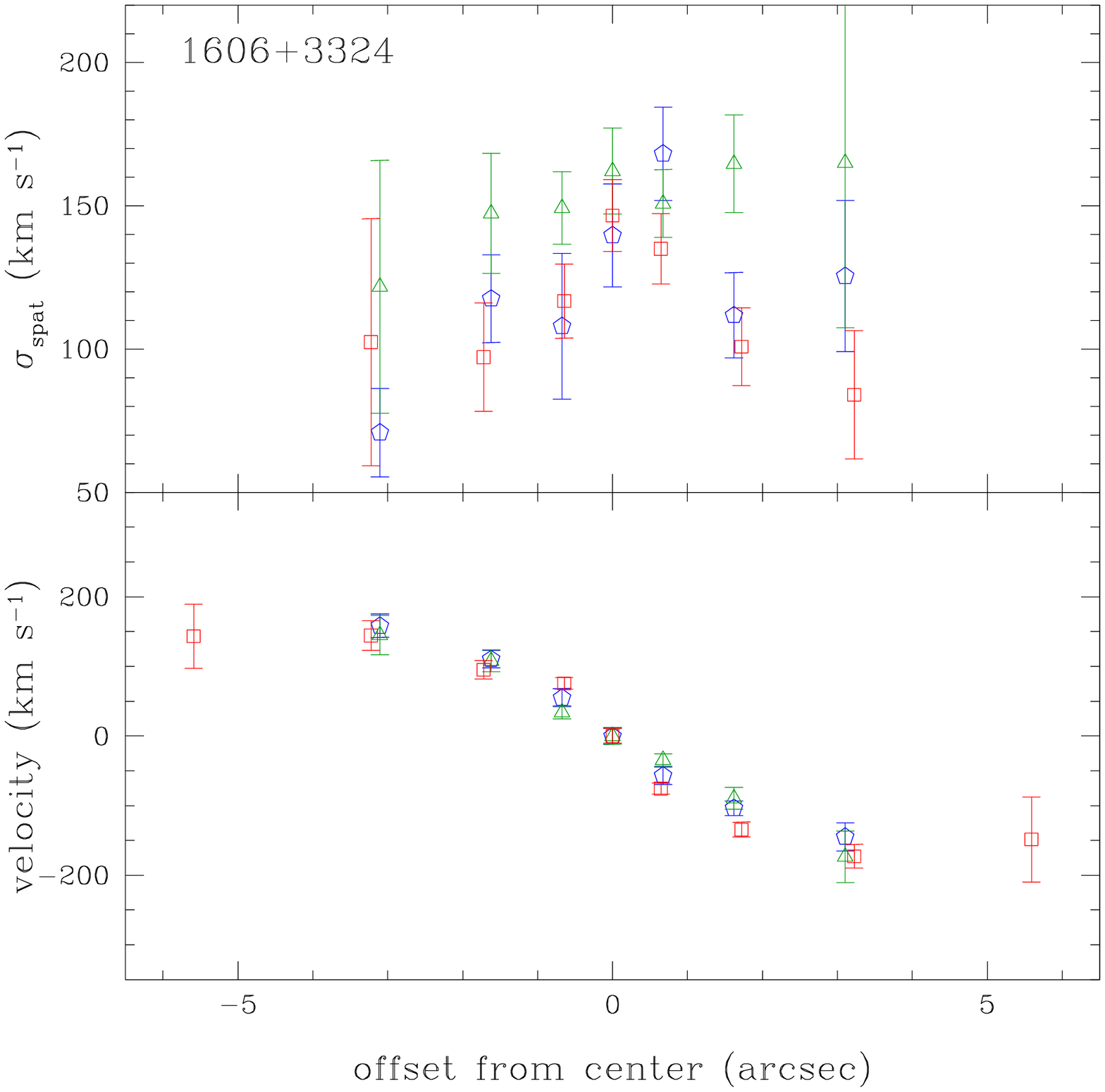}
\includegraphics[scale=0.16]{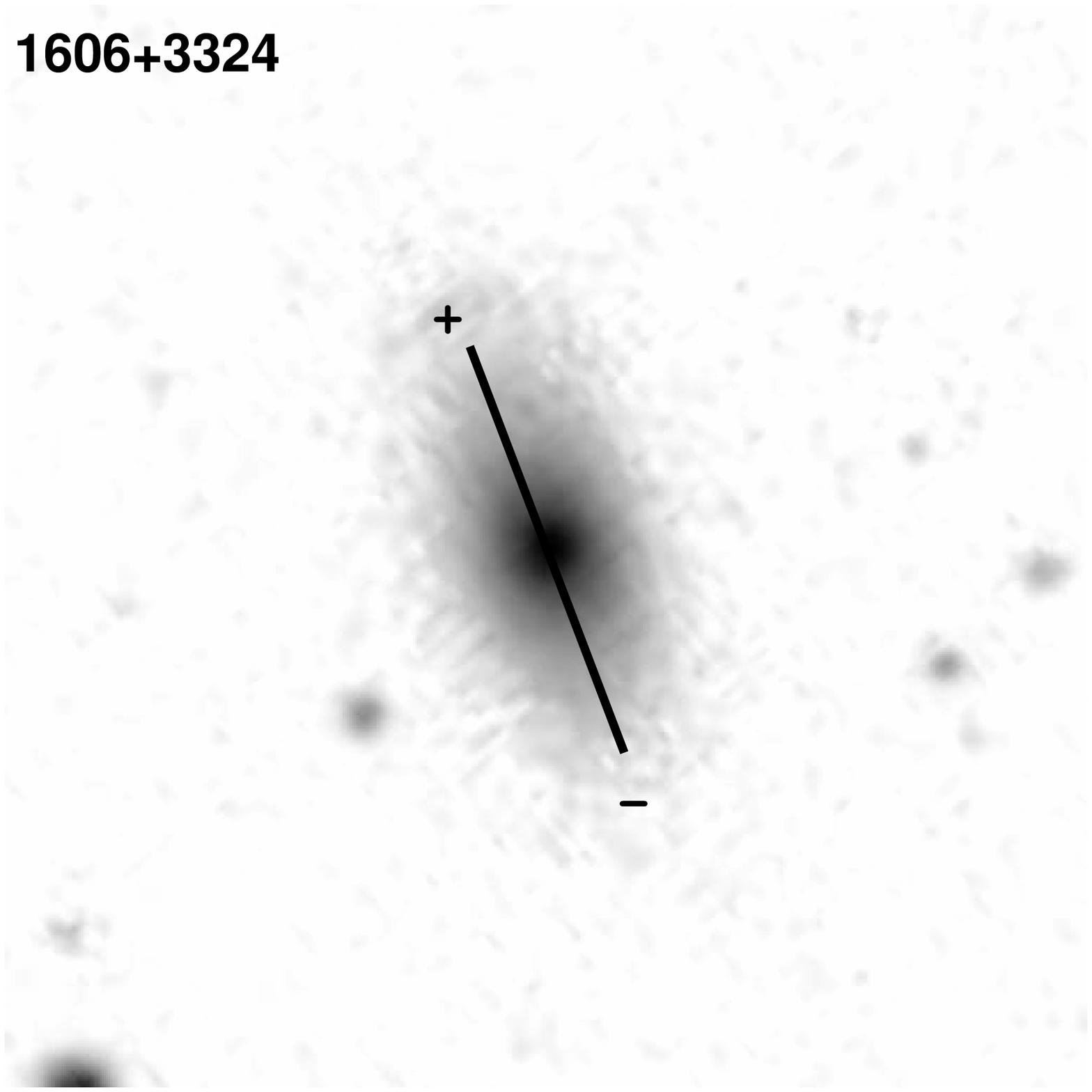}\hspace*{0.48cm}
\includegraphics[scale=0.23]{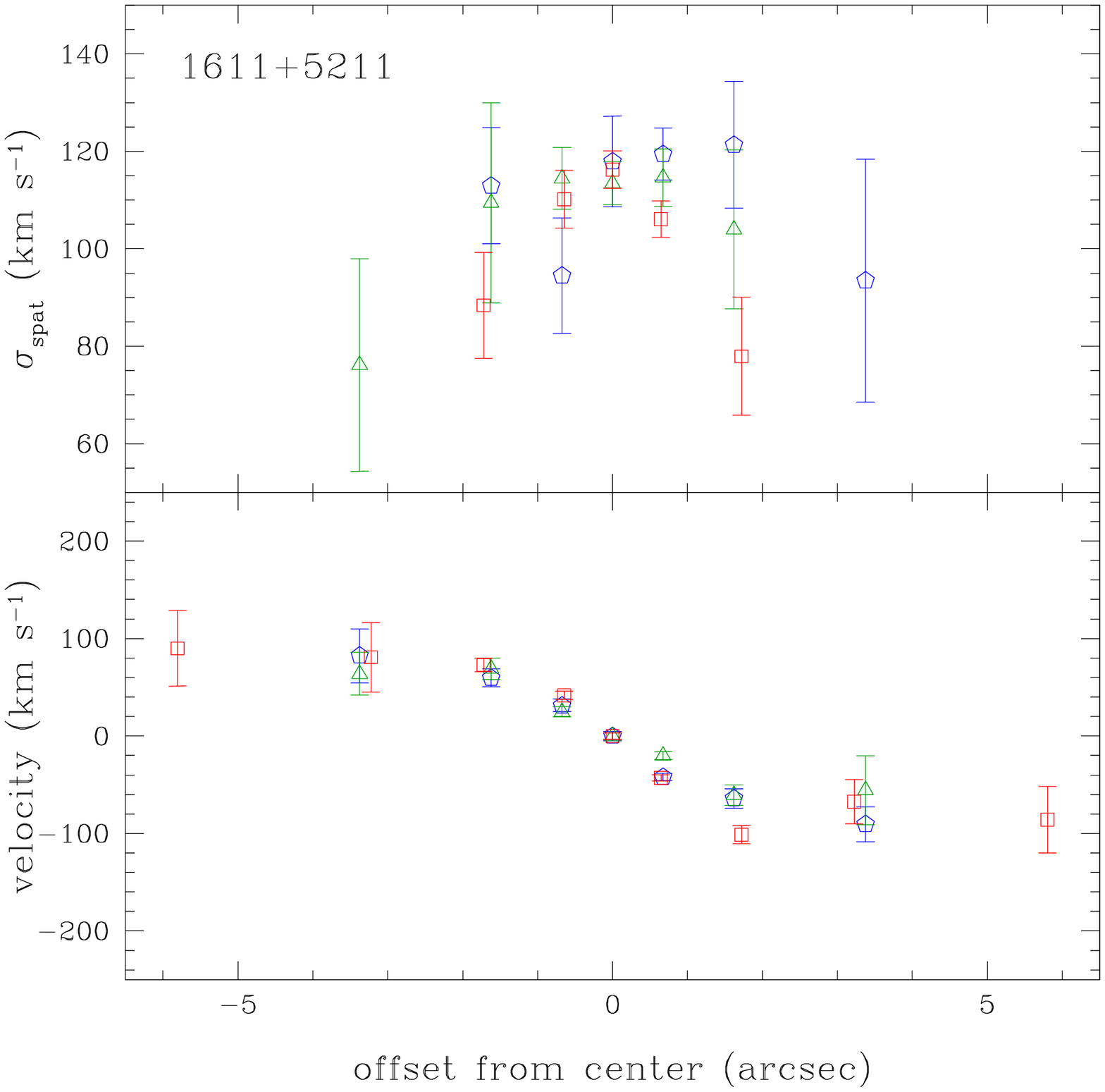}
\includegraphics[scale=0.16]{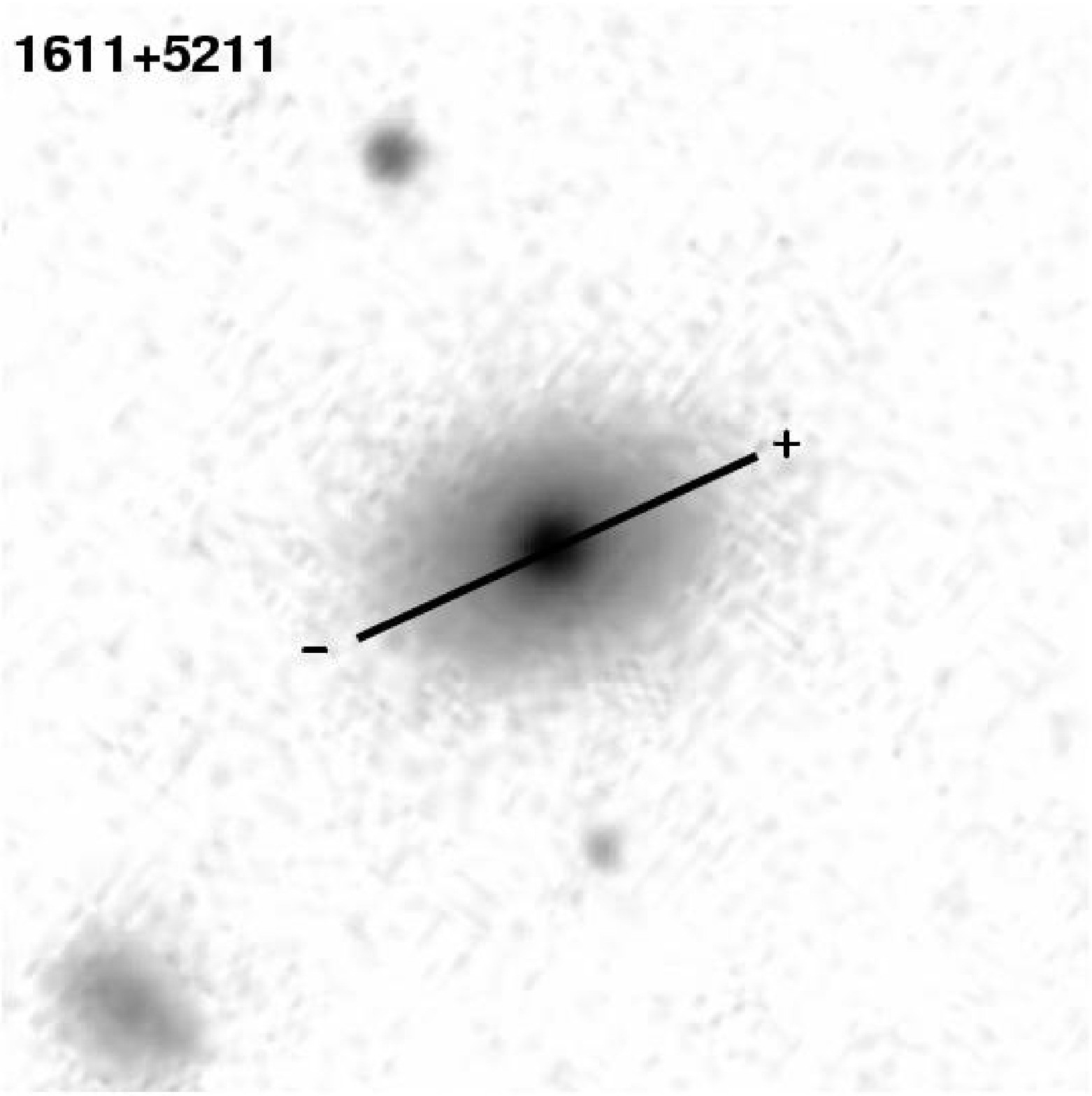}
\caption{The same as in Fig.~\ref{sigma_vel1}.}
\label{sigma_vel3}
\end{figure*}

The code used to determine the stellar-velocity dispersion
also gives the line-of-sight velocity.
(Note that we set the central velocity to zero.)
The resulting measurements for
both stellar-velocity dispersion as well as velocity as a function
of distance from the center are shown in Figures~\ref{sigma_vel1}-\ref{sigma_vel3}.

The error bars are often higher in the center due to AGN contamination.
This is also the reason why the error bars for velocity dispersions
determined from the MgIb or CaHK region are higher,
since in the blue, the contamination by the AGN powerlaw continuum
and broad emission lines is more severe than for the CaT region.
On the other hand, in the outermost spectra, the S/N is 
the dominating error source.

For comparison with literature, 
we choose the velocity dispersion determined from the CaT region
as our benchmark,
since this is the region  least affected by template mismatch \citep{bar02}
and AGN contamination from a featureless continuum as well as 
emission lines \citep{gre06b}.
We calculate the
luminosity-weighted line-of-sight velocity dispersion within the spheroid effective radius 
(determined from the surface photometry of the SDSS DR7 images as outlined below):
\begin{eqnarray*}
\sigma_{\rm spat, reff}^2 = \frac{\int\limits_{\rm -reff}^{\rm reff} \sigma_{\rm spat}^2 (r) \cdot I (r) \cdot r \cdot dr}{\int\limits_{\rm -reff}^{\rm reff} I(r) \cdot r \cdot dr}
\end{eqnarray*}
with $I(r)$ = $I (\rm {reff}) \cdot \exp (-7.67 \cdot [(r/r_{\rm reff})^{0.25}-1])$ \citet{dev48} profile.
(The range ``-reff'' to ``+reff'' refers to the fact that we extracted spectra symmetrically around
the center of each object, along the major axis, and measured stellar velocity dispersions
from each of them; see also Figs.~\ref{sigma_vel1}-\ref{sigma_vel3}.)
As our $\sigma_{\rm spat}$ measurements are discrete, we interpolate over the appropriate
radial range using a spline-function.
In the following, we refer to 
the spatially-resolved stellar-velocity dispersion within the spheroid effective radius as
$\sigma_{\rm spat, reff}$. This represents the spheroid-only dispersion
within the effective radius, free from broadening due to a rotating disk component.
Note that the only place where we show
spatially-resolved velocity dispersion {\it at} a certain radius is in 
Figures~\ref{sigma_vel1}-\ref{sigma_vel3} ($\sigma_{\rm spat}$) and in Figure~\ref{sigmacompare}a for the ratios;
otherwise we always refer to the luminosity-weighted spatially-resolved velocity dispersion 
{\it within} the effective radius $\sigma_{\rm spat, reff}$.

We can in principle choose arbitrary integration limits,
to e.g.~determine the stellar-velocity dispersion within one-eighth of the effective radius,
one-half of the effective radius, or 1.5\arcsec~(comparable to SDSS fiber spectra). 
While these are all ``radii'' found in the literature,
we will not be using them for comparison in this paper,
since literature values all refer to aperture data (and not spatially-resolved
as discussed here; see next section).

\subsection{Aperture Stellar-Velocity Dispersion}
\label{subsec:ap}
\begin{figure*}[ht!]
\begin{center}
\includegraphics[scale=0.7]{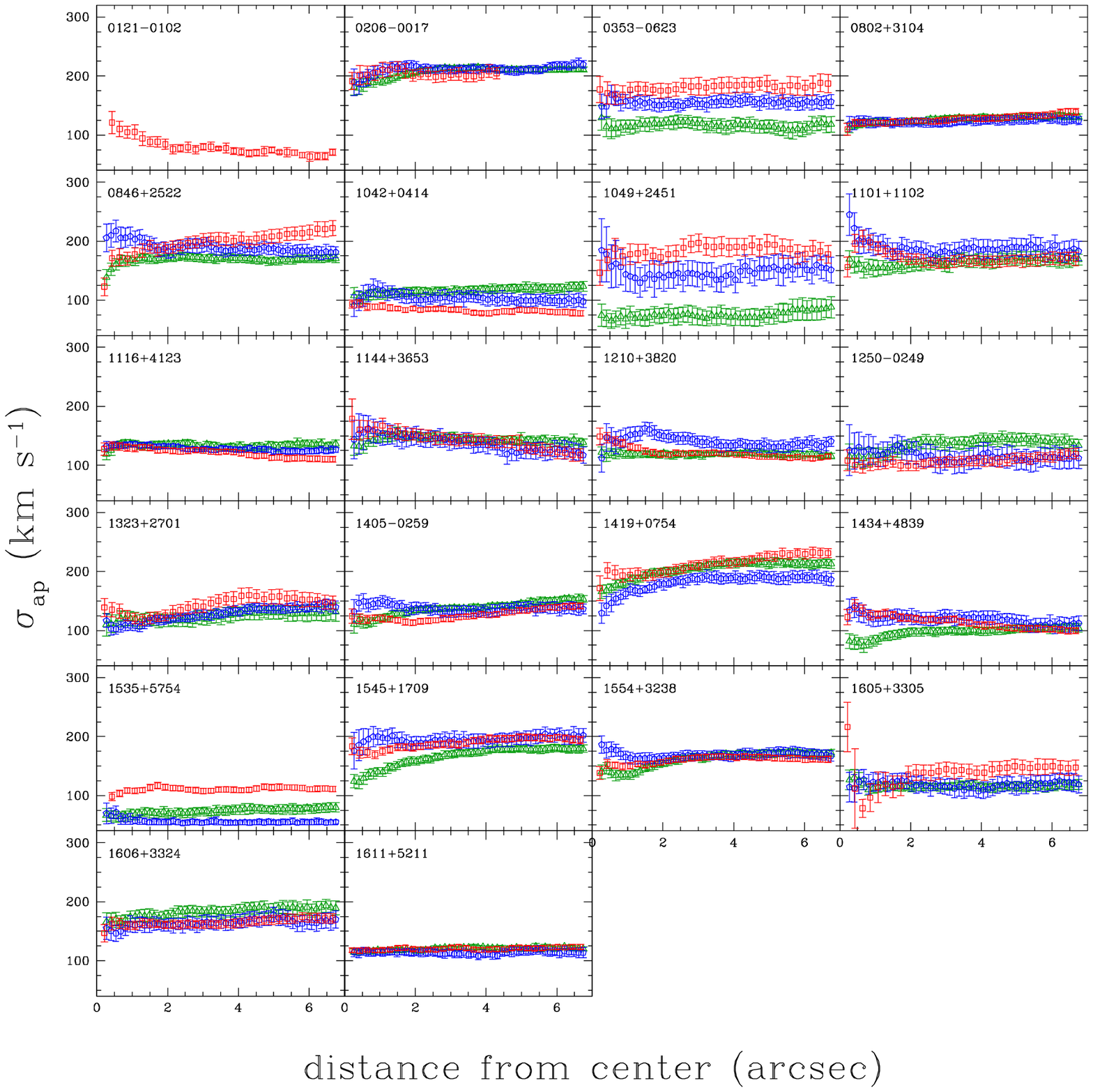}
\end{center}
\caption{Stellar-velocity dispersion as derived from the CaHK region (blue pentagons), the
MgIb region (green triangles), and the CaT region (red squares). 
In contrast to Fig.~\ref{sigma_vel1}-\ref{sigma_vel3},
the unresolved data is shown here, i.e.~using aperture spectra with aperture widths increasing in one pixel steps (corresponding to arcseconds as indicated on the x-axis).}
\label{sigmaall}
\end{figure*}

While we consider $\sigma_{\rm spat, reff}$ the spheroid dispersion within the spheroid effective
radius free of disk contamination, 
comparison with literature data (such as fiber-based SDSS data
or unresolved aperture spectra for more distant galaxies) requires us to also determine
aperture stellar-velocity dispersions.
To do so, we extracted 1D spectra by increasing the width
of the extraction window by one pixel, leaving the centroid
fixed to the central pixel.
We then measured the stellar-velocity dispersion
for each of the extracted spectra using the same procedure as
outlined above. We use the same mask/fitting region as
used for the center for the spatially-resolved spectra
(since here, all spectra will suffer from the AGN contamination).
Also note that broad FeII emission was subtracted for all rows,
if present, fixed to the widths determined in the center.
We refer to the resulting stellar-velocity dispersion from these spectra
as ``aperture'' $\sigma_{\rm ap}$. Note that $\sigma_{\rm ap}$ not only
contains the AGN contamination (continuum and emission
lines) independent of extracted aperture 
(and that is why here the fitting region that we refer to as CaHK
region does not actually include the CaH+K line; see \S~\ref{subsec:spat})
but can also
be broadened by any rotational component present
or biased to lower values in case of contribution of a kinematically
cold disk seen face on.
Also, the resulting $\sigma_{\rm ap, reff}$ is already 
luminosity weighted due to the way the spectra are extracted.
The results are shown in Fig.~\ref{sigmaall}.

We determine an aperture stellar-velocity dispersion
within the effective radius $\sigma_{\rm ap, reff}$ by choosing the aperture
size identical to the spheroid effective radius of a given object.

To compare our results with SDSS fiber spectra,
we determine $\sigma_{\rm ap, 1.5''}$, measured from aperture spectra within the central 1.5\arcsec~radius
as a proxy for what would have been measured with the 3\arcsec~diameter
Sloan fiber. Note, however, that in fact, our $\sigma_{\rm ap, 1.5''}$ corresponds to
a rectangular region with 1.5\arcsec~radius and 1\arcsec~width, given
the width of the long slit.
For eight objects, we can directly compare our results for $\sigma_{\rm ap, 1.5''}$
with those derived from SDSS fiber spectra by \citet{shen08} and \citet{gre06a}.
While individual objects can differ by up to 25\%, slightly larger than the quoted
uncertainties for the SDSS spectra ($\sim$10-15\%), on average, the measurements agree very well
($\sigma_{\rm ap, 1.5''}$/$\sigma_{\rm SDSS}$ = 
1.05 $\pm$ 0.1). 

We summarize the different stellar-velocity dispersion measurements in Table~\ref{sigma},
for the CaT region only.\footnote{Note
that electronic tables with all kinematic measurements
will be presented in the next paper of this series.}

\begin{deluxetable}{lcccc}
\tabletypesize{\scriptsize}
\tablecolumns{5}
\tablewidth{0pc}
\tablecaption{Stellar-Velocity Dispersion Measurements}
\tablehead{
\colhead{Object} & \colhead{$\sigma_{\rm spat, reff}$} & \colhead{$\sigma_{\rm ap, reff}$} & \colhead{$\sigma_{\rm ap, 1.5''}$} & \colhead{$r_{\rm eff, sph}$}\\
& (km\,s$^{-1}$) & (km\,s$^{-1}$) & (km\,s$^{-1}$) & (kpc) \\
\colhead{(1)} & \colhead{(2)} & \colhead{(3)}  & \colhead{(4)} & \colhead{(5)}}
\startdata
0121-0102 &    104$\pm$6  &	  89$\pm$10  &     89$\pm$8  & 1.54\\	   %L11    
0206-0017 &    222$\pm$2  &	 200$\pm$9   &    213$\pm$8  & 3.07 \\	   %L2        
0353-0623 &    159$\pm$6  &	 168$\pm$15  &    179$\pm$11 & 1.29 \\	    %L6       
0802+3104 &     95$\pm$2  &	 128$\pm$5   &    122$\pm$6  & 3.17 \\	    %L1       
0846+2522 &    130$\pm$3  &	 205$\pm$11  &    190$\pm$11 & 4.48 \\	   %L4        
1042+0414 &  	60$\pm$1  &	  85$\pm$4   &     86$\pm$4  & 2.76\\	  %L32     
1049+2451 &    138$\pm$4  &	 189$\pm$15  &    181$\pm$11 & 2.78\\	 %L34	     
1101+1102 &    113$\pm$2  &	 167$\pm$10  &    178$\pm$9  & 4.02\\	 %L35  
1116+4123 &    104$\pm$2  &	 110$\pm$5   &    130$\pm$4  & 3.40\\	 %L13  
1144+3653 &    173$\pm$4  &	 154$\pm$12  &    158$\pm$13 & 1.26\\	 %L15	     
1210+3820 &    137$\pm$3  &	 137$\pm$6   &    124$\pm$5  & 0.43\\	 %L43	     
1250-0249 &    112$\pm$2  &	 102$\pm$8   &    106$\pm$9  & 2.23\\	 %L46	     
1323+2701 &    113$\pm$3  &	 119$\pm$8   &    121$\pm$10 & 1.53\\	 %L49  
1405-0259 &    119$\pm$2  &	 119$\pm$5   &    117$\pm$5  & 1.24\\	    %L51     
1419+0754 &    181$\pm$2  &	 198$\pm$8   &    194$\pm$9  & 2.14\\	    %L53  
1434+4839 &    115$\pm$2  &	 118$\pm$5   &    126$\pm$6  & 2.16\\	    %L54       
1535+5754 &  	99$\pm$1  &	 110$\pm$5   &    114$\pm$5  & 2.09\\	   %L57     
1545+1709 &    157$\pm$2  &	 182$\pm$6   &    182$\pm$6  & 1.73\\	    %L58     
1554+3238 &    136$\pm$3  &	 152$\pm$5   &    152$\pm$5  & 0.66\\	    %L59       
1605+3305 &    134$\pm$6  &	  97$\pm$23  &    115$\pm$12 & 0.79\\	  %L61  
1606+3324 &    130$\pm$3  &	 163$\pm$7   &	  163$\pm$7  & 1.62\\	 %L62	  
1611+5211 &  	95$\pm$1  &	 122$\pm$3   &	  121$\pm$2  & 2.41\\	 %L63	  
\enddata	  					   
\tablecomments{   					   
Col. (1): Target ID (based on RA and DEC).				  
Col. (2): Spatially-resolved stellar-velocity dispersion within spheroid effective radius from CaT region. Random errors are given, while systematic errors are of order 7-15\%.
Col. (3): Aperture stellar-velocity dispersion within spheroid effective radius from CaT region. Random errors are given, while systematic errors are of order 7-15\%.
Col. (4): Aperture stellar-velocity dispersion within 1.5\arcsec~(to compare with SDSS fiber data) from CaT region. Random errors are given, while systematic errors are of order 7-15\%.
Col. (5): Spheroid effective radius (in kpc; semi-major axis) from surface photometry (see Appendix~\ref{sec:mattscode}
and Table~\ref{surface}; fiducial error 0.04 dex).
}
\label{sigma}
\end{deluxetable}

\subsection{Black Hole Mass}
\label{ssec:bh}
Black hole masses are
estimated using the empirically calibrated photo-ionization method,
also sometimes known as ``virial method''
\citep[e.g.,][]{wan99,ves02,woo02,ves06,mcg08}.  
Briefly, the method assumes that the
kinematics of the gaseous region in the immediate vicinity of the BH,
the BLR, traces the gravitational field of the
BH. The width of the broad emission lines (e.g.~H$\beta$) gives the
velocity scale, while the BLR size is given by the continuum
luminosity through application of an empirical relation found from
reverberation mapping (RM)
\citep[e.g.,][]{wan99, kas00, kas05, ben06}. Combining size and
velocity gives the BH mass, assuming a dimensionless coefficient of
order unity to describe the geometry and kinematics of the BLR
(sometimes known as the ``virial'' coefficient).  Generally,
this coefficient is obtained by matching the \mbh-$\sigma$ relation
of local active galactic nuclei (AGNs) to that of quiescent galaxies
\citep{onk04, gre06a, woo10}. Alternatively, the coefficient can be postulated
under specific assumptions on the geometry and kinematics of the
BLR. We adopt the normalizations in \citet{mcg08}, 
which are consistent with those found by \citet{onk04}.
However, since \citet{woo10} find a slightly different f-factor than
\citet{onk04}, causing a decrease in \mbh~by 0.02 dex,
we subtracted 0.02 dex from all BH masses.

\subsubsection{H$\beta$ Widths Measurements}
\begin{figure*}[ht!]
\begin{center}
\includegraphics[scale=0.7]{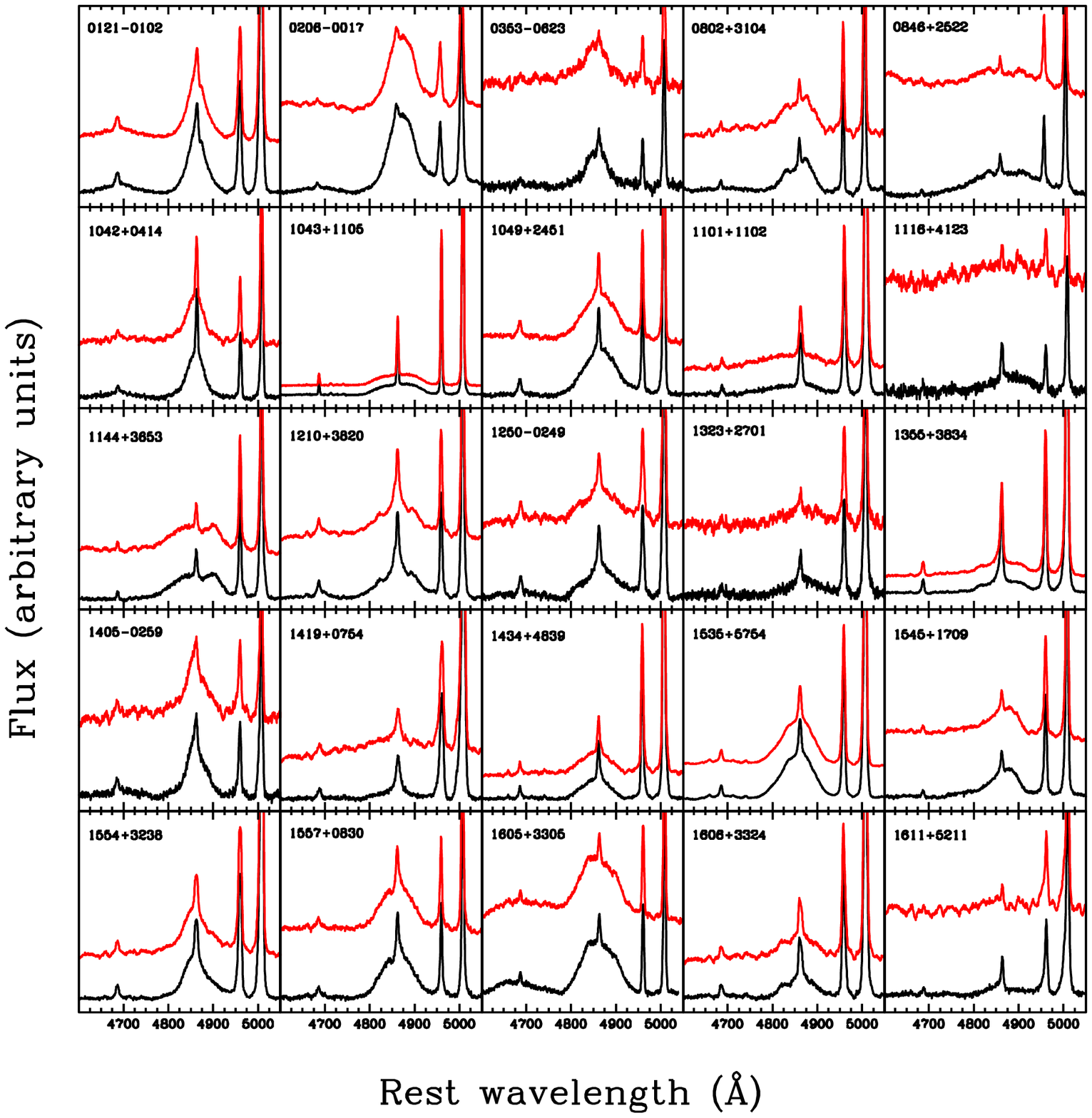}
\end{center}
\caption{Spectra around the broad H$\beta$ emission before (upper spectra; red)
and after subtraction of stellar absorption and continuum (lower spectra; black).}
\label{hbetaorig}
\end{figure*}

\begin{figure*}[ht!]
\begin{center}
\includegraphics[scale=0.7]{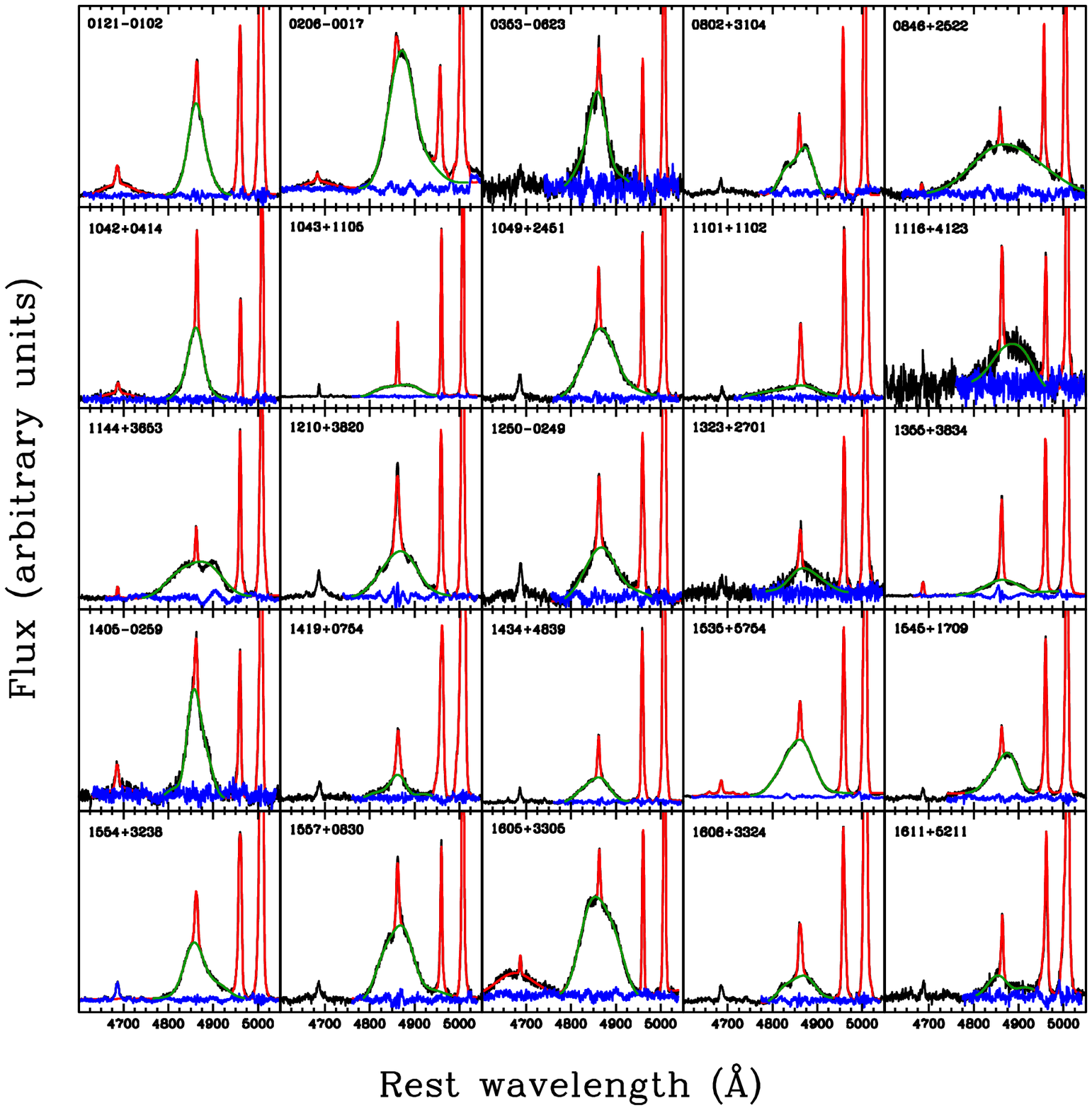}
\end{center}
\caption{Determination of the second moment of the broad H$\beta$ emission.
In addition to a continuum, the narrow H$\beta$ component was fitted
by a Gaussian while the broad H$\beta$ and both [OIII] lines
were fitted by Gaussian-Hermite polynomials (see text for details).
In some cases, strong broad and narrow HeII emission overlaps with the broad
H$\beta$ component and was
fitted additionally with a broad and narrow Gaussian.
The observed spectrum is shown in black, the total fit in red 
and the residual in blue (only in the region that was fitted).
The region of the fit in which the second moment of the broad H$\beta$ component
was determined is shown as green line. 
Note the variety of H$\beta$ profiles.
For 1535+5754, prominent [FeIII]\,$\lambda$4658\AA~and [ArIV]\,$\lambda\lambda$4711,4740\AA~emission is present
in the spectrum and each of these emission lines was fitted by a Gaussian component.}
\label{hbeta1}
\end{figure*}

To measure the width of the broad H$\beta$ emission, we
use the central blue spectrum, extracted with a size of 1.08\arcsec$\times$1\arcsec.
First, underlying broad FeII emission was removed (if needed) as described in \S~\ref{sec:obs}.
Then, a stellar template was subtracted to correct for stellar absorption lines 
underlying the broad H$\beta$ line in the following manner: 
We fixed 
the stellar-velocity dispersion and velocity to the values
determined in the region $\sim$5050-5450\AA~(i.e.~the MgIb region that is mostly free of AGN emission)
and then derived a best fitting-model in the region $\sim$4500-5450\AA, including a
polynomial continuum and outmasking the H$\beta$ and [OIII]\,$\lambda\lambda$\,4959,5007\AA~(hereafter [OIII]) emission lines.
The resulting stellar-absorption free spectra 
in the region around H$\beta$ are shown in Fig.~\ref{hbetaorig}.
Finally, we modeled 
the spectra by a combination of (i) a linear continuum, (ii) a Gaussian at the location of
the narrow H$\beta$ line, (iii) Gauss-Hermite polynomials for both 
[OIII] lines, with a fixed
flux ratio of 1:3 and a fixed wavelength difference, and (iv) Gauss-Hermite polynomials for
the broad H$\beta$ line.
A truncated Gauss-Hermite series \citep{van93, mcg08} 
has the advantage (over symmetrical Gaussians) of taking into
account asymmetries in the line profiles that are often present in the case
of [OIII] and broad H$\beta$ (Fig.~\ref{hbeta1}). 
The coefficients of the Hermite polynomials ($h_3$, $h_4$, etc.)
can be derived by straightforward linear minimization; the
center and width of the Gaussian are the only two non-linear parameters.
For [OIII], we allow coefficients up to $h_{12}$, for H$\beta$ up to $h_5$.

From the resulting fit, the second moment of the broad H$\beta$ ($\sigma_{\rm H\beta}$)
component is measured within a truncated region 
that contains only the broad H$\beta$ line
as determined interactively for each object (green line in Fig.~\ref{hbeta1}). 
Note that the line dispersion is defined as follows.
The first moment of the line profile is given by
\begin{eqnarray*}
\lambda_0 = \int \lambda P (\lambda) d\lambda / \int P (\lambda) d\lambda
\end{eqnarray*}
and the second moment is
\begin{eqnarray*}
\sigma^2_{\rm H\beta} = \left<\lambda^2\right> - \lambda_0^2 = \left[ \int \lambda^2 P (\lambda) d\lambda / \int P (\lambda) d\lambda\right] - \lambda_0^2 .
\end{eqnarray*}
The square root is the line dispersion $\sigma_{\rm H\beta}$ or root-mean-square (rms) width of the line
(see also \citealt{pet04}).

We estimate the uncertainty in $\sigma_{\rm H\beta}$
taking into account the three main sources of error involved
(i) the difference between the fit and the data in the region of the broad H$\beta$ component
(to account for uncertainties by asymmetries not fitted by the Gauss-Hermite polynomials);
(ii) the systematic error involved when determining the size of the fitting region
(which we determined empirically to be of the order of 5\%),
and (iii) the statistical error determined by repeated fitting using the
same fitting parameters (of order 1\%).
Note that due to the very high S/N,
the line dispersion inferred from the Gauss-Hermite polynomial fit is
virtually indistinguishable from that measured directly from
the data (on average, fit/data=1$\pm$0.02 and at most, the difference is 5\%).
We also compare $\sigma_{\rm H\beta}$ with 
that inferred from the FHWM assuming a Gaussian profile:
$\sigma_{\rm H\beta}$/(FHWM/2.355) = 1.11$\pm$0.3.
The average difference of 10\%, expected because broad lines are known not to be Gaussian in shape, 
corresponds to a systematic difference of 0.04 dex, 
negligible for an individual object, and small
compared to the uncertainty
on the BH mass that we assume (0.4 dex),
but potentially a significant source of bias for accurate measurements based on large
samples.

Fig.~\ref{hbeta1} shows the fit for all objects.
The variety of broad H$\beta$ profiles is interesting, with
only 6/25 objects revealing symmetric line profiles,
8/25 objects having more than one peak, and
the majority of objects (11/25) having asymmetric
line profiles, thus showing the need for Gauss-Hermite polynomials.
While the line profile can in principle provide
insights into BLR geometry and kinematics, this is beyond
the scope of the present paper. 
Note that the narrow H$\beta$/[OIII]\,$\lambda$5007\AA~ratio
ranges between 6-28\%, in agreement with other studies
\citep[e.g.,][]{marz03,woo06}.

\subsubsection{Luminosity at 5100\AA~and BH masses}
We use the SDSS images
to simultaneously fit the AGN by a 
point-spread function (PSF) and the host galaxy
by spheroid and disk (if present). The next section and
Appendix~\ref{sec:mattscode} describe the surface photometry in detail.
The resulting PSF g'-band magnitude is corrected for Galactic extinction
(subtracting the SDSS DR7 ``extinction\_g''' column),
and then extrapolated to
5100\AA, assuming a powerlaw of the form $f_{\nu} \propto \nu^{\alpha}$
with $\alpha$=-0.5.
(Literature values of $\alpha$ range between -0.2 and -1; \citealt{mal83, neu87, cri90, fra91, zhe97,van01};
see also \citealt{ben10} and D. Szathm{\'a}ry et al. 2011, submitted).

We calculated BH masses according to the following formula \citep{mcg08}:
\begin{eqnarray*}
\log M_{\rm BH} = 7.68 + 2 \log \left(\frac{\sigma_{\rm H_\beta}}{1000\,{\rm km\,s^{-1}}}\right)
+ 0.518 \log \left(\frac{\lambda L_{5100}}{10^{44}\,{\rm erg\,s^{-1}}}\right)
\end{eqnarray*}
The results are given in Table~\ref{mbh}.

We assume a nominal uncertainty of the BH masses
measured via the virial method of 0.4 dex.
Note that we do not correct for possible effects of radiation pressure
\citep[e.g.,][see, however, \citealt{net09,net10}]{mar08,mar09}.

\begin{deluxetable}{lccccc}
\tabletypesize{\scriptsize}
\tablecolumns{6}
\tablewidth{0pc}
\tablecaption{BH Mass Measurements}
\tablehead{
\colhead{Object} & \colhead{$\sigma_{\rm H\beta}$} & \colhead{$\lambda$$L_{\rm 5100}$} & \colhead{log $M_{\rm BH}/M_{\sun}$}\\
& (km\,s$^{-1}$) & (10$^{44}$\,erg\,s$^{-1}$)\\
\colhead{(1)} & \colhead{(2)} & \colhead{(3)}  & \colhead{(4)}}
\startdata
0121-0102 &   1317$\pm$66  &  0.24 & 7.58 \\	   %L11    
0206-0017 &   1991$\pm$100 &  0.61 & 8.15 \\	  %L2	     
0353-0623 &   1694$\pm$85  &  0.09 & 7.58 \\	   %L6       
0802+3104 &   1472$\pm$74  &  0.05 & 7.31 \\	   %L1       
0846+2522 &   4547$\pm$227 &  0.24 & 8.66 \\	  %L4	     
1042+0414 &   1252$\pm$63  &  0.04 & 7.12 \\	  %L32     
1043+1105 &   1910$\pm$96  &  0.16 & 7.81 \\	  %L33  
1049+2451 &   2252$\pm$113 &  0.18 & 7.98 \\	 %L34	     
1101+1102 &   2900$\pm$145 &  0.05 & 7.91 \\	 %L35  
1116+4123 &   2105$\pm$105 &  0.03 & 7.51 \\	 %L13  
1144+3653 &   2551$\pm$128 &  0.01 & 7.50 \\	 %L15	     
1210+3820 &   2377$\pm$119 &  0.03 & 7.64 \\	 %L43	     
1250-0249 &   2378$\pm$119 &  0.06 & 7.79 \\	 %L46	     
1323+2701 &   2133$\pm$107 &  0.02 & 7.12 \\	 %L49  
1355+3834 &   3110$\pm$156 &  0.09 & 8.10 \\	 %L50	  
1405-0259 &   1343$\pm$67  &  0.05 & 7.22 \\	  %L51     
1419+0754 &   1932$\pm$97  &  0.08 & 7.65 \\	  %L53  
1434+4839 &   1572$\pm$79  &  0.14 & 7.62 \\	  %L54       
1535+5754 &   2019$\pm$101 &  0.26 & 7.97 \\	 %L57	  
1545+1709 &   1604$\pm$80  &  0.06 & 7.42 \\	  %L58     
1554+3238 &   1988$\pm$99  &  0.11 & 7.77 \\	  %L59       
1557+0830 &   2019$\pm$101 &  0.08 & 7.69 \\	 %L60	     
1605+3305 &   1981$\pm$99  &  0.23 & 7.92 \\	  %L61  
1606+3324 &   1737$\pm$87  &  0.03 & 7.36 \\	  %L62     
1611+5211 &   1843$\pm$92  &  0.04 & 7.49 \\	  %L63     
\enddata	   					   
\tablecomments{    					   
Col. (1): Target ID (based on RA and DEC). 		      	   	  
Col. (2): Second moment of broad H$\beta$.			  
Col. (3): Rest-frame luminosity at 5100\AA~determined from SDSS g' band surface photometry (fiducial error 0.1 dex).
Col. (4): Logarithm of BH mass (solar units) (uncertainty of 0.4 dex).
}
\label{mbh}
\end{deluxetable}

\subsection{Surface Photometry}
\label{ssec:phot}
To obtain a host-galaxy free 5100\AA~luminosity (for an accurate BH mass measurement)
as well as a good estimate of the spheroid effective radius (to measure the stellar-velocity dispersion at
the effective radius), we performed surface photometry of archival SDSS DR7 images.
In our previous papers, we ran the 2D galaxy fitting program GALFIT \citep{pen02}
on high-spatial resolution HST images for this purpose.
Here, we lack space-based images, but -
compared to our previous studies - the objects
are at much lower redshifts ($z$$\simeq$0.05 compared to $z$$\simeq$0.4 and 0.6).
(Note that the average seeing ranges between
$\sim$1.2\arcsec~in the z'-band to $\sim$1.4\arcsec~in the g'-band for our sample.) 
Moreover, the SDSS images come in five different filters (u',g',r',i',z'),
allowing us to determine the host-galaxy properties
by simultaneously fitting the multiple bands while imposing
certain constraints between the parameters of each band.
Since this is beyond the scope of GALFIT,
we developed a new image analysis code.
The advantage of a joint multi-band analysis
is that it enables us to more easily distinguish between the AGN which dominates
in the blue bands from the host galaxy which is dominant in the redder filters.
The code is described in detail in Appendix~\ref{sec:mattscode}, including
a comparison with GALFIT.
The results are summarized in Table~\ref{surface}.

Note that as a sanity check, we checked the classic galaxy scaling relations
\citep[see e.g.,][]{hyd09a,hyd09b}, such as fundamental plane (FP)
or stellar mass vs $\sigma_{\rm ap, reff}$ and stellar mass
vs $r_{\rm eff, sph}$. 
Taking into account the
small dynamic range and sample size (especially when
considering elliptical host galaxies only),
the results are consistent within the errors.
We will show and discuss these galaxy scaling relations when presenting the full sample in the upcoming papers of this series.

\begin{deluxetable*}{lcccccccccccccccccccccc}
%\rotate
\setlength{\tabcolsep}{0.0in} 
\tabletypesize{\scriptsize}
\tablecolumns{15}
\tablewidth{0pc}
\tablecaption{Results from Surface Photometry}
\tablehead{
\colhead{Object} & \multicolumn{4}{c}{PSF} & \multicolumn{4}{c}{Spheroid} & \multicolumn{4}{c}{Disk} & $\log L_{\rm sph, V}/L_{\odot}$ & $r_{\rm eff, sph}$\\
& g' & r' & i' & z' & g' & r' & i' & z' & g' & r' & i' & z'  \\
& (mag) & (mag) & (mag) & (mag) & (mag) & (mag) & (mag) & (mag) & (mag) & (mag) & (mag) & (mag) & & (kpc)   \\
\colhead{(1)} & \colhead{(2)} & \colhead{(3)}  & \colhead{(4)} & \colhead{(5)}  & \colhead{(6)}  & \colhead{(7)}  & \colhead{(8)} & \colhead{(9)} & \colhead{(10)} & \colhead{(11)} & \colhead{(12)}  & \colhead{(13)} & \colhead{(14)}  & \colhead{(15)}}
\startdata
0121-0102 &  17.01 & 17.51 & 17.07 & 17.24 & 16.39 & 16.15 & 15.77 & 15.56 & 15.52  &14.93  &14.64   & 14.52   &10.19$\pm$0.07 & 1.54 \\       %L11
0206-0017 &  15.47 & 15.75 & 15.65 & 15.72 & 14.68 & 13.98 & 13.59 & 13.32 & 15.39  &14.63  &14.23   & 14.00   &10.80$\pm$0.07 & 3.07 \\      %L2
0353-0623 &  18.80 & 18.99 & 18.52 & 18.64 & 17.53 & 16.72 & 16.38 & 16.17 & 18.03  &17.32  &16.95   & 16.74   &10.20$\pm$-0.06 & 1.29 \\	%L6 
0802+3104 &  18.13 & 18.28 & 17.77 & 17.67 & 15.76 & 15.13 & 14.79 & 14.57 &\nodata &\nodata& \nodata& \nodata &10.30$\pm$0.06 & 3.17 \\       %L1
0846+2522 &  16.86 & 16.85 & 16.67 & 16.61 & 16.01 & 15.37 & 15.00 & 14.78 &\nodata &\nodata &\nodata &\nodata &10.03$\pm$-0.06 & 4.48 \\	%L4
1042+0414 &  18.94 & 19.26 & 18.80 & 18.97 & 16.82 & 16.09 & 15.63 & 15.36 &\nodata &\nodata& \nodata& \nodata &10.15$\pm$0.06 & 2.76 \\      %L32 
1043+1105 &  17.14 & 17.37 & 16.91 & 17.08 & 16.87 & 16.50 & 16.14 & 16.10 &\nodata &\nodata& \nodata& \nodata & 9.90$\pm$0.06 & 3.55 \\      %L33 
1049+2451 &  17.36 & 17.56 & 17.09 & 17.12 & 16.45 & 15.83 & 15.40 & 15.19 &\nodata &\nodata& \nodata& \nodata &10.30$\pm$0.06 & 2.78 \\     %L34
1101+1102 &  17.77 & 17.51 & 17.32 & 17.31 & 16.43 & 15.35 & 15.07 & 14.71 & 16.74  &16.19  &15.73   & 15.50   &10.04$\pm$0.20 & 4.02  \\     %L35  
1116+4123 &  56.05 & 18.96 & 18.80 & 17.88 & 14.97 & 14.21 & 13.85 & 13.63 &\nodata &\nodata& \nodata& \nodata &10.13$\pm$0.40 & 3.40 \\     %L13  
1144+3653 &  19.36 & 21.32 & 22.10 & 52.97 & 16.47 & 15.59 & 15.19 & 14.90 & 16.12  &15.39  &15.10   & 14.88   &10.03$\pm$0.10 & 1.26  \\     %L15
1210+3820 &  17.28 & 16.70 & 17.21 & 16.91 & 15.81 & 15.13 & 14.74 & 14.45 & 15.28  &14.55  &14.21   & 13.97   & 9.84$\pm$0.36 & 0.43  \\     %L43
1250-0249 &  18.15 & 18.23 & 17.89 & 17.77 & 17.53 & 16.54 & 16.08 & 15.79 & 15.97  &15.29  &14.91   & 14.64   & 9.84$\pm$0.06 & 2.23  \\     %L46
1323+2701 &  19.60 & 19.12 & 18.74 & 18.28 & 19.07 & 18.25 & 17.62 & 17.43 & 17.93  &17.14  &16.78   & 16.49   & 9.36$\pm$0.06 & 1.53 \\      %L49
1355+3834 &  17.93 & 17.92 & 17.21 & 17.58 & 16.55 & 16.15 & 15.73 & 15.60 &\nodata &\nodata& \nodata& \nodata &10.11$\pm$0.06 & 2.06 \\     %L50
1405-0259 &  18.79 & 19.02 & 18.49 & 18.62 & 17.72 & 16.82 & 16.54 & 16.19 & 16.68  &15.92  &15.53   & 15.21   & 9.84$\pm$0.06 & 1.24  \\      %L51 
1419+0754 &  18.32 & 18.18 & 17.48 & 17.49 & 16.85 & 15.71 & 15.21 & 14.84 & 15.51  &14.86  &14.49   & 14.29   &10.32$\pm$0.06 & 2.14 \\      %L53 
1434+4839 &  16.67 & 16.73 & 16.61 & 16.61 & 15.83 & 15.15 & 14.81 & 14.57 & 16.10  &15.47  &15.16   & 14.94   &10.18$\pm$0.11 & 2.16  \\      %L54 
1535+5754 &  15.61 & 15.61 & 15.67 & 15.62 & 15.34 & 14.76 & 14.44 & 14.30 &\nodata &\nodata& \nodata& \nodata &10.20$\pm$0.24 & 2.09 \\     %L57
1545+1709 &  18.33 & 18.08 & 17.69 & 17.35 & 17.41 & 16.77 & 16.31 & 16.15 & 17.69  &16.90  &16.50   & 16.17   & 9.80$\pm$0.06 & 1.73  \\      %L58 
1554+3238 &  17.57 & 17.58 & 17.30 & 17.15 & 17.60 & 16.73 & 16.30 & 16.12 & 16.38  &15.71  &15.37   & 15.09   & 9.79$\pm$0.06 & 0.66  \\      %L59 
1557+0830 &  17.92 & 17.92 & 17.59 & 17.51 & 17.18 & 16.64 & 16.30 & 16.16 &\nodata &\nodata& \nodata& \nodata & 9.81$\pm$0.06 & 1.18 \\     %L60
1605+3305 &  17.03 & 16.42 & 16.20 & 16.07 & 17.08 & 16.40 & 16.19 & 16.20 &\nodata &\nodata& \nodata& \nodata & 9.98$\pm$0.06 & 0.79 \\      %L61 
1606+3324 &  19.37 & 19.77 & 18.72 & 19.14 & 17.46 & 16.54 & 16.10 & 15.78 & 17.61  &16.85  &16.45   & 16.27   &10.05$\pm$0.06 & 1.62  \\      %L62 
1611+5211 &  18.23 & 17.74 & 17.47 & 17.18 & 16.15 & 15.43 & 15.03 & 14.80 &\nodata &\nodata& \nodata& \nodata &10.18$\pm$0.06 & 2.41 \\      %L63     
\enddata	   					 
\tablecomments{    					 
Col. (1): Target ID (based on RA and DEC). 			  
Col. (2-5): Extinction-corrected g', r', i', and z' PSF magnitudes (with an uncertainty of 0.5 mag).
Col. (6-9): Extinction-corrected g', r', i', and z' spheroid magnitudes (with an uncertainty of 0.2 mag). 
Col. (10-13): Extinction-corrected g', r', i', and z' disk magnitudes (with an uncertainty of 0.2 mag). 
Col. (14): Logarithm of spheroid luminosity in rest-frame V (solar units).
Col. (15): Spheroid effective radius (in kpc; semi-major axis).
}
\label{surface}
\end{deluxetable*}

\subsection{Stellar and Dynamical Spheroid Mass}
\label{ssec:mass}
Our surface photometry code gives spheroid, disk, and total host-galaxy magnitudes for four
different SDSS filters (g', r', i', z') which can in turn be used to estimate
stellar spheroid masses. For this purpose, \citet{aug09}
have developed a Bayesian stellar mass estimation code that we use here. 
The code allows informative priors to be placed on the age, metallicity, and
dust content of the galaxy and uses a MCMC sampler
to explore the full parameter space and quantify
degeneracies between the stellar population parameters.
We use a Chabrier initial-mass function (IMF).

Also, with the knowledge of $\sigma_{\rm ap, reff}$ (as determined from the CaT region)
and $r_{\rm eff, sph}$, we
can calculate a dynamical mass:
\begin{eqnarray*}
M_{\rm sph, dyn} = k r_{\rm eff, sph} \sigma_{\rm ap, reff}^2 / G
\end{eqnarray*}
\label{dynamic}

with $G$ = gravitational constant.
For comparison with literature \citep[in particular][]{mar03},
we use $k=3$. For the same reason, we choose $\sigma_{\rm ap, reff}$ 
instead of $\sigma_{\rm spat, reff}$.
The results are given in Table~\ref{mass}.

\begin{deluxetable*}{lccccc}
\tabletypesize{\scriptsize}
\tablecolumns{6}
\tablewidth{0pc}
\tablecaption{Dynamical And Stellar Masses}
\tablehead{
\colhead{Object} & \colhead{$\log M_{\rm sph, vir}/M_{\odot}$} & \colhead{$\log M_{\rm sph, \star}/M_{\odot}$} & \colhead{$\log M_{\rm disk, \star}/M_{\odot}$} & \colhead{$\log M_{\rm host, \star}/M_{\odot}$} \\
\colhead{(1)} & \colhead{(2)} & \colhead{(3)}  & \colhead{(4)} & \colhead{(5)}}
\startdata
0121-0102     & 9.866 & 10.12$\pm$0.24  &    10.60$\pm$0.24  &    10.75$\pm$0.23 \\	  %L11 
0206-0017     & 10.93 & 10.95$\pm$0.23  &    10.71$\pm$0.22  &    11.17$\pm$0.23 \\	 %L2   
0353-0623     & 10.41 & 10.33$\pm$0.22  &    10.08$\pm$0.23  &    10.52$\pm$0.22 \\	  %L6  
0802+3104     & 10.56 & 10.38$\pm$0.23  &     \nodata	 &	   \nodata   \\   %L1  
0846+2522     & 11.12 & 10.50$\pm$0.23	&    \nodata	 &	   \nodata   \\ %L4
1042+0414     & 10.15 & 10.32$\pm$0.23  &     \nodata	 &	   \nodata   \\      %L32  
1043+1105$^a$ & \nodata     	&  9.83$\pm$0.24  &	\nodata    &	      \nodata  \\      %L33  
1049+2451     & 10.84 & 10.41$\pm$0.23  &     \nodata	 &	   \nodata   \\     %L34   
1101+1102     & 10.89 & 10.33$\pm$0.22  &     9.89$\pm$0.23  &    10.46$\pm$0.22  \\	 %L35  
1116+4123     & 10.46 & 10.20$\pm$0.22  &     \nodata	 &	   \nodata   \\     %L13   
1144+3653     & 10.32 & 10.26$\pm$0.22  &    10.21$\pm$0.24  &    10.54$\pm$0.23  \\	 %L15  
1210+3820     & 9.753 &  9.94$\pm$0.24  &    10.13$\pm$0.24  &    10.35$\pm$0.23  \\	 %L43  
1250-0249     & 10.21 & 10.14$\pm$0.22  &    10.50$\pm$0.22  &    10.65$\pm$0.22  \\	 %L46  
1323+2701     & 10.30 &  9.65$\pm$0.22  &    9.93$\pm$0.23   &    10.10$\pm$0.23  \\	 %L49  
1355+3834$^a$ & \nodata		& 10.11$\pm$0.23  &	\nodata    &	     \nodata   \\     %L50   
1405-0259     & 10.09 & 10.04$\pm$0.23  &    10.42$\pm$0.22  &    10.58$\pm$0.22  \\	  %L51 
1419+0754     & 10.77 & 10.73$\pm$0.21  &    10.78$\pm$0.24  &    11.00$\pm$0.24 \\	 %L53  
1434+4839     & 10.33 & 10.30$\pm$0.24  &    10.13$\pm$0.24  &    10.53$\pm$0.23  \\	  %L54 
1535+5754     & 10.19 & 10.24$\pm$0.24  &     \nodata	 &	   \nodata   \\     %L57   
1545+1709     &  10.6 &  9.92$\pm$0.22  &     9.93$\pm$0.22  &    10.23$\pm$0.22  \\	  %L58 
1554+3238     & 10.03 & 10.00$\pm$0.23  &    10.32$\pm$0.23  &    10.50$\pm$0.23  \\	  %L59 
1557+0830$^a$ & \nodata		&  9.82$\pm$0.23  &	\nodata    &	     \nodata   \\     %L60   
1605+3305     & 9.712 &  9.95$\pm$0.23  &     \nodata	 &	   \nodata   \\      %L61  
1606+3324     & 10.48 & 10.33$\pm$0.22  &    10.06$\pm$0.24  &    10.50$\pm$0.23  \\	  %L62 
1611+5211     &  10.4 & 10.33$\pm$0.22  &     \nodata	 &	   \nodata   \\      %L63  
\enddata  		       
\tablecomments{    					 
Col. (1): Target ID (based on RA and DEC). 			  
Col. (2): Dynamical spheroid mass calculated from $r_{\rm eff, sph}$ and $\sigma_{\rm ap, reff}$ (determined from CaT region; 
fiducial error 0.1 dex). 
Col. (3): Stellar spheroid mass (using Chabrier as IMF).
Col. (4): Stellar disk mass (if present).
Col. (5): Stellar host mass (only listed if disk present, i.e.~if different from (3)).
$^a$: For three objects, dynamical masses could not be determined as $\sigma_{\rm ap, reff}$ could not be reliably measured.
}
\label{mass}
\end{deluxetable*}

\section{COMPARISON SAMPLES}
\label{sec:comp}
For the \mbh-scaling relations,
we compile comparison samples from the literature,
including local inactive galaxies \citep{mar03,har04,gue09}
and local active galaxies \citep{gre06a,woo10}.
Note that while for the inactive galaxies, BH masses have been
derived from direct dynamical measurements,
the BH masses for active galaxies are calibrated
masses either from reverberation mapping or from
the virial method.

\subsection{\mbh-\s~Relation}
For the \mbh-\s~relation, we use the data from
\citet{gue09} (local inactive galaxies), \citet{gre06a}
(local active galaxies), and \citet{woo10} (local RM AGNs).
In all cases, the stellar-velocity dispersion measurements
correspond to luminosity-weighted stellar-velocity dispersions
within a given aperture $\sigma_{\rm ap}$. For \citet{gue09},
 the aperture
is typically the effective radius, but as $\sigma_{\rm ap}$ is compiled
from the literature, there are also cases where it is $\sigma_{\rm ap, 1/8 reff}$
or $\sigma_{\rm c}$. However, \citet{gue09} conclude
that the systematic differences are small compared
to other systematic errors. 
The BH masses were determined from direct dynamical measurements
(stellar or gaseous kinematics or masers). In total,
data are available for 49 objects with $z<0.04$.
For \citet{gre06a}, $\sigma_{\rm ap, 1.5''}$ was determined from the fiber-based
SDSS spectra and is thus within  an aperture of 1.5\arcsec~radius. 
From their sample of 56 Seyfert-1 galaxies with $z<0.1$, 
\citet{woo08} chose a sub-sample of 49 objects for which they measured
BH mass using the line dispersion of H$\beta$ and the H$\alpha$ luminosity
consistently calibrated to our BH mass measurements \citep{mcg08}.
These 49 objects have 5 objects in common with our sample,
so we use our results instead, leaving us with 44 local SDSS AGNs.
(We will perform a comparison for the objects in common to
both samples once we have our
full sample available for which we expect to have a total of $\sim$20 objects in common.)
Finally, we include 24 local Seyfert-1 galaxies ($z<$0.09) 
for which the BH mass has been determined via RM \citep{woo10}.
For these objects, the $\sigma_{\rm ap}$ measurements
were measured within an aperture of typically $\sim$1\arcsec$\times$1.5\arcsec~to
$\sim$1.5\arcsec$\times$3\arcsec.

To compare our results with these literature data
which all consist of luminosity-weighted
stellar-velocity dispersions within some aperture,
we use $\sigma_{\rm ap, 1.5''}$ as determined
from the CaT region (which is considered the benchmark). 
(Note that choosing $r_{\rm eff}$ instead
as aperture size does not change the results within the errors; see Table~\ref{sigma}.)

\subsection{\mbh-\ls~and \mbh-\ms~Relations}
For the \mbh-\ls~relation, we again use the local
inactive sample from \citet{gue09}, here limited
to 35 elliptical and S0 galaxies with a reliable spheroid-disk decomposition.

For the \mbh-\mss~relation,
we use the J, H, and K magnitudes from \citet{mar03} for their group 1
(i.e.~with secure BH masses and reliable spheroid luminosities)
to calculate stellar masses in the same way we calculated
our stellar masses.
Also, we updated the BH masses using those listed in \citet{gue09}.
This leaves us with a sample of 18 objects.

Finally, for the \mbh-\msv~relation,
we compile local inactive galaxies using BH masses from \citet{gue09},
and calculate dynamical masses using $r_{\rm reff, sph}$ given by
\citet{mar03} and $\sigma_{\rm ap}$ measurements by \citet{gue09}.

\section{RESULTS AND DISCUSSION}
\label{sec:res}
Here, we describe and discuss our results.
After a brief section on host-galaxy morphologies,
merger rates and rotation curves, we
focus on the different methods to derive
stellar-velocity dispersions and perform
a quantitative comparison. Finally, we present
the different BH mass scaling relations and compare
our results to literature data.
Since the aim of this paper is 
to outline the methodology 
and present the results of our pilot study,
we postpone any detailed quantitative conclusions to the upcoming papers,
once the full sample is available.

\subsection{Host-Galaxy Morphologies, Merger Rates, And Rotation Curves}
\label{ssec:host}
Using the multi-filter SDSS images shown in Fig.~\ref{sdss}, we can
determine the overall host-galaxy morphologies.  Given the low spatial
resolutions, we divide the sample into three categories: ellipticals
(E), S0/a, and spirals later than Sa (S).  11/25 objects can then be
classified as S, 9/25 as E, and 5/25 as S0/a.  One object with a
spiral-like host galaxy morphology is clearly undergoing a merger
(0206-0017). 1419+0754 shows irregular structure differing from normal
spiral arms and might also be in the process of merging. 

The fraction
of ellipticals (36$\pm$11\%) is somewhat higher than expected, given
that these are (almost all radio-quiet) Seyfert galaxies for which the
majority has typically been found to reside in spirals or S0
($\sim$80\%, e.g.~\citealt{hun99} and references therein). However,
due to the low-resolution ground-based imaging data, and the small
number statistics, there is still some uncertainty in this
classification, with some objects potentially falling in the
neighboring category.  Also, we cannot exclude that in a few cases,
the images are too shallow to reveal the presence of the disk.
Indeed, the majority of objects ($\sim$13/22) show rotation curves
with a maximum velocity between 100 and 200 km\,s$^{-1}$.  Also the
object with a clear merger signature (0206-0017) shows a prominent
rotation curve with a maximum of 200 km\,s$^{-1}$ rotational velocity,
hinting at a spiral galaxy experiencing a merger event.  Both, the
variety of host-galaxy morphologies, in particular with a substantial
fraction of host galaxies  having prominent spiral arms
and disk, Hubble types Sa and later (44$\pm$5\%), and the rotation curves
underscore that their kinematic structure is complex and indeed
spatially-resolved information is necessary.

The merger rate (0.06$\pm$0.02) is lower than for our higher redshift
objects \citep[0.29$\pm$0.1 at $z\simeq0.4$;][]{ben10} and more
comparable to inactive galaxies in the local Universe
\cite[e.g.,][see, however, \citealt{tal09}]{pat02}.  The merger rate
is likely to be a function of galaxy mass, with higher (major) merger
rates for higher mass objects \cite[e.g.,][]{hop10}.  Indeed, the
local sample studied here has, on average, lower host-galaxy
luminosities (25 objects; $\log L_{\rm host; L_{\odot}}$ =
10.33$\pm$0.05; rms scatter: 0.29) than the one at $z\simeq0.4$ (34
objects; $\log L_{\rm host; L_{\odot}}$ = 10.54$\pm$0.03; rms scatter:
0.18; \citet{ben10}), indicating lower mass objects.  However, we
suffer from low number statistics and will get back to this
discussion, once we have analyzed our full sample of $\sim$100 local
Seyfert-1 galaxies.

Note that the image quality does not allow us to determine the fraction of 
bars present in the host galaxies to study the effect of bars
on the \mbh-\s~scaling relation \citep[e.g.,][and references therein]{gra10}.

\subsection{Stellar-Velocity Dispersions}
\label{ssec:sigma}
We here compare the various stellar-velocity dispersions;
first spatially-resolved vs aperture stellar-velocity dispersions,
then the results from three different spectral regions.

\subsubsection{Spatially-resolved vs. aperture stellar-velocity dispersions}
Figs.~\ref{sigma_vel1}-\ref{sigma_vel3} show the spatially-resolved
velocity dispersions for the sample. We are tracing the velocity
dispersion for the central 2-6\arcsec~radius, depending on the object.
For the majority of objects ($\sim$17/22),
the overall behavior of $\sigma_{\rm spat, CaT}$ can be described as decreasing from
a central value of $\sim$130-200 km\,s$^{-1}$ to a
value of $\sim$50-100 km\,s$^{-1}$ in the outer parts.
For the remaining objects (5/22), $\sigma_{\rm spat, CaT}$ is roughly
constant with radius, within the errors.

A different picture emerges when looking at
the {\rm aperture stellar-velocity dispersions} in Fig.~\ref{sigmaall}.
Here, categorizing the overall behavior of $\sigma_{\rm ap, CaT}$ as a function
of ``radius'' (which is here the increase in width of the aperture spectrum),
splits the sample into three categories, with the majority of objects
(9/22) showing a constant $\sigma_{\rm ap, CaT}$, and the rest distributes with roughly
even numbers on either decreasing (6/22) or increasing (7/22) $\sigma_{\rm ap, CaT}$ with radius.
Looking at individual objects, 6 objects have a decreasing $\sigma_{\rm ap, CaT}$ with radius
in the spatially-resolved spectra but shift towards an apparently increasing
$\sigma_{\rm ap, CaT}$ in the aperture spectra due to the rotational support (as reflected in
the rotation curve). However, overall, the aperture dispersions change only slowly with
radius, as has already been noted by \citet{cap96} and \citet{geb00} for inactive galaxies.

\begin{figure}[ht!]
\begin{center}
\includegraphics[scale=0.3]{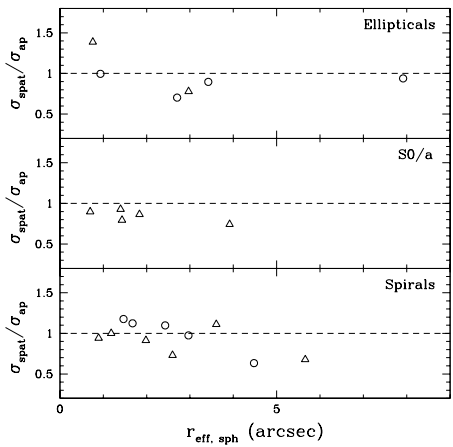}
\end{center}
\caption{Comparison between different stellar-velocity dispersion measurements, in all cases 
determined from the CaT region.
Ratio between $\sigma_{\rm spat, reff}$, i.e.~luminosity-weighted 
stellar-velocity dispersion within effective radius as determined from 
spatially-resolved spectra and $\sigma_{\rm ap, reff}$
as determined from spectra extracted with an aperture equal to the effective radius,
as function of effective radius.
The upper panel shows all objects for which the host galaxy
has been classified as ellipticals,
the middle panel objects with S0/a-type host-galaxy morphology,
and the lower panel objects hosted by spirals.
The triangles (circles) correspond to galaxies seen edge-on (face-on).
The error bars are omitted for clarity, measurement errors on the ratios range between
5-10\%.
}
\label{sigma_spat_ap}
\end{figure}

We can make a more direct comparison between $\sigma_{\rm spat, reff}$
determined from the spatially-resolved spectra with that determined
from the aperture spectra $\sigma_{\rm ap, reff}$, choosing the
stellar-velocity dispersion determined from the CaT region (see also
Table~\ref{sigma}).  We divide the sample into three sub-categories
according to the host-galaxy morphology (Elliptical, S0/a, Spiral) and
additionally distinguish between host galaxies seen face-on and
edge-on.  Fig.~\ref{sigma_spat_ap} shows the result both as function
of effective radius and $\sigma_{\rm spat, reff}$.  The general trend
conforms to our expectations: If a spiral galaxy is seen edge-on, the
rotation component can bias $\sigma_{\rm ap, reff}$ towards higher
values and thus, ``triangles'' are expected to lie below the unity
line.  However, since the disk is kinematically cold, it can also
result in the opposite effect, i.e.~biassing $\sigma_{\rm ap, reff}$
towards smaller values, if the disk is seen face on, so ``circles''
are expected to lie above the unity line. On average, face-on spiral host
galaxies objects have $\sigma_{\rm spat,
reff}$/$\sigma_{\rm reff, ap}$ = 1.02$\pm$0.05 (rms scatter: 0.24) and
edge-on objects have $\sigma_{\rm spat, reff}$/$\sigma_{\rm reff, ap}$
= 0.90$\pm$0.03 (rms scatter: 0.16).  If we calculate the average for
the whole sample (all morphologies and orientations), $\sigma_{\rm
spat, reff}$/$\sigma_{\rm ap, reff}$ = 0.93$\pm$0.04 (rms scatter:
0.2).  There is no obvious trend with either bulge mass, BH mass, or
effective radius.

To compare the stellar-velocity dispersion measurements with 
what would be measured from SDSS fiber spectra,
we use $\sigma_{\rm ap, 1.5''}$.
For $\sigma_{\rm spat, reff}$/$\sigma_{\rm ap, 1.5''}$,
the same trend persists as for $\sigma_{\rm spat, reff}$/$\sigma_{\rm ap, reff}$,
showing that choosing 1.5\arcsec~instead of 
effective radius does not have a large effect on the resulting
stellar-velocity dispersion. This can be attributed 
both to the luminosity weighting with a steep central surface-brightness profile \citep{dev48}
and to the fact that the average effective
radius of our sample is close to 1.5\arcsec~(2.6$\pm$0.07\arcsec; rms scatter: 1.8).

For objects at higher redshift, the effect
can be more pronounced as different sizes are involved. Considering
our $z$$\sim$0.4 Seyfert-1 sample for which
we study the evolution of the \mbh-scaling relations
\citep{tre04,woo06,tre07,woo08,ben10},
the typical extraction window to determine
$\sigma_{\rm ap}$ is 1 square-arcseconds,
with 1\arcsec~corresponding to 5 kpc at that redshift \citep[][J.-H. Woo et al. 2011, in preparation]{woo08}.
This is a factor of $\sim$2.8 larger than the actual effective
radius determined from surface-brightness photometry for these objects \citep[][excluding objects
with only upper limits of the spheroid radius]{ben10}.
If we make another comparison, $\sigma_{\rm spat, reff}$/$\sigma_{\rm ap, 2.8\cdot reff}$
= 0.91$\pm$0.05 (rms scatter: 0.2): On average, aperture spectra can overestimate 
the spheroid-only stellar-velocity dispersion by 0.03$\pm$0.02 dex.
This is attributable to a rotational broadening in edge-on objects.
Note that the fraction of spiral host galaxies in the sample at $z$$\sim$0.4 ($\sim$14/34)
is comparable to the fraction in the local sample studied here
($\sim$9/25) and thus, such a comparison is straightforward.

However, the bias introduced by measuring stellar-velocity dispersions
from aperture spectra, even with extraction windows much larger than
the effective radius, cannot explain the observed offset of
$z$$\sim$0.4 Seyfert-1 galaxies from the \mbh-\ms~scaling relation
seen in \citet{woo08}: For a given BH mass, $\sigma_{\rm ap}$ is too
low for the high-z Seyfert galaxies - the opposite effect to the
average bias determined here.  $\sigma_{\rm ap}$ can be underestimated
in case of face-on spiral galaxies with a contribution of the
dynamically cold disk, but this effect is too small (on average less
than 0.01 dex when considering face-on spirals only; see above).  We
performed the same comparisons also for the other spectral fitting
regions (since the region around MgIb was used in \citealt{woo08}),
finding similar results.  To conclude, aperture effects can introduce
a small bias in the $\sigma$ measurements but cannot explain the
offset seen in the \mbh-\s~relation for higher-redshift objects
\citep{woo08}.  

Another more recent study that benefits from our comparison
between stellar-velocity dispersions derived from aperture
spectra to those derived from spatially-resolved spectra 
is the one by \citet{gre10}:
They report that a sample of water megamaser residing in spiral galaxies
in the local Universe ($\sim 0.01 < z < 0.03$)
for which BH masses were derived directly
from the dynamics of the H$_2$O masers 
fall below the \mbh-\s~relation defined by inactive elliptical galaxies.
As pointed out by the authors,
given the nature of the host galaxies
of these megamasers - early-to-mid-type spirals -
a bias of the stellar velocity
dispersion measurements from aperture spectra
due to the presence of the disk is expected.
In principle, a rotational component could bias
the stellar-velocity dispersion measurements to higher values
and result in the observed offset.
However, out of their eight objects, only three are significant outliers (their Figs.~8 and 9), 
namely NGC\,2273, NGC\,6323, and NGC\,2960.
Two of these are seen close to face-on (NGC\,2273 and NGC\,2960; their Fig. 6 and 7)
in which case the effect of the disk component on the
stellar-velocity dispersion measured from aperture spectra
cannot explain the observed offset.
Only for NGC\,6323, for which the disk is seen close to edge-on (their Fig. 6)
could the observed offset indeed be due to rotational broadening.
From our Fig.~\ref{sigma_spat_ap}, we estimate that the stellar velocity
dispersion measured from aperture spectra can overestimate that
from spatially resolved measurements by up to $\sim$40\%
in case of rotational broadening by a disk component seen edge on -
large enough to move NGC\,6323 close to the local relation defined by ellipticals. 
To conclude, while for individual objects, the effect of a disk
on the derived stellar-velocity dispersion from aperture spectra can be significant, it cannot explain the offset observed  by \citet{gre10}
for their full sample of megamasers.

\subsubsection{Stellar-velocity dispersions from different spectral regions}
\label{subsec:comp}
\begin{figure*}[ht!]
\begin{center}
\includegraphics[scale=0.4]{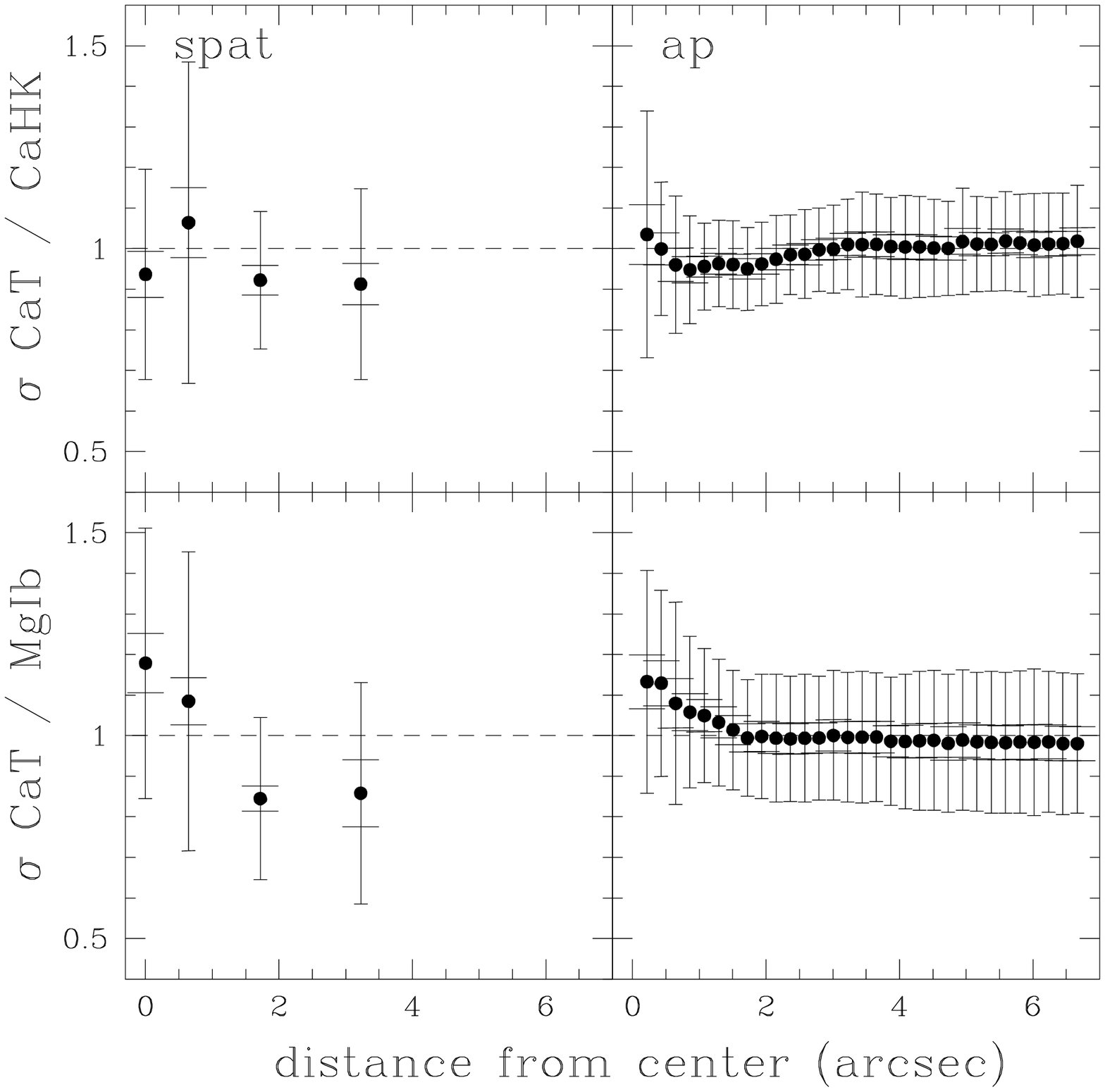}
\includegraphics[scale=0.4]{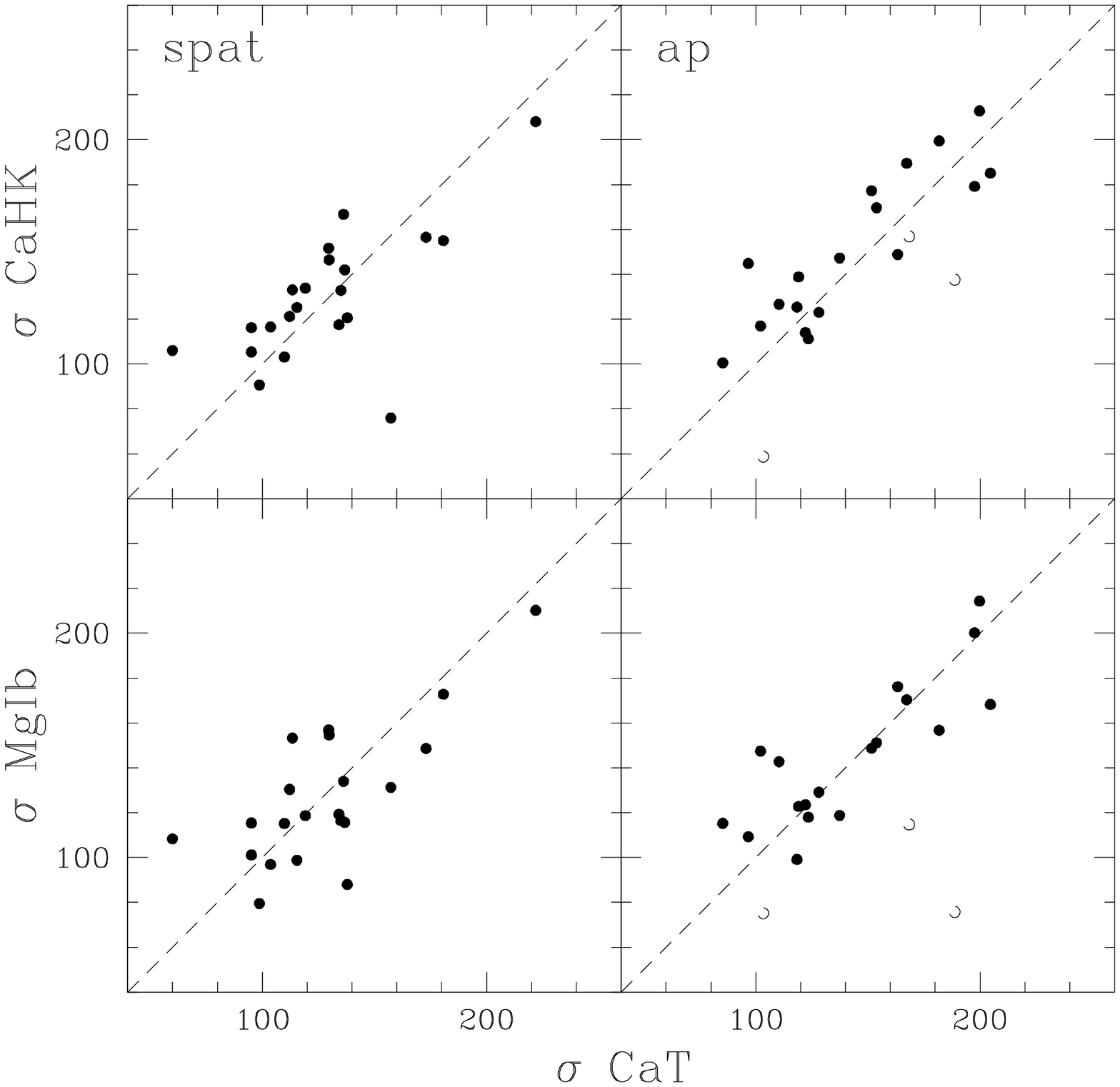}
\end{center}
\caption{Comparison of stellar-velocity dispersion as measured from
CaT, CaHK, and MgIb region.  {\bf a:} Ratio of stellar velocity
dispersions measured from CaT to MgIb (lower panels) and CaHK (upper
panels) as a function of distance from the center.  The mean is shown
at each location including the rms error (shorter - in x - horizontal bar) and
the error on the mean (longer - in x - horizontal bar).  Note that the rms
scatter is due to both intrinsic scatter and measurement errors which
contribute at the level of 7-15\%.  The dashed line corresponds to a
ratio of 1.  From spatially-resolved spectroscopy in the left panels
($\sigma_{\rm spat}$; i.e.~the centroid of the extracted spectra from
which the ratio was measured is at the distance from center that is
indicated on the x-axis with aperture sizes as discussed in
\S~\ref{subsec:spat} and with the measurement on both sides of
the nucleus averaged); from aperture spectra in the right panels
($\sigma_{\rm ap}$; i.e.~the spectra from which the ratio was measured
are always centered on the nucleus but the width of the extracted
aperture increases as indicated on the x-axis).  {\bf b:} One-to-one
comparison for the luminosity-weighted stellar-velocity dispersion
within the effective radius: CaT versus MgIb (lower panels) and CaT
versus CaHK (upper panels) for spatially-resolved spectra in the left
panels ($\sigma_{\rm spat, reff}$) and for aperture spectra
($\sigma_{\rm ap, reff}$) in the right panels, respectively.
The open symbols in the left panels indicate three outliers
that were excluded in Fig.~\ref{sigmacompare}a.
}
\label{sigmacompare}
\end{figure*}

When comparing stellar-velocity dispersion measurements from the three different spectral regions,
the overall picture is that they agree within the errors. 
A more quantitative comparison is
shown in Fig.~\ref{sigmacompare}.
For the aperture data, three extreme outliers were excluded in this figure 
due to contaminating broad AGN emission lines and featureless continuum
swamping the blue spectral region
(namely 0353-0623, 1049+2451, 1535+5754; see also Fig.~\ref{sigmaall}).
These outliers are shown in the one-to-one plot 
of $\sigma_{\rm CaT}$ vs. $\sigma_{\rm MgIb}$
and $\sigma_{\rm CaHK}$ within the effective radii
as open symbols (Fig.~\ref{sigmacompare}b).
(Note that none of the objects was excluded for the spatially-resolved
data, explaining the higher scatter.)

In general, for the spatially-resolved stellar-velocity dispersions
(Fig.~\ref{sigmacompare}a, left panels)
both ratios $\sigma_{\rm spat, CaT}$/$\sigma_{\rm spat, MgIb}$
and  $\sigma_{\rm spat, CaT}$/$\sigma_{\rm spat CaHK}$
scatter at the 20-30\% level.
The average $\sigma_{\rm spat, CaT}$/$\sigma_{\rm spat, MgIb}$ ratio 
shows a slight dependence on radius
in the sense that measuring $\sigma_{\rm spat}$ from the MgIb region 
in the center tends to underpredict the ``true'' $\sigma_{\rm spat}$ 
(here assumed to be $\sigma_{\rm spat, CaT}$) by on average 5-10\%
while it gets overpredicted in the outerparts by on average $\sim$10-15\%.
This trend with radius 
can be attributed to the AGN contamination from emission lines
and featureless continuum that is
only present in the inner most spectra, and results
not only in an increased error in the determined
velocity dispersion but also in a possible  bias.
The ratio $\sigma_{\rm spat, CaT}$/$\sigma_{\rm spat, CaHK}$,
on the other hand,
shows  no clear trend with radius; 
generally, measuring $\sigma_{\rm spat}$ from the CaHK region
overpredicts the ``true'' $\sigma_{\rm spat}$ by on average 5-10\%.
Note that there is no obvious trend with galaxy morphology.

For the average ratio from the aperture spectra
(Fig.~\ref{sigmacompare}a, right panels),
the trend is similar to the spatially-resolved
stellar-velocity dispersions for small apertures.
At larger radii, the average ratio for both
$\sigma_{\rm ap, CaT}$/$\sigma_{\rm ap, CaHK}$
and  $\sigma_{\rm ap, CaT}$/$\sigma_{\rm ap, MgIb}$ 
approaches unity.
For $\sigma_{\rm ap, CaT}$/$\sigma_{\rm ap, MgIb}$ 
this is probably due to the fact that the dependence
on radius seen in the spatially-resolved ratio
cancels out.

For the stellar velocity dispersion within the effective radius
(Fig.~\ref{sigmacompare}b),
all ratios are unity within the errors
($\sigma_{\rm spat, CaT}$/$\sigma_{\rm spat, CaHK}$ = 1.01$\pm$0.06;
rms scatter = 0.28; $\sigma_{\rm spat, CaT}$/$\sigma_{\rm spat, MgIb}$ = 1.03$\pm$0.04;
rms scatter = 0.22; $\sigma_{\rm ap, CaT}$/$\sigma_{\rm ap, MgIb}$ = 0.98$\pm$0.03;
rms scatter = 0.15), except for
the stellar-velocity dispersion measured in the CaHK region
from aperture spectra that tend to overpredict 
$\sigma_{\rm ap, CaT}$ by on average 6\%
($\sigma_{\rm ap, CaT}$/$\sigma_{\rm ap, CaHK}$ = 0.94$\pm$0.03;
rms scatter = 0.12).

To estimate the effect of the potential bias induced
when using the MgIb region to determine
the stellar-velocity dispersion 
for the study of the evolution of the 
\mbh-\ms~scaling relation (as done for 
our $z$$\sim$0.4 Seyfert-1 sample; \citealt{woo08}),
we take into account that typically, an aperture much larger
than the actual effective radius (a factor of $\sim$2.8 for \citealt{woo08}) 
is used for extraction of the spectra (see above).
We find no bias
($\sigma_{\rm ap, CaT}$/$\sigma_{\rm ap, MgIb}$ = 0.98$\pm$0.04;
rms scatter = 0.17).

The general conclusion we can draw from this comparison
is that while the CaT region is the cleanest region
to determine stellar-velocity dispersions, both the MgIb region,
appropriately corrected for FeII emission,
and the CaHK region, although often swamped by the
blue AGN powerlaw continuum and strong AGN emission lines, 
can also give accurate
results within a few percent, given high S/N spectra.
This is an important
improvement over the study by \citet{gre06b}
who use fiber-based SDSS spectra (i.e.~aperture spectra
with a radius of 1.5\arcsec) for a sample of 40 type-1 AGNs
for a similar comparison but find a bias of the order of 20-30\%
(in both MgIb, not corrected for FeII emission, and CaHK).

Furthermore, spatially-resolved spectra are very helpful as they allow to
eliminate the AGN contamination (powerlaw continuum and strong emission lines)
especially prominent in the blue spectra (CaHK and MgIb region)
when extracting spectra outside of the nucleus.

\subsection{\mbh~Scaling Relations}
\label{ssec:mbhs}
\begin{figure*}[ht!]
\begin{center}
\includegraphics[scale=0.4]{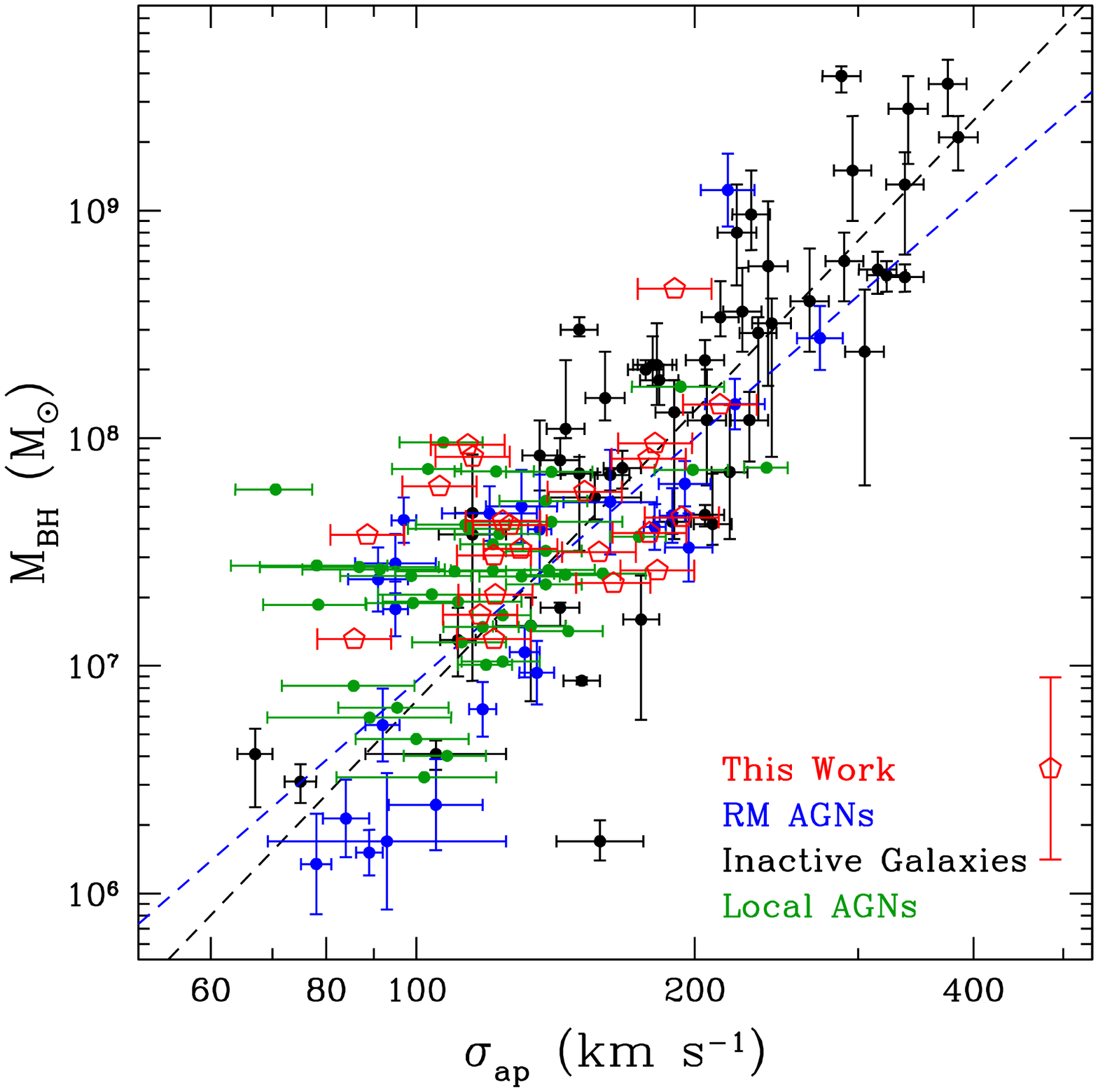}
\includegraphics[scale=0.4]{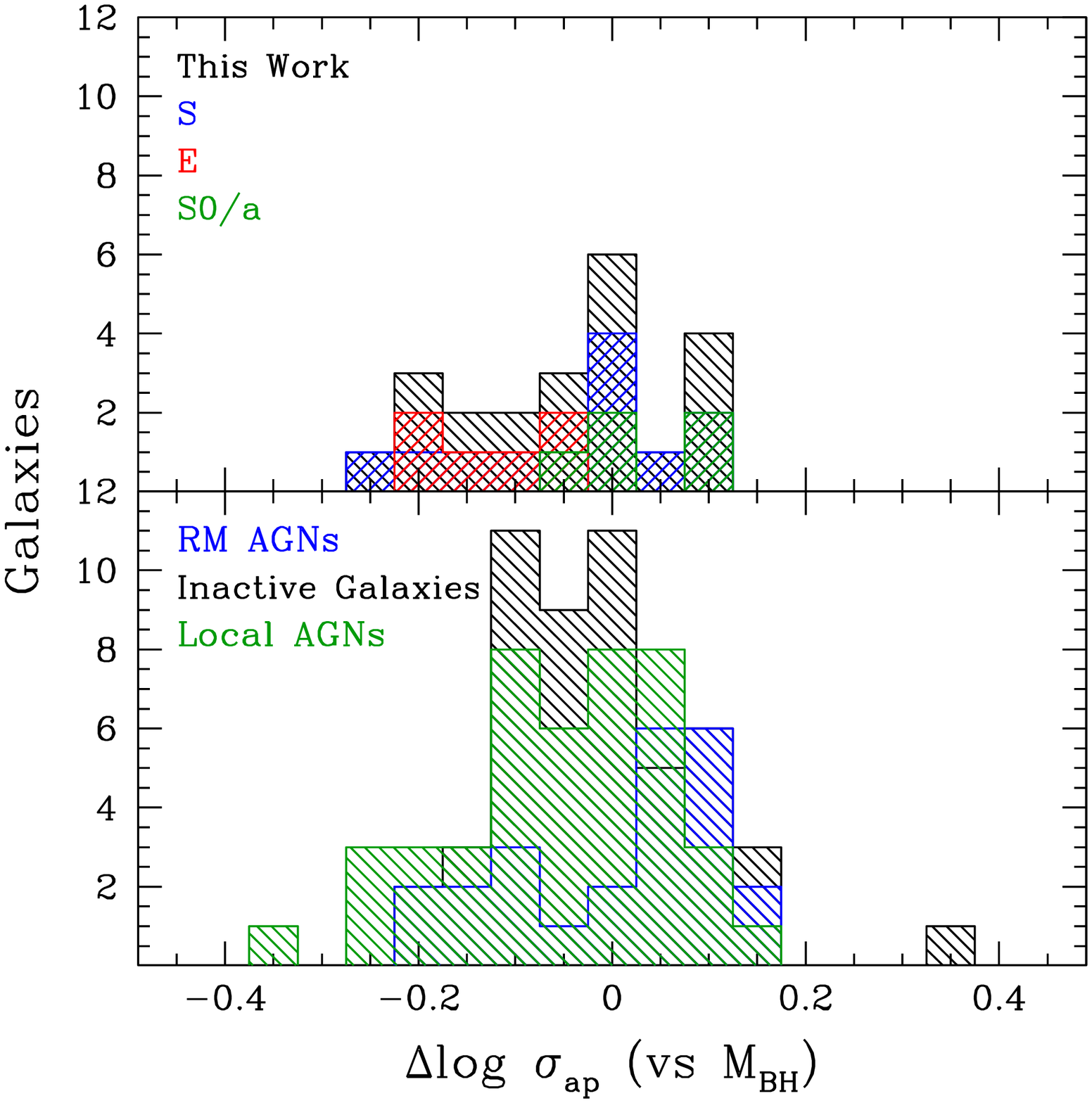}
\end{center}
\caption{{\bf Left panel:} \mbh-\s~relation for our sample (red open pentagons),
the local RM AGNs \citep[blue;][with the blue dashed line their best fit]{woo10},
and a local sample
of AGNs for which $\sigma$ was measured from SDSS \citep[green;][]{gre06a}
and BH masses taken from \citet{woo06}.
The black datapoints correspond to inactive local galaxies
from \citet{gue09} (with the black dashed line their best fit;
see text for details). $\sigma_{\rm ap}$ corresponds to the
luminosity-weighted stellar
velocity dispersions within a given aperture, depending
on the sample (see text for details). The error on the BH mass for both our sample and
the local sample of AGNs (green data points) is 0.4 dex
and shown as a separate point with error bar in the legend, to reduce confusion of data points.
{\bf Right panel:}
Distribution of residuals with respect to the fiducial
local relation of \citet{gue09}. Lower panel: literature data; upper panel:
our sample (black: full sample; blue: spirals, red: ellipticals,
green: S0/a)}
\label{mbh_s}
\end{figure*}

\begin{figure*}[ht!]
\includegraphics[scale=0.26]{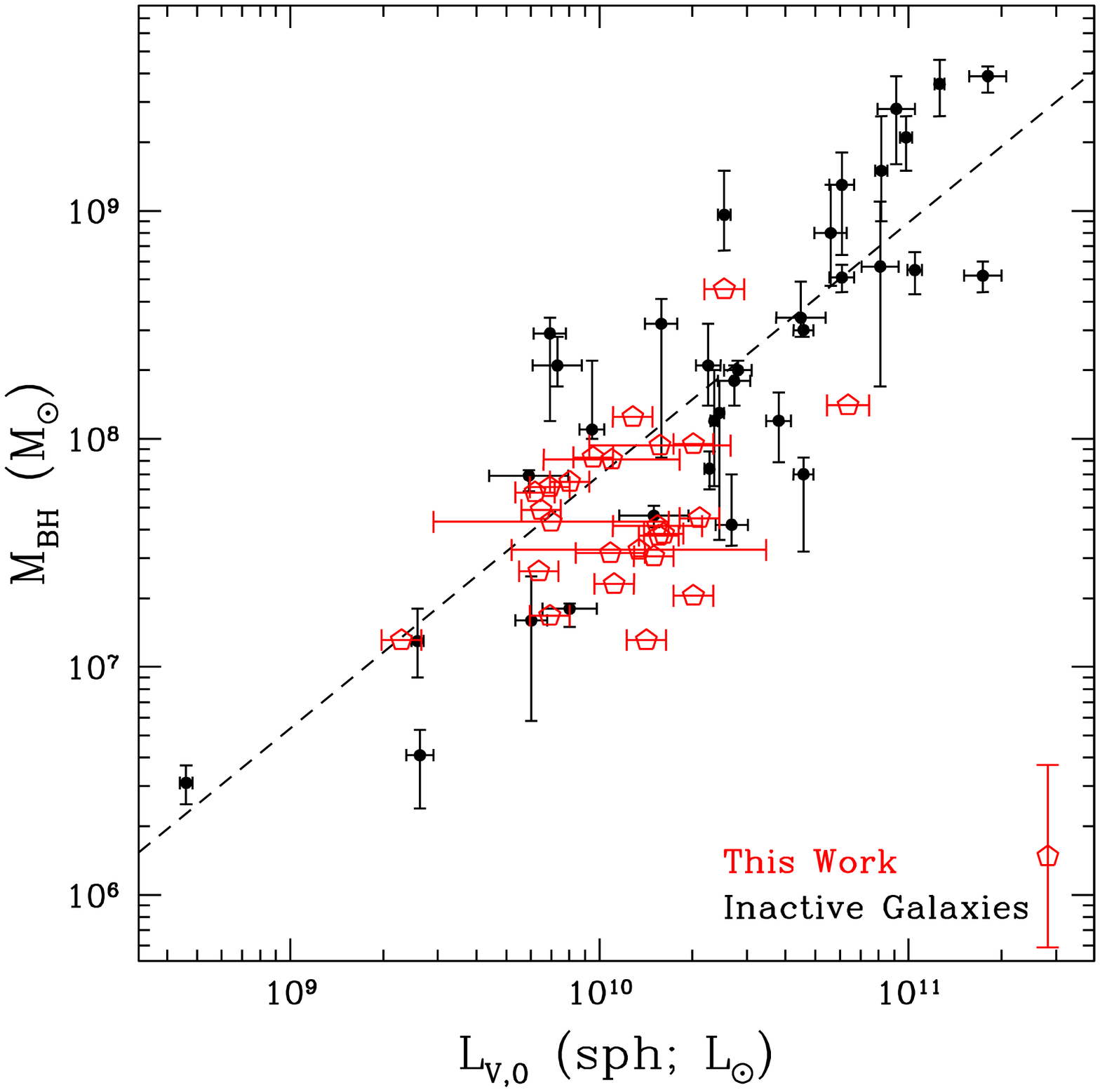}
\includegraphics[scale=0.26]{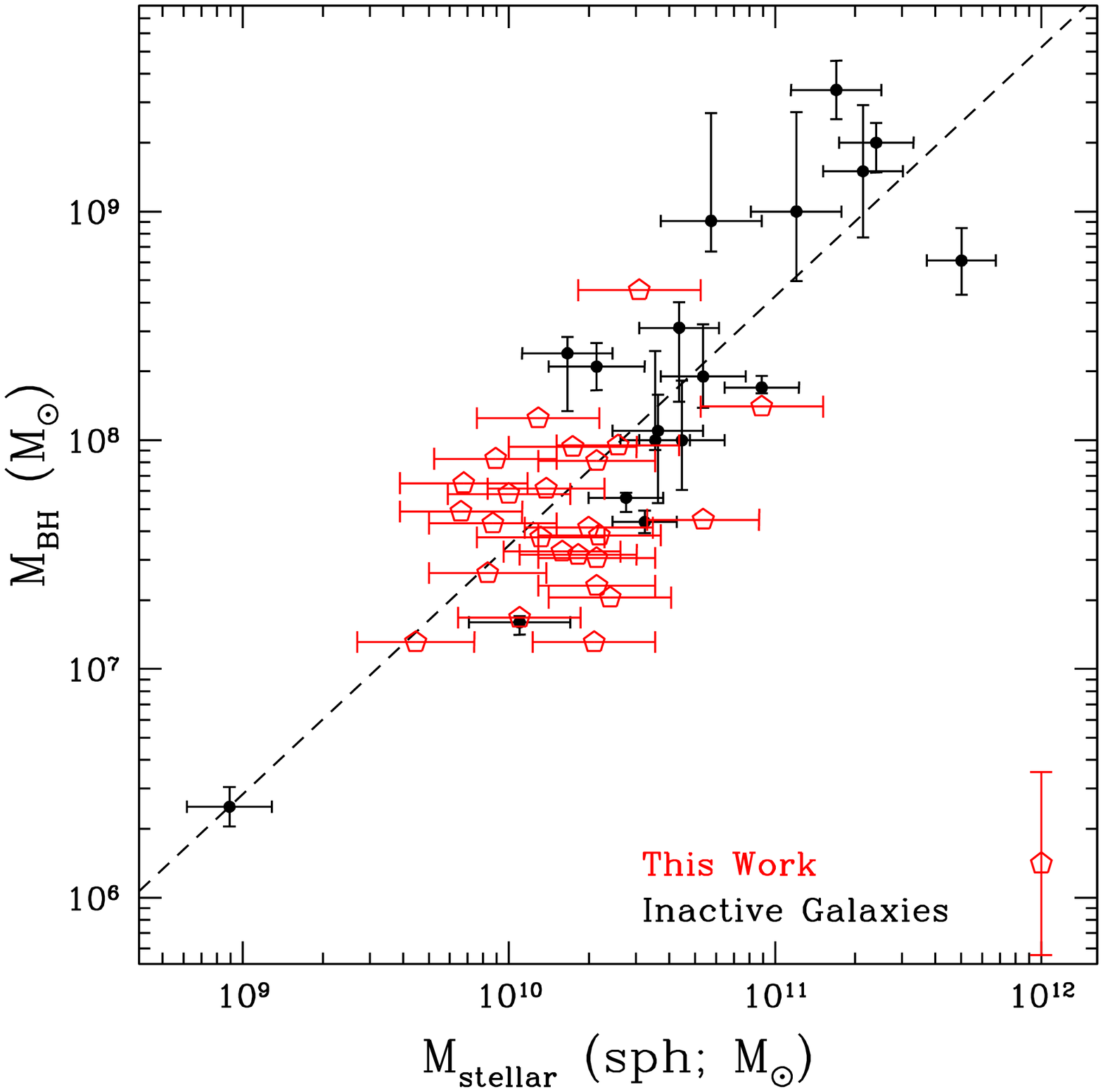}
\includegraphics[scale=0.26]{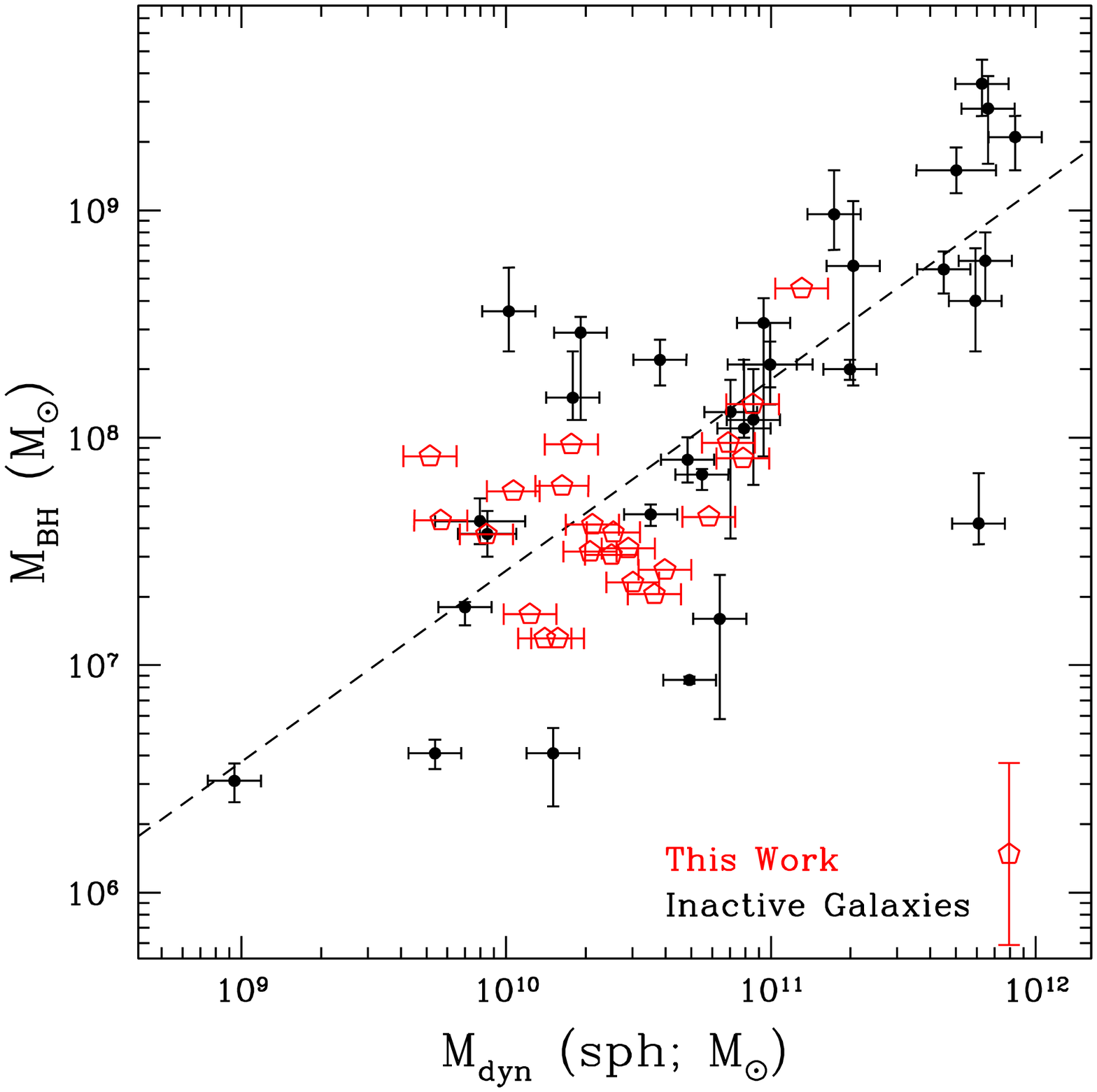}\\
\includegraphics[scale=0.26]{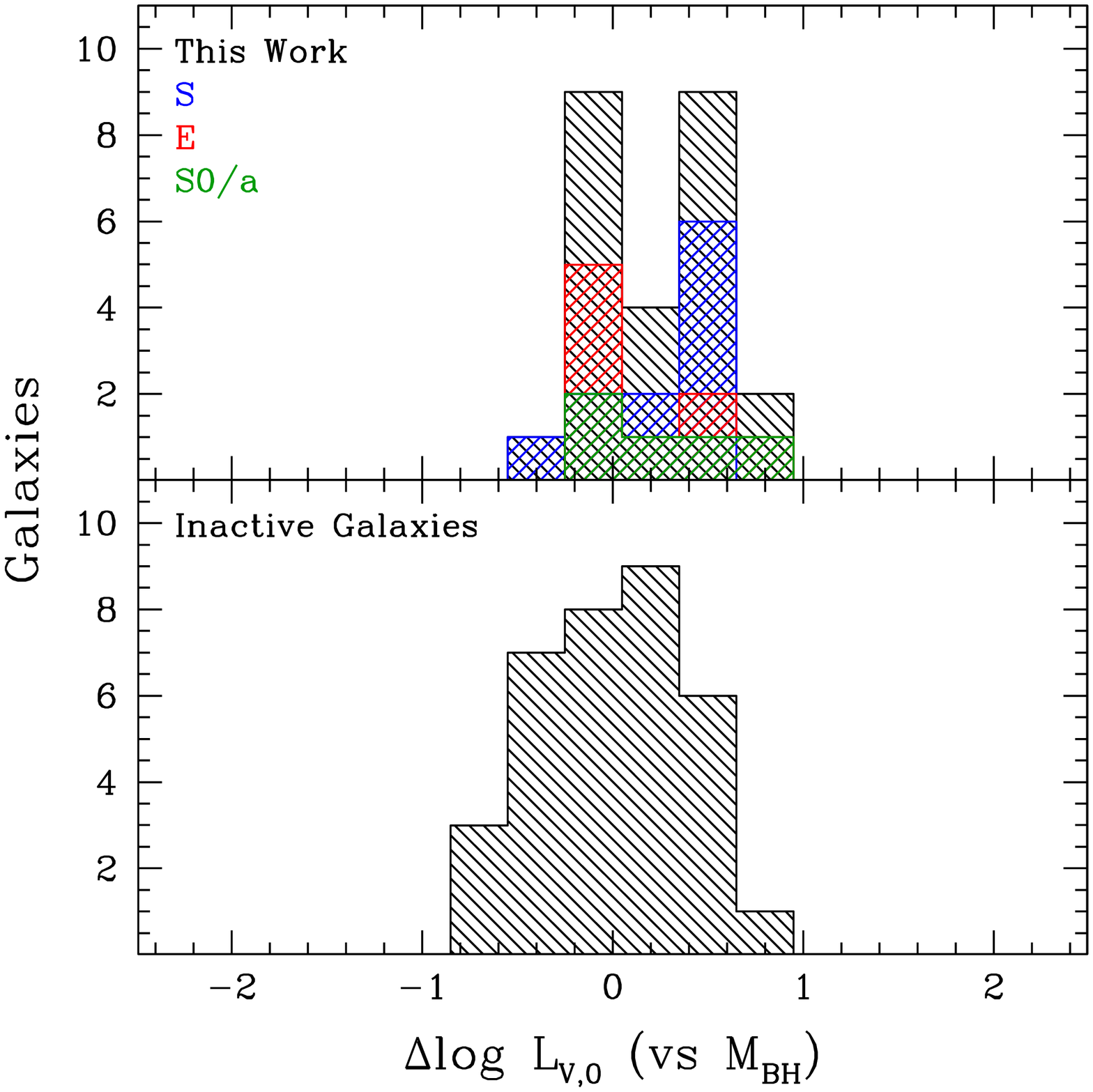}
\includegraphics[scale=0.26]{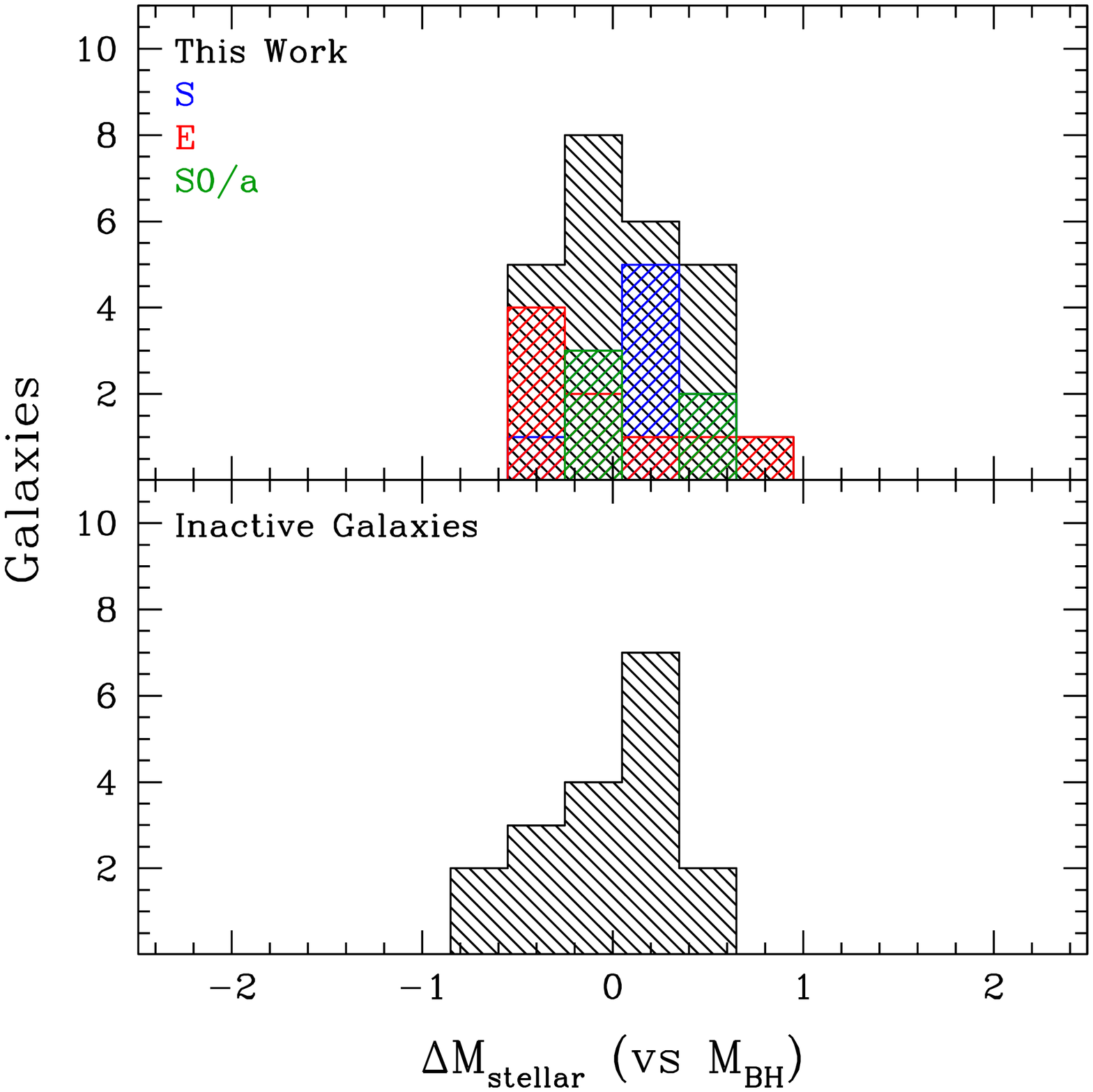}
\includegraphics[scale=0.26]{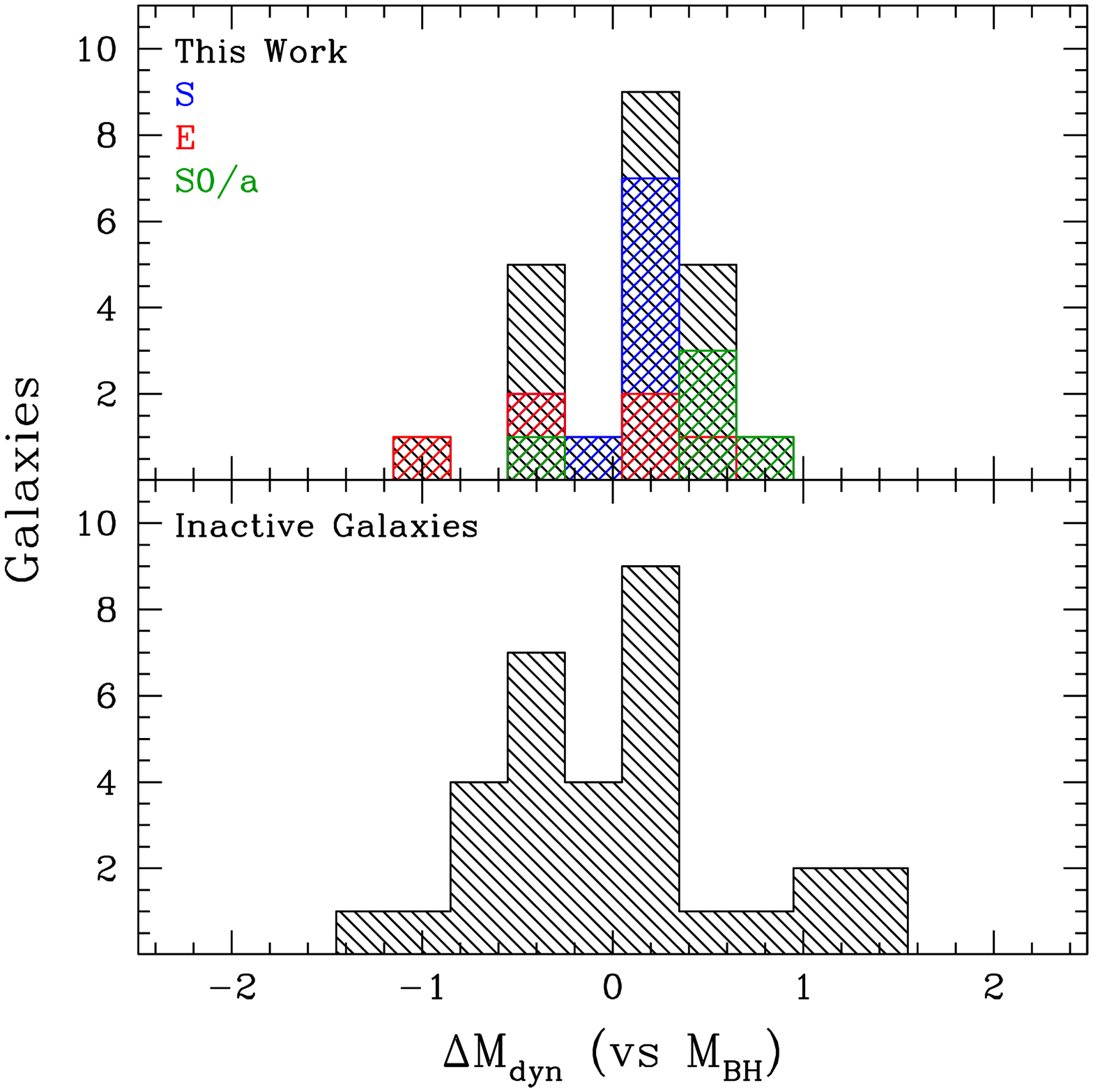}
\caption{The same as in Fig.~\ref{mbh_s}, but 
for the other three  \mbh~scaling relations, namely \mbh-\ls~(left panels),
\mbh-\mss~(middle panels), and \mbh-\msv~(right panels).
{\bf Left panels:}
For the \mbh-\ls~relation, we use the local
inactive sample from \citet{gue09}, here limited
to 35 elliptical and S0 galaxies with a reliable spheroid-disk decomposition.
{\bf Middle panels:}
For the \mbh-\mss~relation,
stellar masses were calculated from the
J, H, and K magnitudes from \citet{mar03} for their group 1
(see text for details).
Also, BH masses were updated using those listed in \citet{gue09}.
{\bf Right panels:}
For the \mbh-\msv~relation,
we compile local inactive galaxies using BH masses from \citet{gue09},
and calculate dynamical masses using $r_{\rm reff, sph}$ given by
\citet{mar03} and $\sigma_{\rm ap}$ measurements by \citet{gue09}.
}
\label{mbh_scaling}
\end{figure*}

We can now create four
different BH mass scaling relations, namely
\mbh-\s, \mbh-\ls, \mbh-\mss, and \mbh-\msv~and compare
our results with literature data (\S~\ref{sec:comp}).
The resulting relations are shown in Figs.~\ref{mbh_s} and~\ref{mbh_scaling}.
The distribution of residuals with respect to the fiducial
local relations (Table~\ref{fits}) are shown as histograms.

In Fig.~\ref{mbh_s}, 
we plot $\sigma_{\rm ap, 1.5''}$ from aperture spectra within
1.5\arcsec as stellar-velocity dispersion of our sample, for comparison
with literature data, for which all measurement
were derived from apertures spectra
(with different sizes, see \S~\ref{sec:comp}; we here use 1.5\arcsec~to be comparable
to SDSS fiber spectra).
Overall, our sample follows the same \mbh-\s~relation  as that
of the other active local galaxies and also that of inactive galaxies.
For the local AGNs with stellar velocity
dispersion measurements from SDSS fiber spectra (green data points in Fig.~\ref{mbh_s}),
\citet{gre06a} already noticed that these
objects seem to follow a shallower \mbh-\s~relation
with an apparent offset at the low-mass end
in the sense that the stellar velocity dispersion is
smaller than expected. The same trend may also to be visible
for the RM AGNs (blue data points, \citealt{woo10})
and our local active galaxies (red data points).
However, for our data points, this trend can be attributed
to five objects (0121-0102, 0846+2522, 1250-0249, 1535+5754, and 1605+3305)
that might simply be outliers, with strong AGN contamination,
especially in the aperture spectra.
However, at this point, the available
dynamic range is too small to distinguish
between a real offset/change in slope or simply a rising scatter.

The \mbh-\ls~relation indicates that our sample of active galaxies
resides in host galaxies that are overluminous compared to the
inactive galaxies (on average by 0.15$\pm$0.08 dex; rms scatter: 0.4).
One potential bias here could be that, due to the shallow images, we
are missing the disk contribution and thus overestimating the bulge
luminosity for objects that we classified as ellipticals and fitted by
a spheroidal component only (see also \S~\ref{ssec:host}).
However, the distribution of residuals with respect to the fiducial
relation of inactive galaxies shows that especially host galaxies
classified as spirals contribute to this offset (offset 0.2
dex$\pm$0.07 dex; rms scatter: 0.37), arguing against such a bias.
In fact, such an offset might not be too surprising for two reasons:
(i) the \citet{gue09} sample only includes ellipticals and S0/a in the
luminosity plot and (ii) the enhanced luminosity might be due to
starformation triggered from a same event that triggered the AGN.
That AGNs are often hosted by actively star-forming galaxies
has been found in various studies at different redshifts
\citep[e.g.,][]{kau03,jah04,hic09,mer10}.

However, we cannot exclude that at least some of the spiral
galaxies have pseudobulges which are characterized by
surface-brightness profiles closer to exponential profiles
\citep[e.g.,]{kor04,fis08}. As discussed in Appendix~\ref{sec:mattscode},
using a S{\'e}rsic index of $n = 1$
instead of $n = 4$ for the spheroid component can
decrease the spheroid luminosity by  $\sim$0.2 dex,
thus accounting for at least some of the offset.
We will explore this effect further when analysing
the full sample.

For both the \mbh-\mss~and \mbh-\msv~relations,
our objects seem to follow the relations determined
by the inactive galaxies.

Note that we have a small sample size and also a small dynamic
range in the parameters covered 
and all four BH mass scaling relations
presented here show a large scatter. 
Thus, we refrain from discussing the results any further at this point
but will get back to the local BH mass scaling relations
in more detail when we have the full sample available.

\begin{deluxetable*}{llcccc}
\tabletypesize{\scriptsize}
\tablecolumns{6}
\tablewidth{0pc}
\tablecaption{Fits to the Local Scaling Relations}
\tablehead{
\colhead{Relation} & \colhead{Sample} & \colhead{$\alpha$} & \colhead{$\beta$} & \colhead{Scatter} & \colhead{Reference}\\
\colhead{(1)} & \colhead{(2)} & \colhead{(3)}  & \colhead{(4)} & \colhead{(5)} & \colhead{(6)}}
\startdata
$\log (M_{\rm BH}/M_{\odot}) = \alpha + \beta \log (\sigma_{\rm ap} / 200 {\rm km\,s}^{-1})$ & RM AGNs           &   8$\pm$0.14     & 3.55$\pm$0.60   & 0.43$\pm$0.08  & Woo+10$^a$\\
                                                                                             & inactive galaxies &   8.12$\pm$0.08  & 4.24$\pm$0.41   & 0.44$\pm$0.06  & G\"ultekin+09\\
$\log (M_{\rm BH}/M_{\odot}) = \alpha + \beta \log (L_{\rm sph, V} / 10^{11} L_{\odot, V})$  & inactive galaxies &   8.95$\pm$0.01  & 1.11$\pm$0.18   & 0.38$\pm$0.09  & G\"ultekin+09\\
$\log (M_{\rm BH}/M_{\odot}) = \alpha + \beta \log (M_{\rm sph, \star} / M_{\odot})$         & inactive galaxies &  -3.34$\pm$1.91  & 1.09$\pm$0.18   & 0.38$\pm$0.1   & here\\
$\log (M_{\rm BH}/M_{\odot}) = \alpha + \beta \log (M_{\rm sph, dyn} / M_{\odot})$           & inactive galaxies &  -0.98$\pm$1.31  & 0.84$\pm$0.12   & 0.54$\pm$0.08  & here\\
\enddata
\tablecomments{
Relations plotted as dashed lines in Figs.~\ref{mbh_s}-~\ref{mbh_scaling} and used as fiducial relation when calculating residuals.
Col. (1): Scaling relation.
Col. (2): Sample used for fitting. (Note that we do not fit our local sample as the scatter is too large.) 
Col. (3): Mean and uncertainty on the best fit intercept.
Col. (4): Mean and uncertainty on the best fit slope.
Col. (5): Mean and uncertainty on the best fit intrinsic scatter.
Col. (6): References for fit. ``here'' means determined in this paper independently.\\
$^a$ Assuming the virial coefficient log f = 0.72$\pm$0.10 \citep{woo10}.}
\label{fits}
\end{deluxetable*}

\section{SUMMARY}
\label{sec:sum}
To create a local baseline of the BH mass-scaling relations
for active galaxies, we selected a sample of $\sim$100 local 
(0.02 $\le$ $z$ $\le$ 0.1) Seyfert-1 galaxies from the SDSS (DR6)
with \mbh~$>10^{7}$M$_{\odot}$. All objects were observed with
Keck/LRIS, providing us with high-quality longslit spectra.
These data allow us to determine, for the first time, spatially-resolved
stellar velocity dispersions.
Here, we present the methodology and first results of a pilot
study of 25 objects. The full sample will be presented
in the forthcoming papers of this series.

From the Keck spectra, we
measure both spatially-resolved stellar-velocity dispersion
and aperture stellar-velocity dispersions
in three different spectral regions:
around CaHK, around MgIb (after subtraction of 
underlying broad FeII emission), and around CaT.
We present a detailed comparison between 
spatially-resolved and aperture stellar-velocity dispersions
as well as stellar-velocity dispersions from different spectral regions.
Also, we determine the width of the H$\beta$ emission line
(after subtraction of broad FeII emission and stellar absorption).

On archival SDSS images  (g', r', i', z'), we perform surface photometry,
using a newly developed code
that allows a joint multi-band analysis.
We determine the spheroid effective radius, spheroid luminosity,
and the host-galaxy free 5100\AA~AGN continuum luminosity.

Combining the results from spectroscopy and imaging
allows us to estimate BH masses via the empirically calibrated photo-ionization
method from the width of the H$\beta$ emission line
and the host-galaxy free 5100\AA~AGN continuum luminosity.
The spheroid effective radius is used to determine the luminosity-weighted stellar-velocity dispersion
within $r_{\rm reff, sph}$.
The spheroid luminosities in four different bands are used to
calculate stellar masses.
Also, our results allow us to estimate dynamical masses.
We can thus study four different BH mass scaling relations:
\mbh-\s, \mbh-\ls, \mbh-\mss, and \mbh-\msv.

The main results for the pilot study can be summarized as follows.

\begin{itemize}
\item{The host galaxies show a wide variety of morphologies
with a significant fraction of spiral galaxies and prominent
rotation curves. This underscores the need for spatially-resolved
stellar-velocity dispersions.}
\item{We find a lower merger rate than for our higher redshift study,
comparable to inactive galaxies in the local Universe.}
\item{Determining stellar-velocity dispersions from aperture spectra
(such as SDSS fiber spectra or unresolved data from distant galaxies)
can be biased, depending on the size
of the extracted region, AGN contamination, and
the host-galaxy morphology. An overestimation of
the stellar velocity dispersion
from aperture spectra is due to broadening from an underlying
rotation component (if seen edge-on), an underestimation can originate
from the contribution of the dynamically cold disk (if seen face on).
However, comparing with the higher-redshift Seyfert-1 sample
of \citet{woo08}, we find that, on average, such a bias is small
($<$0.03 dex) and, moreover, in the opposite direction
to explain the offset seen in the \mbh-\s~relation.}
\item{The CaT region is the cleanest region to determine stellar-velocity
dispersion in AGN hosts. However, it gets shifted out of the optical
wavelength regime to be used beyond redshifts of $z\simeq0.1$.
Alternatively, both the MgIb region,
appropriately corrected for FeII emission, and the 
CaHK region, although often swamped by the
blue AGN powerlaw continuum and strong AGN emission lines, 
can also give accurate
results within a few percent, given high S/N spectra. 
Spatially-resolved data are very helpful to eliminate the AGN
contamination by extracting spectra outside of the nucleus.}
\item{The BH mass scaling relations of our pilot sample
agree in slope and scatter with those of other local active galaxies
as well as inactive galaxies for a canonical choice of the
normalization of the virial coefficient.}
\end{itemize}

\acknowledgments

We thank the anonymous referee for carefully reading
the manuscript and for useful suggestions.
V.N.B. is supported
through a grant from the National Science Foundation (AST-0642621) and
by NASA through grants associated with HST proposals GO 11208, GO 11341,
and GO 11341. T.T. acknowledges support from the
NSF through CAREER award NSF-0642621, and
from the Packard Foundation.  
Data presented in this paper were obtained at the W.M.
Keck Observatory, which is operated as a scientific
partnership among Caltech, the University of California,
and NASA. The Observatory was made possible by the generous financial
support of the W.M. Keck Foundation. The authors wish to
recognize and acknowledge the very significant cultural role
and reverence that the summit of Mauna Kea has always had
within the indigenous Hawaiian community. We are most fortunate
to have the opportunity to conduct observations from this mountain.
This research has made use of the public archive of the Sloan Digital Sky Survey
and the NASA/IPAC Extragalactic Database (NED) which is operated by the Jet
Propulsion Laboratory, California Institute of Technology, under
contract with the National Aeronautics and Space Administration.

{\it Facilities:} \facility{Keck:I (LRIS)}

\appendix
\section{SURFACE PHOTOMETRY}
\label{sec:mattscode}
The imaging results of this paper rely on a new surface photometry code that we have developed (\S~\ref{ssec:phot}). Here we describe the code in detail and discuss the SDSS photometric data and how the code was used to fit surface brightness models to these data.

\subsection{SPASMOID: A New Surface Photometry Code}
Surface Photometry and Structural Modeling of Imaging Data, or SPASMOID, written by MWA, 
is an image analysis code designed to supersede the functionality of GALFIT \citep{pen02}. 
The code employs a Bayesian framework wherein the various model parameters (e.g., the centroid, total magnitude, effective radius, ellipticity, position angle, and/or S{\'e}rsic index) can be tied together by priors (i.e., constraints) \emph{between model components or across different filters}. For example, it is straightforward to impose a prior that the redder bands have smaller effective radii than the bluer bands, that the AGN has blue colors related by a power-law of frequency, or that the relative position between the AGN and the bulge is the same in all bands and the offset between these components must be small. The code also allows for a different PSF model for extended objects and point sources (this is useful if the different components have different colors), and can even use linear combinations of PSFs for the AGN to account for PSF-mismatch, which may be particularly important for HST imaging.

The code implements a MCMC sampler to explore degeneracies between the parameters and provide robust error estimates. A set of reduced data images is provided by the user along with variance images, image masks, meta data (e.g., the photometric zeropoints and the pixel scale in each image), and a starting guess for the parameters (this can be substantially different than the best parameters, although a closer guess to the ``true'' value leads to more efficient sampling). A likelihood function is defined assuming Gaussian uncertainties on the pixel values as described by the variance images, e.g.,
$$
{\rm log} P = \sum_{\rm images} \sum_{\rm pixels} \frac{({\rm image} - {\rm model})^2}{\sigma^2_{\rm image}} - \frac{1}{2}{\rm log} 2\pi\sigma^2_{\rm image}.
$$
Priors on all relevant model parameters and -- for more complicated models -- hyperparameters are defined by the user, and the code uses the PyMC python module to explore the posterior. We have tested our code on simulated and real data and have compared our results with those derived from GALFIT. We find that, for priors that approximate the implicit GALFIT priors, we are able to reproduce the GALFIT results (\S~\ref{ssec:mbhs}).

A code as flexible as SPASMOID is ideally suited for space-based data;
the spatial resolution and depth of the SDSS photometry limits the utility of such a code,
however, it still has its advantages.
We now describe how the SDSS data are prepared and subsequently modeled with SPASMOID.

\subsection{Preparation of SDSS images}
First, we determine the magnitude zeropoint (zp) following the recipe
described on the SDSS DR7 webpage\footnote{http://www.sdss.org/DR7/algorithms/fluxcal.html\#counts2mag}:
\begin{eqnarray*}
zp = -1 \cdot (a + b*{\rm airmass} - 2.5 \cdot \log({\rm exptime}))
\end{eqnarray*}
with $a$ = zeropoint count rate and
$b$ = extinction coefficient taken from the ``tsfield'' header keywords for
a given field and filter, and exptime = 53.91 sec.
Then, the sky was subtracted using either the sky value in the image header (if present)
or as determined independently directly from the image (note that the 1000-counts bias was also subtracted).
We created noise images (in counts) according to
\begin{eqnarray*}
{\rm noise} = \sqrt{\frac{\rm data+sky}{\rm gain}+\frac{\rm dark\_variance}{\rm gain^2}}
\end{eqnarray*}
with data images in counts and gain and dark\_variance
taken from the ``tsfield'' header keywords.
Then, for convolution with the PSF of the Sloan telescope
optics, we use a Gaussian with the parameters given in the ``ObjAll'' table of
SDSS for a given object and filter (``mRrCcPSF\_filter'', ``mE1PSF\_filter'', ``mE2PSF\_filter'').
The average seeing $s$ for the 25 objects was 
$s_g$ = 1.35$\pm$0.15, $s_r$ = 1.24$\pm$0.15, $s_i$=1.14$\pm$0.15, $s_z$=1.16$\pm$0.14.

We only use the four filters g', r', i', and z', as u' is generally
too faint. 

\subsection{Running GALFIT for comparison}
For comparison, we ran GALFIT on all objects,
in a very similar fashion as described in detail in \citet{ben10}. 
In short, we first fit the central AGN component with a PSF, thus
determining the center of the system which was subsequently
fixed to all components.
The PSF used was created as a circular Gaussian with
an FWHM corresponding to the seeing, derived from the parameter ``mRrCcPSF\_filter''
by $\sqrt({\rm mRrCcPSF\_filter}/2)*2.355$.
We then fitted a two component
model only, consisting of a PSF
and a \citet{dev48} profile. The starting parameters for the GALFIT
runs were taken from the SDSS DR7 catalog, i.e.~PSF magnitude (``psf\_mags'')
and \citet{dev48} magnitude  (``deVmag'') for the spheroid.
The minimum radius of the \citet{dev48} profile was set to 2 pixels ($\sim$0.8\arcsec),
i.e.~the minimum resolvable size given the seeing.

\subsection{Fitting with SPASMOID}
Using the GALFIT results (PSF magnitude, location of PSF, spheroid magnitude,
spheroid effective radius, ellipticity and position angle)
as starting parameters,
we assumed the following AGN/host galaxy fitting procedure.

We fit the host galaxy by either a single \citet{dev48} profile or by a \citet{dev48} 
plus an exponential profile to account for a disk, while the AGN point source is modeled 
as the Gaussian described for the convolution PSF in the previous section. 
All of the components are concentric but the centroids may vary between bands to account 
for imperfect registration of the images. We also fix the effective radius to be 
the same in all bands (i.e., our photometry is similar to the modelMag photometry of SDSS) 
as well as the position angle and axis ratio, although these are free to vary between the bulge and disk component.
Again, the minimum radius of the \citet{dev48} profile was set to 2 pixels.

The normalizations of the profiles -- that is, the magnitudes of each component -- 
are determined by generating models given the structural parameters (centroid, effective 
radius, axis ratio, and position angle) and finding the best coefficients of a linear fit of 
these models to the data. In principle the magnitudes could be free parameters of the MCMC sampler, 
but taking advantage of the linear nature of the fit allows us to very quickly find 
the `optimal' magnitudes for each proposed set of structural parameters.

Finally, depending on the images, residuals, and the $\chi^2$ statistics,
we decide whether a given object is best fitted by three components
(PSF, spheroid, disk) or two components (PSF + spheroid).
(This procedure is similar to the one adopted in \citealt{tre07} and \citealt{ben10}.)

We subsequently applied correction for Galactic extinction
(subtracting the SDSS DR7 ``extinction\_filter''' column).
The resulting AB magnitudes
were transformed to rest-frame optical bands by performing synthetic
photometry on an early-type galaxy template spectrum, a procedure
traditionally referred to as k-correction.  The template spectrum
initially has arbitrary units, and these units were adjusted so that
the synthetic observed frame magnitudes match the magnitudes
from our photometry.  We then evaluated the V-band magnitudes at the
rest-frame of the template; luminosities were determined by correcting
for the distance modulus given our adopted cosmology.  The errors on
extinction and rest-frame transformation are a few hundredths of a
magnitude. We estimate an uncertainty of $<$0.05 mag (using the
scatter in 20 single stellar population templates
with ages ranging from 2 Gyr to 8 Gyr).
Table~\ref{surface} summarizes the results.

Note that our model assumes that the host galaxies can be best fitted
by either a single \citet{dev48} profile or by a \citet{dev48} 
plus an exponential profile.  
However, some of the spiral galaxies may not have classical bulges,
but pseudobulges which are characterized by
surface-brightness profiles closer to exponential profiles
\citep[e.g.,][]{kor04,fis08}.
To test the most extreme systematic
uncertainties in derived spheroid and PSF
magnitude, we re-ran our models for those 14 objects
for which we fit a bulge plus disk component
using a S{\'e}rsic index of $n = 1$
instead of $n = 4$ for the spheroid component.
The results are comparable to what has already been
observed by \citet{ben10} (e.g.~their Fig. 10b):
Decreasing $n$ from 4 to 1 decreases the spheroid luminosity
by on average 0.4 mag and increases the nuclear luminosity
by on average 0.6 mag. At the same time, the disk luminosity increases
by on average 0.1 mag.
Thus, the extreme systematic effect would move those objects
up in BH mass by on average $\sim$0.1 dex 
(small, but not negligible compared to the assumed error of 0.4 dex)
and towards lower spheroid luminosities by $\sim$0.2 dex.
However, the image quality does not allow to determine the best S{\'e}rsic index, 
especially since the objects are complex in nature
due to the presence of the AGN for which a perfectly matching
PSF fit cannot always be achieved and can result in
degeneracies between PSF and spheroid.

\subsection{Comparison between GALFIT and SPASMOID Photometry}
\label{ssec:comparison}
\begin{figure*}[ht!]
\begin{center}
\includegraphics[scale=0.5]{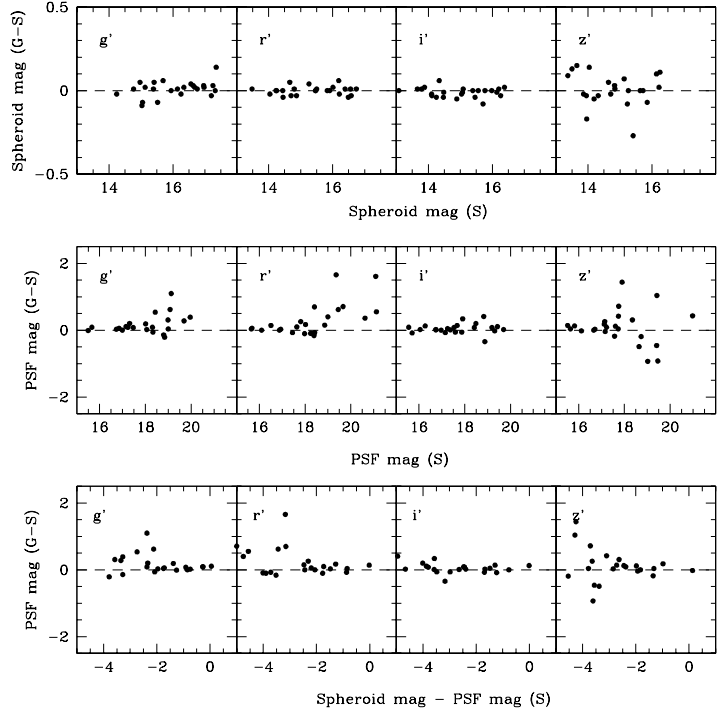}
\end{center}
\caption{Comparison between the parameters obtained using GALFIT (G) and SPASMOID (S). {\bf First row:}
Difference between spheroid magnitude determined using GALFIT and SPASMOID as a function of spheroid magnitude
(SPASMOID) for the four different SDSS filters. {\bf Second row:} The same as in the first row,
but for the PSF magnitude. {\bf Third row:} The same as in the second row, but as a function of difference
between spheroid and PSF magnitude.}
\label{mattgalfit1}
\end{figure*}

\begin{figure*}[ht!]
\begin{center}
\includegraphics[scale=0.5]{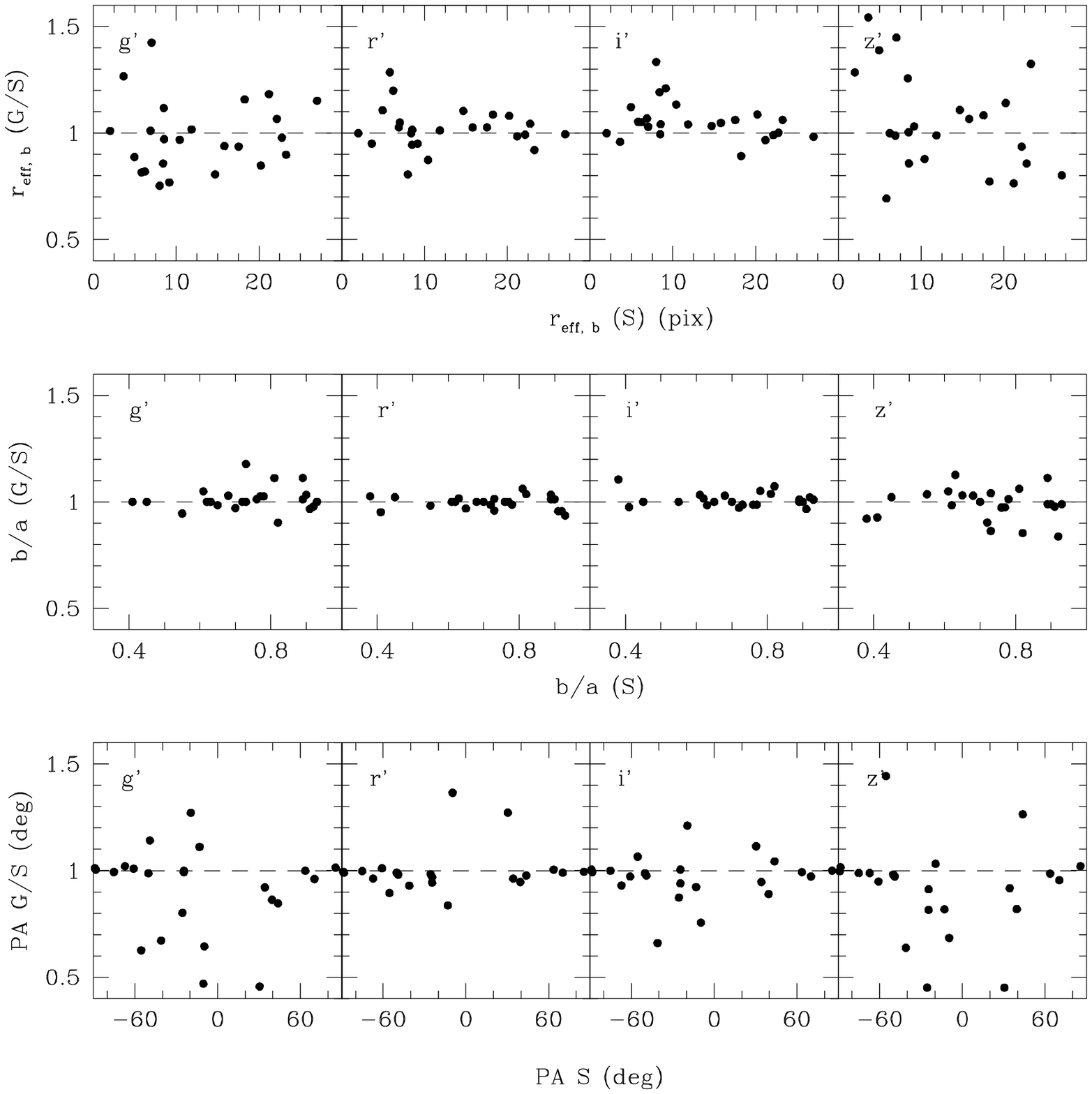}
\end{center}
\caption{Comparison between the parameters obtained using GALFIT (G) and SPASMOID (S). {\bf First row:}
Ratio of spheroid effective radius determined by GALFIT vs. SPASMOID 
as a function of spheroid effective radius determined by SPASMOID for the
four different SDSS filters. Note that only for GALFIT is the effective radius different
between the four different filters, but the same for all bands in SPASMOID.
 {\bf Second row:} The same as in the first row,
but for axis ratio b/a. {\bf Third row:} The same as in the second row, but for position angle.}
\label{mattgalfit2}
\end{figure*}

A comparison between the results from our fit and GALFIT is shown in Figs.~\ref{mattgalfit1}-\ref{mattgalfit2},
for the AGN+spheroid decomposition. Overall, the agreement is good, showing that our code
performs as expected. In detail, the spheroid magnitudes agree to within 
0.01$\pm$0.02 magnitudes (rms scatter: 0.08)
with the largest difference at the faint end. For the PSF magnitude, the difference
can be larger, due to a fainter PSF compared to the spheroid;
on average 0.36$\pm$0.01 (rms scatter: 0.33). 
We use this comparison for a conservative estimate of our error bars,
i.e.~spheroid and disk magnitude of 0.2 mag, PSF magnitude of 0.5 mag, and total
host magnitude of 0.1 mag. 
Note that we apply constraints between the different filters
which is not the case for GALFIT which is responsible for some of
the discrepancies, e.g.~in effective radius (Fig.~\ref{mattgalfit2}). 
While setting these constraints is less important for an AGN+spheroid fit,
it helps to solve the degeneracies for a three component fit.
Also, the PSF differs slightly between our code and GALFIT as we use a circular
Gaussian for GALFIT but allow for an elliptical Gaussian in our code.

\end{document}